\documentclass[aps,pra,twocolumn,superscriptaddress,10pt,longbibliography]{revtex4-2}
\usepackage{amsmath,amssymb,bm}
\usepackage{graphicx}
\usepackage{epstopdf}
\usepackage{latexsym}
\usepackage[normalem]{ulem} 
\usepackage[caption=false]{subfig}
\usepackage{hyperref}
\usepackage{mathtools}
\usepackage{natbib}
\usepackage{soul}
\usepackage{bbold}
\usepackage{physics}
\usepackage{lipsum}
\usepackage{enumitem}   
\usepackage{comment}
\usepackage{booktabs}
\usepackage{nicematrix}
\usepackage{makecell}
\usepackage[dvipsnames]{xcolor}

\begin{document}

\newcommand{\expt}[1]{\langle #1 \rangle}
\renewcommand{\mod}[1]{\lvert #1 \rvert}
\newcommand{\modsq}[1]{\mod{#1}^2}
\newcommand{\partialD}[2]{\frac{\partial #1}{\partial #2}}
\newcommand{\ns}{\mathcal{N}_{\mathrm{s}}}
\newcommand{\sinc}{\mathrm{sinc}}
\newcommand{\nb}{\mathcal{N}_{\mathrm{b}}}

\definecolor{cellgreen}{RGB}{153, 209, 112}
\definecolor{cellred}{RGB}{230,97,97}
\newcommand{\cellgreen}{\cellcolor{cellgreen!30}}
\newcommand{\cellred}{\cellcolor{cellred!30}}

\DeclarePairedDelimiter{\ceil}{\lceil}{\rceil}

\newcommand{\warn}[1]{{\color{red}\textbf{* #1 *}}}
\newcommand{\warntoedit}[1]{{\color{blue}\textbf{EDIT: #1 }}}
\newcommand{\warncite}[1]{{\color{green}\textbf{cite #1}}}

\newcommand{\Rev }[1]{{\color{black}{#1}\normalcolor}} 
\newcommand{\Com}[1]{{\color{red}{#1}\normalcolor}} 
\newcommand{\AddRef}[1]{{\color{Plum}{[REF]}\normalcolor}} 
\newcommand{\LBCom}[1]{{\color{blue}{LB: #1}\normalcolor}} 

\makeatletter
\def\maketitle{
\@author@finish
\title@column\titleblock@produce
\suppressfloats[t]}
\makeatother

\newcommand{\mytitle}{Optimal Displacement Sensing with Spin-Dependent Squeezed States}

\title{\mytitle}
\date{\today}

\newcommand{\affA}{Institute of Physics, University of Amsterdam, Science Park 904, 1098 XH Amsterdam, the Netherlands}
\newcommand{\affB}{QuSoft, Science Park 123, 1098 XG Amsterdam, the Netherlands}
\newcommand{\affC}{CWI, Science Park 904, 1098 XH Amsterdam, the 
Netherlands}
\newcommand{\affD}{Department of Physics, Indian Institute of Technology Madras, Chennai 600036, India}
\newcommand{\affE}{School of Physics, University of Sydney, NSW 2006, Australia}
\newcommand{\affF}{Sydney Nano Institute, University of Sydney, NSW 2006, Australia}
\newcommand{\affG}{Center for Quantum Information, Communication and Computing, Indian Institute of Technology Madras, Chennai 600036, India}

\title{\mytitle}
\date{\today}

\author{Liam~J.~Bond}\email[Contact author:]{L.J.Bond@uva.nl}\affiliation{\affA}\affiliation{\affB}
\author{Christophe~H.~Valahu}\affiliation{\affE}\affiliation{\affF}
\author{Athreya~Shankar}\affiliation{\affD}\affiliation{\affG}
\author{Ting~Rei~Tan}\affiliation{\affE}\affiliation{\affF}
\author{Arghavan~Safavi-Naini}\affiliation{\affA}\affiliation{\affB}

\begin{abstract}
Displacement sensing is a fundamental task in metrology. 
However, the development of quantum-enhanced sensors that fully utilize the available degrees of freedom in many-body quantum systems remains an outstanding challenge. 
We propose many-body displacement sensing schemes that use spin-dependent squeezed (SDS) states -- hybrid spin-boson states whose bosonic squeezed quadrature is conditioned on an auxiliary spin. 
We prove that SDS states are \emph{optimal}, i.e. their quantum Cram\'{e}r-Rao bound saturates the Heisenberg limit.
We propose explicit measurement sequences that can be readily implemented in systems such as trapped ions. 
We also introduce a scalable state-preparation protocol and numerically demonstrate the preparation of $8.7$~dB of spin-dependent squeezing $15$ times faster than the standard approach using second-order sidebands in trapped ions. 
The potential applications of our sensing protocols range from measuring single-photon scattering to searches for dark matter. 
\end{abstract}

\maketitle  

\section{Introduction}
Measuring the response of a system to estimate an unknown parameter underpins the development of science and technology. For example, the detection of gravitational waves at the Laser Interferometer Gravitational-Wave Observatory (LIGO) confirmed one of the most important predictions of the general theory of 
relativity~\cite{aasiEnhancedSensitivityLIGO2013,AbbottObservation}, while precision spectroscopy in atomic clocks has revolutionized timekeeping with fractional uncertainties now below $10^{-18}$~\cite{aeppliClock8102024,ludlowOpticalAtomicClocks2015}. At the same time, significant efforts have been devoted to using quantum systems for sensing theorized couplings between Standard Model fields and dark matter candidates such as dark photons, millicharged dark matter or axions~\cite{TURNER199067,budkerMillichargedDarkMatter2022,zhengQuantumenhancedDarkMatter2025}. Such couplings would induce ultra-small bosonic \emph{displacements} of a microwave field~\cite{duSearchInvisibleAxion2018,braggioQuantumEnhancedSensingAxion2025,zhengQuantumenhancedDarkMatter2025,dixitSearchingDarkMatter2021,backesQuantumEnhancedSearch2021} or a mechanical oscillator~\cite{kolkowitzCoherentSensingMechanical2012,schrepplerOpticallyMeasuringForce2014,gilmoreAmplitudeSensingZeroPoint2017,gilmoreQuantumenhancedSensingDisplacements2021,budkerMillichargedDarkMatter2022,delaneyMeasurementMotionQuantum2019}, which could be measured by a sufficiently sensitive displacement sensor. 

The sensitivity of the quantum sensor crucially depends on its initial reference state. Using a classical, uncorrelated reference state, the estimation variance is lower bounded by the so-called standard quantum limit (SQL). For displacement sensing, the SQL is defined by bosonic coherent states and is constant, $\rm{SQL} \propto \mathcal{O}(1)$. As such, the sensitivity cannot be improved by simply increasing the amplitude of the coherent state. By instead using non-classical reference states and quantum measurement protocols, the estimation variance is ultimately limited by Heisenberg scaling, $\rm{HS} \propto 1/\langle \hat{n} \rangle$, where $\langle \hat{n} \rangle$ is the reference state's average bosonic occupation~\cite{degenQuantumSensing2017}. This is known as \emph{quantum-enhanced sensing}. 

Previous works have considered a range of different bosonic reference states for quantum-enhanced displacement sensing. Examples of states that can achieve sub-SQL sensitivities are Fock states~\cite{wolfMotionalFockStates2019a,dengQuantumenhancedMetrologyLarge2024} and their (optimal) superpositions~\cite{grochowskiOptimalPhaseinsensitiveForce2025}, phase-twirled squeezed states~\cite{goreckiQuantumMetrologyNoisy2022} and two- and four-component cat states~\cite{munroWeakforceDetectionSuperposed2002,zhengQuantumenhancedDarkMatter2025}. Simultaneous sub-SQL sensing of a displacement's real and imaginary phase-space component can also be achieved using, for example, single-mode grid states~\cite{valahuQuantumenhancedMultiparameterSensing2025,duivenvoordenSingleModeDisplacementSensor2017,labarcaQuantumSensingDisplacements2025} and two-mode squeezed states~\cite{cardosoSuperpositionTwomodeSqueezed2021,genoniOptimalEstimationJoint2013,liMultiparameterQuantumMetrology2023}. 

However, the aforementioned reference states do not take full advantage of the available degrees of freedom in \emph{many-body} experimental platforms, which feature exquisite control over both spin and bosonic degrees of freedom~\cite{shawErasureCoolingControl2025,campagne-ibarcqQuantumErrorCorrection2020,fluhmannEncodingQubitTrappedion2019,sanerGeneratingArbitrarySuperpositions2024,penasaMeasurementMicrowaveField2016}. Such many-body systems are ideal quantum sensors as they offer the possibility of \emph{collective enhancement} for faster reference state preparation, or for amplification of the displacement signal. For example, spin-boson cat states are prepared $\sqrt{N}$ times faster in a trapped-ion system with $N$ ions, while electric field sensing benefits from a $\sqrt{N}$ amplification of the displacement signal when using $N$ ions~\cite{gilmoreQuantumenhancedSensingDisplacements2021}. While there are existing protocols for single-parameter displacement sensing in many-body systems~\cite{gilmoreQuantumenhancedSensingDisplacements2021,lewis-swanCavityQEDProtocolPrecise2020,barberenaAtomlightEntanglementPrecise2020}, there is a lack of sensing protocols that utilize the advantages of many-body quantum systems for estimating a displacement's amplitude, or jointly estimating its real and imaginary components.  

Here, we solve this problem by introducing many-body displacement sensing schemes using spin-dependent bosonic squeezed states -- hybrid spin-boson states whose bosonic squeezed quadrature is conditioned on the state of an auxiliary spin. 
We prove that such states are \emph{optimal} reference states for both amplitude and joint-parameter displacement sensing, with quantum Cram\'{e}r-Rao bounds (QCRBs) that saturate the Heisenberg limit. 
We develop explicit measurement protocols that follow Heisenberg scaling and can be readily implemented on existing quantum hardware platforms. We propose a scalable protocol for preparing spin-dependent squeezed states in trapped ions that requires only global addressing, and numerically demonstrate that for $N = 20$ ions our protocol achieves $8.7$~dB of spin-dependent squeezing $15$ times faster than the usual approach of driving the second-order sidebands. Our state-preparation protocol also has immediate applications to quantum computing with both discrete~\cite{katzBodyInteractionsTrapped2022b,katzDemonstrationThreeFourbody2023,shapiraRobustTwoQubitGates2023} and continuous~\cite{ayyashDrivenMultiphotonQubitresonator2024,hopePreparationConditionallysqueezedStates2025} variable systems. 
Finally, we prove that the metrological utility of sds states is insensitive to single-spin dephasing, and from numerics observe robustness to motional dephasing and heating. Although our explicit measurement protocols are less robust, when taking into account the collective enhancements that are available in trapped ions, we identify parameter regimes where the metrological utility of our protocols increases with the number of ions, realizing a collective enhancement. 

\begin{figure*}
    \centering
    \includegraphics[scale=1]{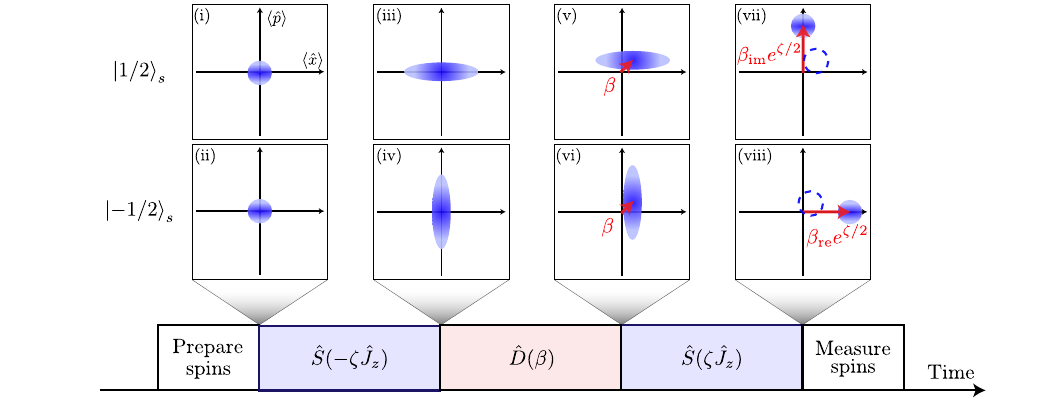}
    \caption
    {Protocol steps for phase-insensitive sensing of a displacement amplitude using spin-dependent squeezed states. Top (bottom) row shows for $N =1 $ spin, schematic representations of the bosonic Wigner function associated with the spin $\ket{1/2}_s$ ($\ket{-1/2}_s$) state. After initialization to $\ket{0}_b \ket{\psi_0}_s \cong \ket{0}_b (\ket{-1/2}_s + \ket{1/2}_s)$, (panels (i) and (ii)), spin-dependent squeezing $\hat{S}\mathopen{(}-\zeta \hat{J}_z\mathclose{)}$ squeezes the bosonic mode conditioned on the spin state (panels (iii) and (iv)). The bosonic mode then undergoes an unknown displacement $\hat{D}(\beta)$ (panels (v) and (vi)). Finally, the spin-dependent squeezing is reversed (panels (vii) and (viii)). The signal's imaginary part, $\beta_{\rm im}$, is exponentially amplified in the $\ket{1/2}_s$ Hilbert space sector, while the real part, $\beta_{\rm re}$, is amplified in the $\ket{-1/2}_s$ sector. Unlike the phase-insensitive amplification protocol of Ref.~\cite{burdExperimentalSpeedupQuantum2024}, here the signal amplification is spin-dependent. 
    Phase-insensitive information about the displacement is accessed via a spin measurement. Without spin-dependent squeezing, the final state is a purely bosonic displaced state (blue dashed circles). 
    }
    \label{fig:phase_insensitive_schematic}
\end{figure*}

\section{Displacement sensing}\label{sec:Sensing}

Consider a system of $N$ spins and a single bosonic mode, prepared in a reference state, $\ket{\psi_{\rm ref}}$. The reference state is subjected to a small, coherent displacement by an unknown amount $\beta \in \mathbb{C}$, described by the displacement operator, $\hat{D}(\beta) = \exp(\beta \hat{a}^\dagger - \beta^* \hat{a})$. The resulting displaced state is,
\begin{align}
    \ket{\psi(\bm{\theta})} = \hat{D}(\beta) \ket{\psi_{\rm ref}},
\end{align}
which encodes a set of $d$ real parameters, $\bm{\theta} = \{\theta_1, \theta_2, \dots, \theta_d\}$. Here, $\bm{\theta}$ is related to $\beta$. We review quantum multi-parameter metrology in more detail in the Supplemental Material~\cite{supp}\nocite{armanGeneratingOverlapCompass2024,bressaniniMultiparameterQuantumEstimation2024,demkowicz-dobrzanskiMultiparameterEstimationQuantum2020,Fadel_2025,Gottesman2001,sandersSuperpositionTwoSqueezed1989,toscanoSubPlanckPhasespaceStructures2006a}. 

Our goal is to estimate $\bm{\theta}$ from a set of $M$ independent measurement outcomes, $\bm{x} = \{x_1,\dots,x_M\}$. The uncertainty of the estimation is quantified by the covariance matrix, $\bm{V}(\bm{\theta}) = \sum_{\bm{x}} p(\bm{x}|\bm{\theta})[\bm{\theta} - \tilde{\bm{\theta}}(\bm{x})][\bm{\theta} - \tilde{\bm{\theta}}(\bm{x})]^T$, where $p(\bm{x}|\bm{\theta})$ are the measurement probabilities and $\tilde{\bm{\theta}}(\bm{x})$ an unbiased estimator. 

For a chosen reference state and measurement observable, minimizing over all possible unbiased estimators lower bounds $\bm{V}$ according to the classical Cram\'{e}r-Rao bound (CCRB),
\begin{align}
    \Tr(\bm{V}) \geq \Tr(\bm{F}^{-1}),
\end{align}
where we set $M = 1$ and where the classical Fisher information matrix $\bm{F}$ is, 
\begin{align}
    F_{ij} = \sum_k \frac{1}{p(x_k|\bm{\theta})} [\partial_i p(x_k|\bm{\theta})][\partial_j p(x_k|\bm{\theta})],
\end{align}
with $\partial_i = \partial/\partial \theta_i$. Further optimizing over all possible measurement observables, $\bm{V}$ is lower bounded by the QCRB~\cite{pezzeAdvancesMultiparameterQuantum2025,liuQuantumFisherInformation2020},
\begin{align}
    \Tr(\bm{V}) \geq \Tr(\bm{Q}^{-1}),
    \label{eq:QCRB}
\end{align}
where the quantum Fisher information matrix $\bm{Q}$ is,
\begin{align}
    Q_{ij} = \frac{1}{2}\bra{\psi(\bm{\theta})} \{\hat{L}_i,\hat{L}_j\} \ket{\psi(\bm{\theta}},
\end{align}
with $\hat{L}_i$ the symmetric logarithmic derivative~\cite{supp}. In general, finding explicit measurement observables whose classical Fisher information matrix saturates the quantum Fisher information matrix is a non-trivial task. 

Typically, displacement sensing is performed in the paradigm of metrology by repetition, in which a single experimental shot -- consisting of reference state preparation, linear interaction with the signal, and measurement -- is repeated many times. It is therefore important to establish whether the relative phase between the reference state and the unknown displacement varies between experimental shots. This leads to two settings:  
\begin{enumerate}[label=(\roman*)]
    \item \textit{Phase-insensitive sensing of the displacement's magnitude}: If the phase varies between experimental shots, we can only estimate the displacement magnitude, so $\bm{\theta} = |\beta|$. 
    \item \textit{Single- or multi-parameter sensing}: If the phase is fixed between shots, we can decompose $\beta = \beta_{\rm re} + i \beta_{\rm im}$. We can estimate a single parameter, so $\bm{\theta} = \{\beta_{\rm re}\}$ or $\bm{\theta} = \{\beta_{\rm 
    im}\}$, or jointly estimate both parameters, so $\bm{\theta} = \{\beta_{\rm re},\beta_{\rm im}\}$. 
\end{enumerate}

Minimizing the QCRB over all possible reference states yields the Heisenberg limit (HL). Here, we take the minimization to be over all possible single-mode bosonic reference states, see the Supplemental Material~\cite{supp}. In terms of the reference state's average mode occupation $\langle \hat{n} \rangle$, for single-parameter displacement sensing ${\rm HL} = 1/(16 \langle \hat{n} \rangle + 4)$, for phase-insensitive amplitude sensing ${\rm HL} = 1/(8 \langle \hat{n} \rangle + 4)$ and for multi-parameter displacement sensing ${\rm HL} = 1/(4 \langle \hat{n} \rangle + 2)$. If a metrology protocol's CCRB or QCRB equals the HL lower bound, we say that it ``saturates the Heisenberg limit'', and is therefore ``optimal''. If it falls inversely with $\langle \hat{n} \rangle$ but with a larger constant prefactor than that of the HL, we say that it ``follows Heisenberg scaling''. 

\subsection{Spin-dependent squeezed states}
In this work, we consider reference states of the form,
\begin{align}
    \ket{\psi_{\rm ref}} = \hat{U}\ket{0}_b \ket{\psi_0}_s.
\end{align}
Here, $\ket{0}_b$ is the bosonic vacuum, and $\ket{\psi_0}_s=\sum_{m=-N/2}^{N/2} c_m \ket{m}_s$ is a collective spin state written with coefficients $c_m \in \mathbb{C}$ in the basis of Dicke states, $\hat J_z\ket{m}_s = m \ket{m}_s$, where the magnetization $m$ is (half-)integer $-N/2\leq m \leq N/2$, and where the collective spin operators are $\hat J_\alpha= \sum_{i=1}^{N} \hat \sigma_{i}^\alpha/2 $ with $\alpha=x, y, z$. The unitary $\hat{U}$ entangles the spin and bosonic degrees of freedom; possible choices that have been previously studied include spin-dependent displacements~\cite{gilmoreQuantumenhancedSensingDisplacements2021,zhengQuantumenhancedDarkMatter2025} or dispersive spin-boson interactions~\cite{barberenaAtomlightEntanglementPrecise2020,lewis-swanCavityQEDProtocolPrecise2020}. 

We ask, for each of the metrology settings defined above, what is the optimal reference state? In Fig.~\ref{fig:phase_insensitive_schematic}(a) we sketch the Wigner function for one such state, which is generated by spin-dependent squeezing~\cite{katzBodyInteractionsTrapped2022b,Sutherland2021PRA},
\begin{align}
        \hat{S}(\zeta \hat{J}_z) = \exp\left(\frac{1}{2}(\zeta^* \hat{a}^2 - \zeta \hat{a}^{\dagger2} ) \hat{J}_z \right)   \label{eq:SDS},
\end{align}
where $\zeta$ is the squeezing parameter. The creation and annihilation operators, $\hat{a}^\dagger$ and $\hat{a}$, are related to position and momentum operators via $\hat{x} = \hat{a}^\dagger + \hat{a}$ and $\hat{p} = i(\hat{a}^\dagger - \hat{a})$, respectively. Spin-dependent squeezing is also known as conditional squeeze~\cite{ayyashDrivenMultiphotonQubitresonator2024,ayyashDispersiveRegimeMultiphoton2025,liuHybridOscillatorQubitQuantum2025} or control-squeeze~\cite{drechslerStateDependentMotional2020,delgrossoControlledsqueezeGateSuperconducting2025}. 
In Sec.~\ref{sec:SDSPreparation} we show how it can be implemented in a fast, scalable manner in trapped-ion systems. 
We write the reference state obtained from spin-dependent squeezing as, 
\begin{align}
    \ket{\psi_{\rm ref}} = \hat{S}(\zeta \hat{J}_z) \ket{0}_b \ket{\psi_0}_s = \sum_{m=-N/2}^{N/2} c_m \ket{\zeta m}_b \ket{m}_s, 
\end{align}
where the bosonic squeezed state is defined as,
\begin{align}
    \ket{\zeta m}_b = \hat{S}(\zeta m)\ket{0}_b =  e^{\zeta m (\hat a^2 - \hat a^{\dagger2})/2} \ket{0}_b
\end{align}
Without loss of generality we set $\zeta \in \mathbb{R}$, which is equivalent to setting the initial phase-space axis. 

For $\ket{\psi_0}_s$, we focus on two particular choices: the Greenberger–Horne–Zeilinger (GHZ) state~\cite{greenbergerGoingBellsTheorem1989} and a spin coherent state (SC) pointing along $x$ of the collective Bloch sphere, 
\begin{subequations}\begin{align}
    \ket{\psi_{\rm GHZ}}_s &= \frac{1}{\sqrt{2}} (\ket{-N/2}_s + \ket{N/2}_s), \label{eq:psiGHZ} \\ 
    \ket{\psi_{\rm SC}}_s &= e^{i \pi/2 \hat{J}_y}  \ket{-N/2}_s = \sum_{m=-N/2}^{N/2} c_m \ket{m}_s,  \label{eq:psicoh}
\end{align}\end{subequations}
where $c_m = 2^{-N/2}\sqrt{\binom{N}{m+N/2}}$ for the spin coherent state. It is instructive to visualize the reference states generated by spin-dependent squeezing from each of these initial collective spin states. For $N = 1$ spin, both spin states are the same. In Fig.~\ref{fig:phase_insensitive_schematic}(a) we separately plot the Wigner functions for the $m = 1/2$ (top row) and $m = -1/2$ (bottom row) Hilbert space sectors. The bosonic state is squeezed along orthogonal directions (panels (iii) and (iv)), which indicates utility for phase-insensitive sensing of a displacement amplitude. 

\subsection{Phase-insensitive displacement sensing}
\label{sec:phase_insensitive_displacement_sensing}
We begin in the setting where the phase between displacement and reference state varies between shots, which limits us to sensing the magnitude, $|\beta|$. The metrological utility is quantified by the quantum Fisher information (QFI), cf. the QCRB of Eq.~\ref{eq:QCRB}. 

We consider any initial collective spin state with symmetric Dicke weights, i.e. $|c_{m}| = |c_{-m}|$. As we derive in the Supplemental Material~\cite{supp}, the QFI is completely independent of $\beta$, 
\begin{align}
    Q_{|\beta|} = 8 \sum_{m>0}^{N/2}|c_m|^2 \cosh(2\zeta m),
\end{align}
and grows exponentially with squeezing $\zeta$, indicating metrological utility. The corresponding QCRB is,
\begin{align}
    V(|\beta|) = \frac{1}{Q_{|\beta|}} \geq \frac{1}{8 \langle \hat{n} \rangle + 4}, \label{eq:QCRB_AbsBeta}
\end{align}
where $\langle \hat{n} \rangle = \langle \psi_{\rm ref} |\hat{n} | \psi_{\rm ref} \rangle$. The Heisenberg limit for the estimation of $|\beta|$ is $1/(8 \langle \hat{n} \rangle + 4)$~\cite{supp}, and therefore the QCRB saturates the Heisenberg limit. As such, as a function of the available resources (i.e. the mode occupation), the reference state is optimal for phase-insensitive estimation of the displacement's amplitude $|\beta|$. 

\subsubsection*{Measurement protocol}

Because the displaced reference state is spin-boson entangled, the metrological advantage is in a mixed spin-boson quadrature, which is often challenging to measure experimentally. This problem is solved by mapping the information to the spin degree of freedom using a time-reversal sequence, shown in Fig.~\ref{fig:phase_insensitive_schematic}(a). The sequence braids to a displacement that depends exponentially on the spin, 
\begin{align}
    \hat{S}(\zeta \hat{J}_z) \hat{D}(\beta) \hat{S}(-\zeta \hat{J}_z) = \hat{D}(\hat{\beta}), \label{eq:braiding}
\end{align}
where $\hat{\beta} = \beta \cosh\mathopen{(}\zeta \hat{J}_z\mathclose{)} - \beta^* \sinh\mathopen{(}\zeta \hat{J}_z\mathclose{)}$. 

We start by considering the sequence for a single spin initialized in,
\begin{align}
\ket{0}_b\ket{\psi_0}_s \simeq \ket{0}_b(\ket{-1/2}_s + \ket{1/2}_s).
\end{align}
The final state is,
\begin{align}
    \ket{\psi_f} \simeq \hat{D}(\beta_-)|0\rangle_b \ket{-1/2}_s + \hat{D}(\beta_+)|0\rangle_b \ket{1/2}_s,
\end{align}
where the spin-dependent displacement amplitudes are $\beta_{-} = \beta_{\rm re} \exp(\zeta/2) + i \beta_{\rm im}  \exp(-\zeta/2)$ and $\beta_{+} = \beta_{\rm re} \exp(-\zeta/2) + i \beta_{\rm im} \exp(\zeta/2)$. The parameter $\beta_{\rm re}$ is exponentially amplified in the spin down Hilbert space sector, while $\beta_{\rm im}$ is exponentially amplified in the spin up sector (see panels (vii) and (viii) of Fig.~\ref{fig:phase_insensitive_schematic}(a)). For small displacements $|\beta| \ll 1$, after a $\pi/2$ spin rotation about $\hat{\sigma}_y$, the probability of measuring $\ket{1/2}_s$ is,
\begin{align}
    P_{1/2} = 1 - |\beta|^2 \sinh(\zeta/2)^2 + \mathcal{O}(\beta^4),
\end{align}
which is phase-insensitive up to third order in $\beta$. In the limit $\beta \rightarrow 0$, the classical Fisher information (CFI) is,
\begin{align}
    F_{|\beta|} = 4\sinh(\zeta/2)^2 = 4\langle \hat{n} \rangle, \label{eq:CFI_sds_1Q}
\end{align}
and therefore the CCRB is,
\begin{align}
    V(|\beta|) \geq \frac{1}{F_{|\beta|}} = \frac{1}{4\langle \hat{n} \rangle},
    \label{eq:1Q_CCRB}
\end{align}
which follows Heisenberg scaling, $\propto 1/\langle \hat{n} \rangle$. 

\begin{figure}
    \centering
    \includegraphics[scale=1]{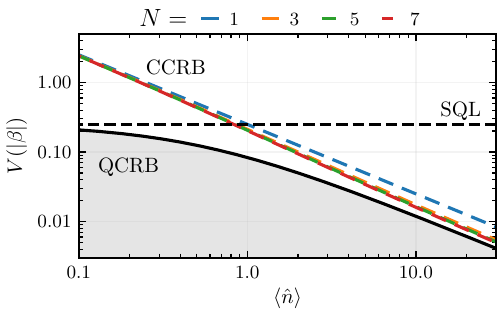}
    \caption{Metrological performance of spin-dependent squeezed spin coherent states, for phase-insensitive estimation of the displacement's magnitude, $|\beta|$. 
    The classical Cram\'{e}r-Rao bound (CCRB) for the time-reversal protocol and collective spin measurement (described in the main text) is plotted for increasing numbers of spins, $N$ (solid colored lines).
    The CCRB is below the standard quantum limit (SQL, dashed black line) for all $N$ at $\langle \hat{n} \rangle \gtrsim 1$. Moreover, the CCRB for $N=1$ is $10$~dB below the SQL at $\langle \hat{n} \rangle = 10$. 
    The SQL is defined as the QCRB for bosonic coherent states, which for sensing of $|\beta|$ is $\rm{SQL} = 1/4$~\cite{supp}. 
    The CCRB is further compared to the quantum Cram\'{e}r-Rao bound (QCRB, solid black line).
    At $\langle \hat{n} \rangle \gtrsim 1$, the CCRB for all $N$ follows Heisenberg scaling, $\propto 1/\langle \hat{n} \rangle$. In this same regime, the CCRB for $N=1$ spin is a factor of two larger than the QCRB. The prefactor decreases to one in the $N \rightarrow \infty$ limit.
    }
    \label{fig:phase_insensitive_metrology}
\end{figure}

The metrological performance of our measurement protocol with a single spin is shown in Fig.~\ref{fig:phase_insensitive_metrology}, where we plot the CCRB (blue line) and compare it against the QCRB (black line) of Eq.~\ref{eq:QCRB_AbsBeta}. The CCRB diverges from the QCRB as $\langle \hat{n} \rangle \rightarrow 0$ because no information is mapped to the spin. For $\langle \hat{n} \rangle \gtrsim 1$, the CCRB follows QCRB scaling but remains a factor of two larger than the Heisenberg limit. When $\langle \hat{n} \rangle \gtrsim 1$ the CCRB falls below the SQL, which is defined by the use of bosonic coherent states as reference states, $\rm{SQL} = 1/4$~\cite{supp}. 

We next generalize the measurement protocol to $N$ spins, which can potentially offer a beneficial collective enhancement. We begin with the case of the spins in a GHZ state. We assume that the GHZ state is prepared using a one-axis twisting (OAT) Hamiltonian evolving for time $\pi/2$, 
\begin{align}
    \hat{U}_{\rm OAT} = \exp( - \frac{\pi}{2} \hat{J}_x^2 ). \label{eq:U_OAT}
\end{align}
For the case of even $N$, after initializing all spins in spin down, evolving under $\hat{U}_{\rm OAT}$ produces $\hat{U}_{\rm OAT} \ket{-N/2}_s = 1/\sqrt{2} ( e^{-i\pi/4} \ket{-N/2} + e^{i \pi/4 + i \pi N/2} \ket{N/2})$, which is the same form as the GHZ state of Eq.~\ref{eq:psiGHZ} but with different relative phases. After applying the time-reversal protocol of Eq.~\ref{eq:braiding}, we reverse the one-axis twisting and measure in the Dicke basis (see Supplemental Material~\cite{supp} for details). The only two states with nonzero measurement probabilities are $P_{N/2}$ and $P_{-N/2} = 1-P_{N/2}$. For small displacements $\beta \ll 1$, they are, 
\begin{align}
    P_{N/2} = |\beta|^2 \sinh(N\zeta/2)^2 + \mathcal{O}(\beta^4).
\end{align}
The corresponding CFI in the limit $\beta \rightarrow 0$ is,
\begin{align}
    F_{|\beta|} = 4 \sinh(N \zeta/2)^2 = 4 \langle \hat{n} \rangle. \label{eq:CFI_sds_GHZ}
\end{align}
As a function of $\langle \hat{n} \rangle$, this many-spin CFI is identical to the single-spin result of Eq.~\ref{eq:1Q_CCRB}. The advantage of using $N$ spins lies in a potential collective enhancement. For example, in trapped ions there is a $\sqrt{N}$ amplification of the displacement signal, as well as a $\sqrt{N}$ reduction in state-preparation time, as we show later in Sec.~\ref{sec:SDSPreparation}.

Next, we consider $N$ spins initialized in the spin coherent state of Eq.~\ref{eq:psicoh}. For small displacements $\beta \ll 1$, after a collective $\pi/2$ spin rotation about $\hat{J}_y$, the probability of measuring each Dicke state is phase-insensitive up to third order in $\beta$ (see Supplemental Material~\cite{supp} for derivation). The exact CFI is of the form, 
\begin{align}
    \label{eq:general_cfi_coherent_spin}
    F_{|\beta|} = \sinh(\zeta/2)^2   \sum_{i=0}^{N-1} a_i \cosh(i \zeta),
\end{align}
where the coefficients $a_i$ are specified in the Supplemental Material~\cite{supp}. The metrological performance is shown in Fig.~\ref{fig:phase_insensitive_metrology}, where we plot the CCRB, $1/F_{|\beta|}$, for increasing $N$, as well as the QCRB (black line) of Eq.~\ref{eq:QCRB_AbsBeta} which is independent of $N$. For $\langle \hat{n} \rangle \approx 1$, the CCRB is a factor of approximately two larger than the QCRB. The CCRB is below the SQL (black dashed line) for $\langle \hat{n} \rangle \gtrsim 1$. Taking the limit $\langle \hat{n} \rangle \rightarrow \infty$, the ratio between the CFI and QFI is, 
\begin{align}
    \frac{Q_{|\beta|}}{F_{|\beta|}} = \frac{2^N}{2^N-1}, \label{eq:Q_F_Ratio}
\end{align}
which approaches one in the limit $N \rightarrow \infty$. Therefore, in this extreme regime, the CCRB for phase-insensitive sensing of the displacement's amplitude with spin-dependent squeezed spin coherent states saturates the Heisenberg limit~\cite{supp}. 

\subsection{Multi-parameter displacement sensing}

Next, we consider the setting where the phase is fixed between experimental shots, and perform joint estimation of $\beta_{\rm re}$ and $\beta_{\rm im}$. 
For any symmetric Dicke weights $|c_{-m}| = |c_{m}|$, we prove in the Supplemental Material~\cite{supp} that the quantum Fisher information matrix is, 
\begin{align}
    \mathbf{Q} = 8  \sum_{m>0}^{N/2} \left[|{c}_m|^2 \cosh(2\zeta m) \right] \mathbb{1} = (8 \langle \hat{n} \rangle + 4)\mathbb{1}. \label{eq:QFIM_MP_SDS}
\end{align}
The corresponding multi-parameter QCRB is, 
\begin{align}
    V(\beta_{\rm{re}}) + V(\beta_{\rm{im}}) \geq \frac{1}{4\langle \hat{n} \rangle +2}, \label{eq:QCRB_MP_SDS}
\end{align}
which saturates the multi-parameter Heisenberg limit, and therefore symmetric Dicke weight spin-dependent squeezed states are optimal reference states. 
In the Supplemental Material~\cite{supp}, we further study the multi-parameter QCRB as a function of the squeezing parameter $\zeta$, and show that reference states obtained from GHZ spin states outperform those obtained from sin coherent states, because the mode occupation $\langle \hat{n} \rangle$ as a function of $\zeta$ is higher.

When jointly estimating two or more parameters, the optimal measurement for each parameter may be different, and the set of optimal measurements may be incompatible. If this is the case, the attainable estimation uncertainty is determined by the Holevo-Cram\'{e}r-Rao bound (HCRB). The HCRB is at most two times larger than the QCRB, ${\rm QCRB} \leq {\rm HCRB} \leq (1+R) {\rm QCRB}$, where $R \in [0,1]$ is the asymptotic incompatibility (see the Supplemental Material~\cite{supp} and Ref.~\cite{carolloQuantumnessMultiparameterQuantum2019}). For symmetric Dicke weight spin-dependent squeezed states, we calculate the asymptotic incompatibility,
\begin{align}
    R = \frac{1}{2\langle \hat{n} \rangle + 1}, \label{eq:MainTextAsymptoticAttainability}
\end{align}
and so the HCRB approaches the QCRB in the infinite squeezing limit because $R \rightarrow 0$ when $\langle \hat{n} \rangle \rightarrow \infty$. 

\subsubsection*{Measurement protocol}

\begin{figure}
    \centering
    \includegraphics[scale=1.0]{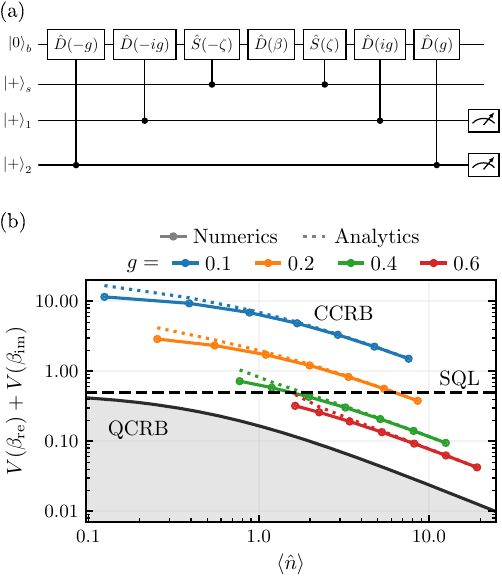}
    \caption{
    Joint estimation of the real and imaginary components of a displacement signal using spin-dependent squeezed states. 
    \textbf{(a)} Example measurement protocol for joint parameter displacement estimation using spin-dependent squeezing, which uses two additional ancillary spins initialized in $\ket{+}_\mathrm{1}$ and $\ket{+}_\mathrm{2}$, respectively.
    Spin-dependent displacements conditioned on each ancilla map information about the real and imaginary components of the displacement signal to the ancillas, which are then individually measured. 
    \textbf{(b)} Metrological performance of the measurement scheme for joint displacement parameter estimation. The classical Cram\'{e}r-Rao bound (CCRB, colored lines) is shown for $N = 1$ spin at various spin-dependent displacement strengths $g$. For each value of $g$, we vary the squeezing parameter in the range $\zeta \in [0.2,1.4]$ in steps of $0.2$. Solid lines are obtained from numerics, dashed lines are analytics of Eq.~\ref{eq:CCRB_MP}. 
    The CCRB is compared to the standard quantum limit (dashed black line), defined by the QCRB for bosonic coherent states, $\rm{SQL}=1/2$~\cite{supp}. 
    For $g = 0.6$, the CCRB is below the SQL for mode occupations $\langle \hat{n} \rangle \gtrsim 1$. 
    For fixed $\langle \hat{n} \rangle$, the CCRB approaches the quantum Cram{\'e}r-Rao Bound (QCRB, solid black line) as $g$ increases.
    }
    \label{fig:multiparam_metrology}
\end{figure}

Similar to the phase-insensitive setting, accessing the metrological advantage of the mixed spin-boson quadrature to perform multi-parameter displacement sensing is challenging. Here, we provide one possible measurement protocol that uses two ancilla qubits. As shown in Fig.~\ref{fig:multiparam_metrology}(a), the sequence is built by wrapping the time-reversal sequence in spin-dependent displacements of strength $g$ that are conditioned on the state of each ancilla. Each ancilla is then measured simultaneously and independently in the Dicke basis. 

Our analysis focuses on the single spin case. When the spin-dependent squeezing is much larger than both the spin-dependent displacement, $\zeta \gg g$, and the displacement signal, $\zeta \gg \beta$, information about the displacement's real component is mapped only to the first ancilla, while information about the imaginary component is mapped only to the second ancilla. The probability of measuring the $i$th ancilla in the $\ket{+}$ state is then (see Supplemental Material~\cite{supp}), 
\begin{align}
    P_{+}^{(i)} = \frac{1}{2} [1 + \cos^2(\phi_i)], 
\end{align}
where $\phi_1 = 2 g \exp(\zeta) \beta_{\rm re}$, and $\phi_2 = -2 g \exp(\zeta) \beta_{\rm im}$. In the limit $\beta \rightarrow 0$, the CFI matrix is $\mathbf{F} = 8 g^2 e^{2 \zeta} \mathbb{1}$. The multi-parameter variance is therefore lower bounded by the CCRB according to, 
\begin{align}
    V(\beta_{\rm re}) + V(\beta_{\rm im}) \geq \frac{1}{4 g^2 e^{2 \zeta}}. \label{eq:CCRB_MP}
\end{align}
In Fig.~\ref{fig:multiparam_metrology}(b), we show for various $g$ the CCRB obtained from the approximate analytic expression of Eq.~\ref{eq:CCRB_MP} (dotted lines) and from numerical simulations of the measurement sequence (solid colored dots). Due to the additional spin-dependent displacements in the time-reversal protocol, the reference state, i.e. the state before exposure to the signal, becomes $\ket{\psi_{\rm ref}} = \hat{S}(-\zeta \hat{J}_z) \hat{D}(-ig \hat{\sigma}_z^{(1)})\hat{D}(-g \hat{\sigma}_z^{(2)})\ket{0}_b \ket{\psi_0}_s \ket{+}_1 \ket{+}_2$, which is the state used to calculate $\langle \hat{n} \rangle$. 
For all curves, when $\zeta \gg g$ we observe excellent agreement between theory and numerics. For increasing $g$, the curves are shifted to the right due to the larger $\langle \hat{n} \rangle$. For $\langle \hat{n} \rangle \gtrsim 1$, the CCRBs follow the Heisenberg scaling of the QCRB (black line), with a prefactor that decreases with increasing $g$. At $g = 0.4$, the estimation uncertainty is below the SQL (black dashed line) when $\langle \hat{n} \rangle \gtrsim 1.5$.

\section{Engineering spin-dependent squeezing}\label{sec:SDSPreparation}

In this section we introduce a protocol for the fast preparation of spin-dependent squeezed states in many-ion systems. We simultaneously drive the first-order red sideband and blue sideband, each with two tones (see Fig~\ref{fig:sds_preparation_protocol}(b). By stroboscopically varying the phases and detunings of each tone (see Fig.~\ref{fig:sds_preparation_protocol}(c) and (d)), we realize spin-dependent squeezing while minimizing undesired error terms. The duration of our protocol scales favorably as $1/\sqrt{N}$, and can be readily realized on current trapped-ion platforms. 

\subsection{Experimental protocol for fast and scalable spin-dependent squeezing}

\begin{figure}
\includegraphics{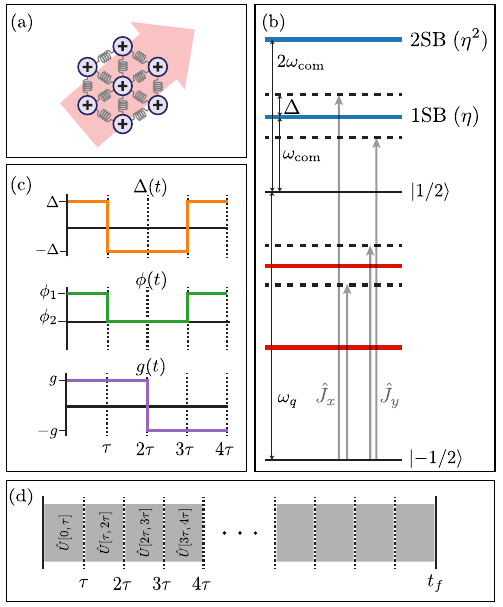}
\caption{
Experimental protocol for engineering fast and scalable spin-dependent squeezing in a trapped ion system.
\textbf{(a)} We consider a system of $N$ trapped ions coupled via laser interactions to their collective motional mode. \textbf{(b)} Energy diagram for $N = 1$ spin, without loss of generality. Multiple laser tones are tuned close to the first-order sidebands (1SB) that are $\pm \omega_{\rm COM}$ from the carrier transition $\omega_q$. A first bichromatic field driving interactions in the $\hat{J}_x$ basis is detuned by $\pm (\omega_{\rm COM} + \Delta)$, while a second bichromatic field driving interactions in the $\hat{J}_y$ basis is detuned by $\pm (\omega_\mathrm{COM} - \Delta)$ (gray lines). The resulting effective spin-dependent squeezing has an interaction strength $\propto \Omega^2 \eta^2/\Delta^2$, offering a speed-up for appropriately chosen $\Delta$ over alternative approaches that use second-order sideband (2SB) interactions with a weaker coupling strength $\propto \Omega \eta^2$. \textbf{(c)} The laser detuning, phase and interaction strengths are dynamically modulated to obtain the desired spin-dependent squeezing interaction and to minimize unwanted errors from higher-order terms. \textbf{(d)} Schematic of the stroboscopic pulse sequence. Each segment of duration $\tau$ requires evolving under $\hat{H}(t)$ of Eq.~\ref{eq:Hamiltonian} with the parameters shown in panel (c). The four segments are repeated $P$ times, with the total protocol duration $t_f$. 
}
\label{fig:sds_preparation_protocol}
\end{figure}

Consider a system of $N$ ions, each with an internal spin-$1/2$ degree of freedom with transition frequency $\omega_q$, see Fig.~\ref{fig:sds_preparation_protocol}(a). The combined action of the confining potentials and the Coulomb repulsion supports a set of motional normal modes, which are coupled to the spin degree of freedom via laser-induced interactions. Setting the laser frequency to $\omega_L = \omega_q + \omega_{\rm COM}$ or $\omega_L = \omega_q - \omega_{\rm COM}$ drives the first-order blue- or red-sideband, respectively, see Fig.~\ref{fig:sds_preparation_protocol}(b). Here, $\omega_{\rm COM}$ is the frequency of the center-of-mass (COM) mode, which describes uniform excitations of the ions. The strength of the first-order sidebands is linearly proportional to the Lamb-Dicke parameter $\eta = k x_0$, where $k$ is the wave vector and $x_0 = \sqrt{1/(2m\omega)}$ is the harmonic oscillator length scale with $m$ the ion mass, and where we set $\hbar = 1$. Setting $\omega_L = \omega_q + 2\omega_{\rm com}$ or $\omega_L = \omega_q - 2 \omega_{\rm com}$ drives the second-order blue or red sideband, respectively, with strength quadratically proportional to the Lamb-Dicke parameter, $\propto \eta^2$. 

Spin-dependent squeezing is commonly realized by simultaneously driving the second-order red and blue sidebands~\cite{Wineland1998,katzBodyInteractionsTrapped2022b,katzDemonstrationThreeFourbody2023}. However, because trapped-ion experiments typically operate in the Lamb-Dicke regime where $\eta \ll 1$, the interaction strength is weak. Here, we realize spin-dependent squeezing with a stronger effective interaction strength by adapting the single-spin protocol of Ref.~\cite{Sutherland2021PRA}, with experimental demonstrations in Refs.~\cite{sanerGeneratingArbitrarySuperpositions2024,bazavanSqueezingTrisqueezingQuadsqueezing2024}, to a many-spin system. 

Our spin-dependent squeezing protocol is realized by simultaneously driving the first-order sidebands at four different frequencies: one pair of red- and blue-sideband interactions with detuning $ \Delta(t)$ and phase of zero, and another pair of red- and blue-sideband interactions with detuning $- \Delta(t)$ and phase $\phi(t)$, see Fig.~\ref{fig:sds_preparation_protocol}(b). Performing a rotating wave approximation and moving to the interaction picture with respect to the spin and motion, the interaction Hamiltonian is~\cite{supp}, 
\begin{align}
    \hat{H}(t) = g(t) \hat{a} \left[ \hat{J}_x e^{-i t\Delta(t)} + \hat{J}_y e^{i t \Delta(t)} e^{-i \phi(t)} \right] + \text{h.c.}, \label{eq:Hamiltonian}
\end{align}
where $g(t) = \eta \Omega(t)/\sqrt{N}$ is the coupling strength between the collective spin and the bosonic mode, and $\Omega(t)$ is the Rabi frequency. To obtain $\hat{H}(t)$, we neglected off-resonant coupling of the sideband interactions to other motional modes, under the assumption that $\Delta(t)$ is small with respect to the frequency spacing between motional modes. 
Furthermore, we note that time-dependent control over $\Delta(t)$ and $\phi(t)$ has been previously demonstrated in trapped-ion experiments, see e.g.~\cite{Leung2018, Milne2020}. 
In our companion work~\cite{Bondinprep}, we prove that $H(t)$ provides universal control over the collective spin-boson Hilbert space, which was similarly shown for the single-spin case in Ref.~\cite{Sutherland2021PRA}. 

The real-time dynamics generated by $\hat{H}(t)$ during evolution from $t_0$ to $t$ are described by the Magnus expansion~\cite{magnusExponentialSolutionDifferential1954},
\begin{align}
    \hat{U}[t_0,t] = \exp \left(-i \hat{\Theta}_1[t_0,t] - \frac{1}{2}\hat{\Theta}_2[t_0,t] + \dots  \right), 
\end{align}
where the first- and second-order terms are, respectively, 
\begin{subequations}
\begin{align}
        \hat{\Theta}_1[t_0,t] &= \int_{t_0}^t dt_1 \hat{H}(t_1), \\ 
        \hat{\Theta}_2[t_0,t] &= \int_{t_0}^t dt_1 \int_{t_0}^{t_1}dt_2 [\hat{H}(t_1),\hat{H}(t_2)].
\end{align}
\end{subequations}

In our protocol, the laser parameters are dynamically modulated to obtain the desired interaction from the Magnus expansion and to minimize unwanted error terms, as discussed below.
The parameters $\Delta(t)$, $\phi(t)$ and $g(t)$ are varied according to a stroboscopic sequence consisting of four segments, each of duration $\tau = 2\pi\ell/|\Delta|$ with $\ell \in \mathbb{Z}^+$ an adjustable parameter. The sequence is shown in Fig.~\ref{fig:sds_preparation_protocol}(c). 

During the first segment, the parameters are set to $\Delta(t) = \Delta$, $\phi(t) = \phi_1$ and $g(t) = g$. At the end of this segment, i.e. at $t = \tau$, the first-order Magnus term integrates to zero, $\hat{\Theta}_1[0,\tau] = 0$, because the integral is performed over a complete loop in phase space. The second-order Magnus term is nonzero, 
\begin{align}
    \hat{\Theta}_2[0,\tau] = \frac{4\pi g^2 \ell}{\Delta^2} \left[ ( e^{-i \phi_1}\hat{a}^2-\text{h.c.})\hat{J}_z - i (\hat{J}_x^2 - \hat{J}_y^2)\right], 
\end{align}
where the first term is the desired spin-dependent bosonic squeezing. The second term is an unwanted error term that, for $N > 1$, describes rotated two-axis twisting~\cite{maQuantumSpinSqueezing2011,kotibhaskarProgrammableXYtypeCouplings2024}, and drives transitions between Dicke states. Its dynamics are undone during the second segment, in which $\Delta(t) = - \Delta$ and $\phi(t) = \phi_2$, while $g(t) = g$ is unchanged. At the end of the second segment, i.e. at $t=2\tau$, the first-order Magnus term again integrates to zero, $\hat{\Theta}_1[\tau,2\tau] = 0$, while the second-order term is, 
\begin{align}
    \hat{\Theta}_2[\tau,2\tau] = -\frac{4 \pi g^2 \ell}{\Delta^2} \left[ ( e^{-i \phi_2}\hat{a}^2-\text{h.c.})\hat{J}_z -
     i (\hat{J}_x^2 - \hat{J}_y^2)\right],
\end{align}
which is identical to $\hat{\Theta}_2[0,\tau]$ up to a global sign change and after the replacement $\phi_1 \rightarrow \phi_2$. The total evolution after the first and second segments is described by the Baker–Campbell–Hausdorff (BCH) formula, 
\begin{align}
    e^{ \hat{\Theta}_2[\tau,2\tau]} e^{\hat{\Theta}_2[0,\tau]} = e^{\hat{\Theta}_2[0,\tau] + \hat{\Theta}_2[\tau,2\tau] + \dots},
\end{align}
from which we see that the phase-independent error-inducing two-axis twisting term always cancels, while the desired spin-dependent squeezing term survives if $\phi_1 \neq \phi_2$. 

The third and fourth segments of the stroboscopic sequence are a repetition of the first two segments, but with $g(t) = -g$~\footnote{Note that $g(t) \rightarrow -g$ can be realized by phase control by sending $\phi_{1,2} \rightarrow \phi_{1,2} + \pi$}. Including these segments results in a many-body echo protocol which, to first order in BCH, cancels all odd-powered Magnus terms. This significantly increases the fidelity of the spin-dependent squeezing, as the leading-order Magnus error term, the third-order term, is canceled. 

Both the Magnus expansion and the BCH formula introduce errors originating from higher-order error terms that are not entirely removed by the stroboscopic sequence.
We first analyze the remaining leading error terms from the Magnus expansion. For a fixed number of ions $N$, the strength of the $n$th-order Magnus term is $\mathcal{O}\left(g^n \ell /\Delta^n \right)$. Scaling $\ell \rightarrow \ell \Delta^2/g^2$ ensures that the second-order Magnus term is independent of $g/\Delta$, while in the limit $g/\Delta \rightarrow 0$ the higher-order terms are small and can be neglected. Under this scaling of $\ell$, the duration of the segment scales as $\tau \rightarrow \tau \Delta/ g^2$, which indicates that reducing the strength of the higher-order Magnus terms requires longer segment durations. For nonfixed $N$, the magnitude of higher-order Magnus terms depends on the spin operators that arise from the nested commutators, because their matrix elements depend on $N$. In the Supplemental Material~\cite{supp}, we study the magnitude of the third- and fourth-order Magnus terms with $N$, and show that they are all either bounded above, constant, or decreasing with $N$. Therefore, we expect our protocol to be scalable. 

The leading-order BCH error term is, assuming a worst-case scaling of $J_z^2 \propto N^2$, $[\hat{\Theta}_2[0,\tau],\hat{\Theta}_2[\tau,2\tau]] \sim g^4 \ell^2 N^2/\Delta^4 \sim \eta^4 \Omega^4/\Delta^4$. This error is small when $g^2 \ell/\Delta^2 \ll 1$, favoring shorter segment durations in the stroboscopic sequence. The amount of spin-dependent squeezing is increased by repeating the full stroboscopic sequence $P$ times. The strength of the $n$th-order Magnus term is then $\mathcal{O}(g^n \ell P/\Delta^n)$, and therefore, to first order in the BCH formula, $P$ plays an identical role to $\ell$ in ensuring that the desired second-order Magnus term is unchanged while $g/\Delta \rightarrow 0$ as required to make the higher-order Magnus terms vanish. That is, instead of scaling $\ell \rightarrow \ell \Delta^2/g^2$, we have $ P \rightarrow P \Delta^2/g^2$. 

The total duration of the spin-dependent squeezing protocol is calculated from the duration of a single segment, $\tau$, and the number of times $P$ that the four-step stroboscopic sequence is repeated, 
\begin{align}
    t_f = 4 P \tau = \frac{8 \pi \ell P}{|\Delta|}. \label{eq:tf}
\end{align}
In the regime where higher-order Magnus and BCH terms can be neglected, i.e. when $g/\Delta \ll 1$ and $\ell g^2 / \Delta^2 \ll 1$, the total evolution is effectively described by the unitary, 
\begin{align}
    \hat{S}(\zeta \hat{J}_z) = \exp\left(\frac{1}{2}(\zeta^* \hat{a}^2 - \zeta \hat{a}^{\dagger2} ) \hat{J}_z \right), \label{eq:USq}
\end{align}
which exactly corresponds to the spin-dependent squeezing operation of Eq.~\ref{eq:SDS} with, 
\begin{align}
    \zeta = -\frac{8 \pi g^2 \ell P(e^{i \phi_1}-e^{i \phi_2})}{\Delta^2}. \label{eq:zeta}
\end{align}
To maximize the magnitude of the squeezing parameter $|\zeta|$, we set $\phi_2 = \phi_1 - \pi$ which results in $e^{i \phi_1} - e^{i \phi_2} = 2e^{i \phi_1}$. The free parameter $\phi_1$ provides control over the squeezing direction in phase space, while $g$, $\ell$, $P$ and $\Delta$ control both the squeezing amplitude and the relative strength of the higher-order Magnus and BCH terms. 
A large coupling strength $g$ leads to faster spin-dependent squeezing, but can also require a large detuning $\Delta$ to ensure the higher-order error terms remain small. On the other hand, the number of phase space loops $\ell$ can be kept small, i.e. $\ell = 1$, while $P$ is increased to achieve a larger squeezing magnitude $|\zeta|$. 

\subsection{Speed limit}

We investigate the smallest achievable duration required to achieve a target squeezing magnitude in an ideal setting. To do so, we ignore higher-order error terms and consider evolution under the exact spin-dependent squeezing operator of Eq.~\ref{eq:USq}.
The duration $t_f$ is minimized when only a single phase loop is performed with $\ell = 1$, and when the stroboscopic sequence is performed only once with $P=1$, giving $t_f = 4\tau$.
Starting from a spin-polarized state, $\ket{-N/2}_s$ or $\ket{N/2}_s$, the bosonic mode is (spin-dependently) squeezed with a squeezing parameter of $z = |\zeta N/2|$. Using Eq.~\ref{eq:tf} and Eq.~\ref{eq:zeta} with $g = \eta \Omega/\sqrt{N}$, the minimum protocol time $t_f^*$ is,
\begin{align}
    t_f^* = \frac{\sqrt{\pi z}}{2g \sqrt{N/2}} = \frac{\sqrt{\pi z}}{\sqrt{2} \eta \Omega}.\label{eq:speedlimit}
\end{align}
For a fixed amount of bosonic squeezing $z$, in terms of the Lamb-Dicke parameter $\eta$ and Rabi frequency $\Omega$ the minimum protocol time $t_f^*$ is therefore independent of $N$. In comparison, the duration required to realize spin-dependent squeezing via the second-order sidebands is $t_f^* = 4z/(\eta^2 \Omega)$~\cite{katzBodyInteractionsTrapped2022b}. When neglecting higher-order terms, our protocol is therefore faster by a factor of $\sqrt{z}/\eta$. Trapped-ion experiments commonly operate in the Lamb-Dicke regime where $\eta \ll 1$, in which our protocol is expected to be significantly faster than using the second-order sidebands. However, higher-order Magnus and BCH terms may prevent the protocol from saturating the speed bound. These terms are negligible only when $g/(\Delta\ell) \ll 1$ and $g^2/\Delta^2 \ll 1$, which requires $t_f \rightarrow t_f \Delta/g^2$. In the following section, we turn to numerics to further study the protocol's performance, including the impact of higher-order terms. 

\subsection{Numerical results}
Here, we numerically study the performance of the protocol in systems of up to $N = 40$ spins using exact numerics. We simulate the dynamics of the Hamiltonian of Eq.~\ref{eq:Hamiltonian} using the complete stroboscopic sequence, without any of the approximations used in our analytics~\footnote{The simulations are performed by numerically solving the Schr{\"{o}}dinger equation using QuantumOptics.jl~\cite{KramerCPC2018} in Julia~\cite{Julia-2017} with a Fock space truncation of $n_{\rm max} = 1660$. This permits us to simulate squeezing amplitudes of up to $|z| \leq 3$ with infidelities due to the truncation of the bosonic Fock space of less than $1 - \mathcal{F} \leq 10^{-4}$. We also simulate the dynamics with $n_{\rm max} = 1801$ and $n_{\rm max} = 2000$ to confirm that the finite truncation of bosonic Hilbert space does not impact our results.}\nocite{KramerCPC2018, Julia-2017}. The performance is quantified using the fidelity $\mathcal{F} = \abs{\braket{\psi_t}{\psi(t_f)}}^2$, where $\ket{\psi(t_f)}$ is the final state after the stroboscopic protocol, and $\ket{\psi_t}$ is the ideal target state. 
The target state is a GHZ spin state that has been subjected to spin-dependent squeezing, $\ket{\psi_t} = \hat{S}(\zeta \hat{J}_z)\ket{0}_b\ket{\psi_{\rm GHZ}}_s$, cf. Eq.~\ref{eq:psiGHZ}. The performance of the protocol starting from spin-polarized states, $\ket{0}_b \ket{-N/2}_s$, is studied in the Supplemental Material~\cite{supp}. 

\begin{figure}
\includegraphics[width=0.95\linewidth]{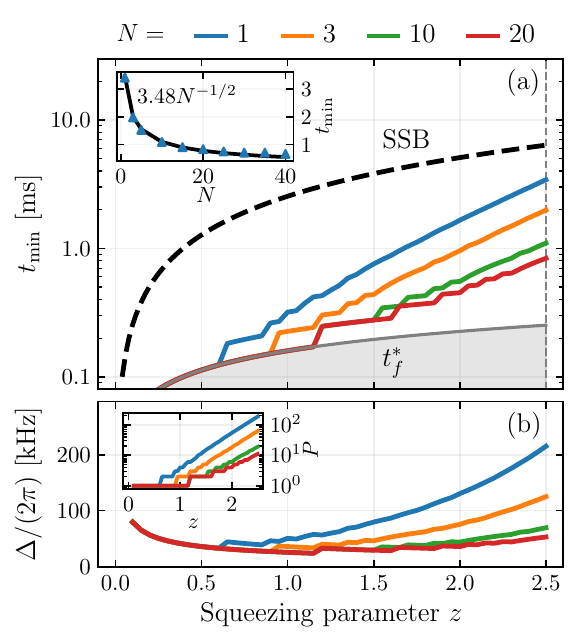}
\caption{
Performance of first-order sideband protocol to engineer spin-dependent squeezing in a crystal of $N$ ions.
\textbf{(a)} Minimum time $t_{\rm min}$ to prepare a spin-dependent squeezed state $\ket{\psi_t} = \hat{S}(\zeta \hat{J}_z) \ket{0}_b \ket{\psi_{\rm GHZ}}_s$ with fidelity $\mathcal{F} > 0.99$ versus squeezing parameter $z = \zeta N/2$. 
We use a typical coupling strength $g = 2\pi {\times} 5/\sqrt{N} \text{ kHz}$ and set the phase-space direction $\phi_1 = \pi$ and the number of phase loops per segment $\ell = 1$.
The minimum time is numerically optimized for increasing numbers of ions, $N$, and is compared to the theoretical speed limit $t_f^*$ (gray line). We further compare against the minimum time using second-order sidebands (2SB, black dashed line) using $\eta = 0.05$. 
In the inset, we fix the squeezing to $z=2.5$ and plot the minimum time $t_{\rm min}$ for increasing $N$, up to $N = 40$ spins. Solid black line is a least-squares fit to $\propto 1/\sqrt{N}$.
\textbf{(b)} Detuning $\Delta$ versus the target squeezing $z$ for increasing number of spins, $N$. The inset plots the number of repetitions, $P$, of the stroboscopic modulation for varying squeezing $z$.
}
\label{fig:sds_preparation_numerics}
\end{figure}

In Fig.~\ref{fig:sds_preparation_numerics}(a), we plot the minimum protocol duration $t_{\rm min}$ (colored lines) required to prepare the target state $\ket{\psi_t}$ with fidelity $\mathcal{F} \geq 0.99$ versus the amount of bosonic squeezing in the target state, $z = |\zeta N/2|$. Because $g$, $\phi_1$, $\ell$ and the target $z$ are fixed, only $P$ and $\Delta$ are left as free parameters in Eq.~\ref{eq:zeta}. To minimize $t_{\rm min}$ for a given target squeezing parameter, we iteratively increase $P$ while obtaining $\Delta$ from Eq.~\ref{eq:zeta}, and report the smallest duration for which $\mathcal{F} > 0.99$. This process is repeated for increasing numbers of ions, $N$. 

We first compare our protocol against the duration required to realize the same target state using the second-order sidebands (2SB, black dashed line). Overall, we find that the duration using our protocol is significantly smaller than 2SB for all target squeezings. At $z = 1$, the variance of the squeezed quadrature is $8.7$~dB below that of a coherent state, calculated using $-10 \log_{10}(e^{-2z})$. At this point, we observe an $\approx 8$-fold reduction in protocol duration for $N = 1$, and a $\approx 15$-fold reduction for $N = 10$ and $N = 20$.

We then compare the minimum duration of our protocol with the speed limit $t_f^*$ (gray line) obtained in Eq.~\ref{eq:speedlimit}. For small amounts of squeezing, $z \lesssim 0.5$, we observe that $t_{\rm min}$ saturates $t_f^*$ for all $N$. In this regime, higher-order Magnus terms are small, permitting $P=1$ while achieving high fidelities. 
For moderate amounts of squeezing, $0.5 \lesssim z \lesssim 1.5$, $t_{\rm min}$ begins to deviate from $t_f^*$, because both $P$ and $\Delta$ increase to ensure the higher-order Magnus terms remain small. The value of $z$ at which $t_{\rm min}$ departs from $t_f^*$ increases with $N$, since the most deleterious higher-order Magnus terms decrease in magnitude with $N$, as we show in the Supplemental Material~\cite{supp}. 
The step-like behavior arises from the dependence of $t_f$ on integer values of $P$, with steps becoming indistinguishably small at large $z$. 

For large squeezing parameters $z \gtrsim 1.5$, we observe that the protocol duration $t_{\rm min}$ decreases with $N$. In the inset of Fig.~\ref{fig:sds_preparation_numerics}(a), we plot $t_{\rm min}$ with increasing $N$ (triangles) for a fixed squeezing of $z = 2.5$ ($21.7$~dB).
We observe excellent agreement with a least-squares fit (black line) to $1/\sqrt{N}$. The $\sqrt{N}$ reduction in protocol duration is a unique feature of the stroboscopic protocol, and is a clear advantage over the 2SB approach. In the large squeezing regime, we observe an exponential growth of $t_{\rm min}$ with $z$. Specifically, for $z \gtrsim 2.5$, $t_{\rm min}$ approaches the 2SB duration, which is particularly prominent for $N = 1$. At extremely large squeezing amplitudes, we expect our protocol to be slower than the 2SB method. 
However, $z = 2.5$ ($\approx 22$~dB of squeezing) is larger than existing experimental demonstrations of motional squeezing in a trapped ion, which typically achieve $\lesssim 20$~dB of squeezing~\cite{Kienzler2015, Burd2019, Matsos2024}. Therefore, we do not yet expect this to be a significant limitation of our protocol, and it can be well-mitigated by increasing the number of ions, $N$. 

In Fig.~\ref{fig:sds_preparation_numerics}(b), we plot the detuning $\Delta$ for each point in panel (a), and plot the corresponding number of stroboscopic repetitions $P$ in the inset. For small squeezing parameters $z \lesssim 0.5$, $\Delta$ decreases with $z$ because higher-order Magnus terms are negligible, allowing small values of $P$. We verify in the inset that $P=1$ for all values of $N$ in this range. 
In contrast, for $z \gtrsim 0.5$, the higher-order Magnus terms are no longer negligible. Their strength is reduced by increasing $\Delta$ to ensure that the desired second-order term remains significant, which is accompanied by an increase in $P$, again visible in the inset. 
Although larger $\Delta$ leads to higher theoretical fidelities, in experimental realizations this benefit should be traded off against the fact that larger $\Delta$ leads to both longer protocol times and larger off-resonant coupling to other motional modes. The $\Delta$ required to achieve a target theoretical fidelity at a given $z$ can be lowered by increasing $N$, while off-resonant coupling can be reduced by employing strategies such as optimal control~\cite{Milne2020, Arrazola2020}. 

\section{Noisy displacement sensing}\label{sec:NoisySensing}
Realistic quantum sensors are subject to technical noise and decoherence which may degrade their metrological performance. We investigate the impact of decoherence on the displacement sensing performance of reference states obtained from spin-dependent squeezing. We focus on the three decoherence sources that are typically the most deleterious in trapped-ion systems: spin dephasing, motional heating and motional dephasing. 

\subsection{Spin dephasing}
We first consider dephasing of the spins, described by $\hat{\sigma}_z$ fluctuations that are identical but independent across all spins. We consider a density matrix, $\rho_s$, that describes a symmetric state of $N$ spins-$1/2$ in the basis of Dicke states. Single-spin dephasing at a rate $\gamma_s$ for time $t$ causes the $(i,j)$th element of $\rho_s$ to evolve as $\rho_s^{(i,j)}(t) = \rho_s^{(i,j)} {\exp(-\gamma_s t |i-j|)}$, where $0 \leq i,j \leq N$. The effect of spin dephasing is most deleterious for GHZ spin states: its density matrix elements $i=0, j = N$ and $i=N, j=0$ give the maximal scaled dephasing rate, where $|i-j| = N $. Below, we focus on this ``worst-case scenario'' and investigate the impact of spin dephasing on reference states obtained from applying spin-dependent squeezing to GHZ states. 

The evolution of a GHZ spin state  $\rho_{\rm GHZ} = \ketbra{\psi_{\rm GHZ}}$ under single-spin dephasing is described by, 
\begin{align}
\begin{split}
    \rho_{\rm GHZ}(\gamma_s t N ) = &\frac{1}{2}(\ket{N/2} \bra{N/2} + \ket{-N/2} \bra{-N/2} \\ 
    &+ e^{- \gamma_s t N }\ket{N/2} \bra{-N/2} + \text{h.c.} ). 
\end{split} \label{eq:rho_GHZ}
\end{align}
Because spin-dependent squeezing is diagonal in the Dicke basis, the state $\hat{S}(\zeta \hat{J}_z) \rho_{\rm GHZ}(\gamma t N) \hat{S}^\dagger(\zeta \hat{J}_z)$ exactly describes the application of spin-dependent squeezing, including spin dephasing, to a pure GHZ state for duration $t$. The same state also approximates the state after both noisy GHZ state preparation and noisy spin-dependent squeezing, where $t$ is now the total preparation time. 

We assume that the unknown displacement is fast with respect to the decay rate $\gamma_s$, and therefore approximate it as noise-free. After this displacement acts on the reference state, the mixed state encoding $\beta$ is, 
\begin{align}
    \rho(\beta) = \hat{D}(\beta) \hat{S}(\zeta \hat{J}_z) \rho_{\rm GHZ}(\gamma_s t N) \hat{S}^\dagger(\zeta \hat{J}_z) \hat{D}^\dagger(\beta). 
\end{align}
For the mixed state $\rho(\beta)$, the QFI matrix for the estimation of $\bm{\theta} = \{ \beta_{\rm re}, \beta_{\rm im} \}$ is (see Supplemental Material for derivation~\cite{supp}), 
\begin{align}
    \bm{Q}[\rho(\beta)] = (8\langle \hat{n} \rangle + 4) \mathbb{1}. 
\end{align}
Surprisingly, the QFI matrix is identical to the pure state case of Eq.~\ref{eq:QFIM_MP_SDS}, and therefore completely insensitive to spin dephasing. The QFI for the estimation of a displacement amplitude, $Q_{|\beta|}[\rho(\beta)]$ is similarly insensitive to spin dephasing. Displacement sensing with spin-dependent squeezed states is therefore robust to spin dephasing. This implies that the metrological utility of spin-dependent squeezed states arises due to the mixture of states that are squeezed orthogonally along $\hat{x}$ and $\hat{p}$, not due to the off-diagonal coherences. 

In contrast, we find that our time-reversal protocol for estimating the displacement's amplitude $|\beta|$, cf. Fig.~\ref{fig:phase_insensitive_schematic}, is sensitive to spin dephasing. The metrological utility is given by the CFI~\cite{supp},
\begin{align}
    \begin{split}
    F_{|\beta|}&[\rho(\beta)] = \frac{4 |\beta|^2 e^{2 |\beta|^2}}{e^{2 \left(|\beta|^2 \cosh (2 z)+\gamma_s t N \right)}-e^{2 |\beta|^2} \cos^2(b)} \\ 
   \times& \left[\sinh(2z) \sin(2\phi) \sin(b) + 2 \sinh ^2(z) \cos(b)\right]^2,
    \end{split} \label{eq:Noisy_CFI}
\end{align}
where we defined $b = |\beta|^2 {\sinh(2z)} \sin(2\phi)$. First taking the limit of vanishingly small dephasing, $\gamma_s t N \rightarrow 0$, and then a vanishingly small displacement amplitude, $|\beta|\rightarrow 0$, we recover the pure state result of Eq.~\ref{eq:CFI_sds_GHZ}, 
\begin{align}
    \lim_{|\beta| \rightarrow 0} \lim_{\gamma_s t N \rightarrow 0} F_{|\beta|}[\rho(\beta)] = 4 \langle \hat{n} \rangle. 
\end{align}
In contrast, without first taking the limit of vanishingly small dephasing, the CFI vanishes in the limit of a vanishingly small displacement amplitude only, 
\begin{align}
     \lim_{|\beta| \rightarrow 0} F_{|\beta|}[\rho(\beta)] = 0.
\end{align}
Similar behaviors have previously been observed in time-reversal protocols~\cite{gilmoreQuantumenhancedSensingDisplacements2021,KoppenhoferRevisitingImpactofDissipation2023}. In our protocol, it arises because spin dephasing results in a degradation in signal contrast which, in the limit $|\beta|\rightarrow 0$, implies that changes in the signal due to the displacement cannot be distinguished from changes due to dephasing. 

In general, the decay of $F_{|\beta|}[\rho(\beta)]$ with ${\exp(\gamma_s t N)}$ suggests that the metrological utility of our measurement protocol decreases with increasing $N$. To investigate whether there are parameter regimes where our protocol's metrological utility survives with increasing $N$, we account for the two collective enhancements that arise for realizations in trapped-ion systems. Firstly, an amplification of the displacement amplitude by $|\beta| \rightarrow \sqrt{N} |\beta|$. Secondly, using our preparation protocol of Sec.~\ref{sec:SDSPreparation}, the duration to realize spin-dependent squeezing scales as $t \rightarrow t /\sqrt{N}$ for a fixed bosonic squeezing parameter $z$. Assuming that the GHZ state preparation time is at most proportional to $\sqrt{N}$~\cite{YinFastAccurateGHZ}, we therefore fix $z$ and scale the spin dephasing exponent as $\gamma_s t N \rightarrow \gamma_s t \sqrt{N}$. 

\begin{figure}
    \centering
    \includegraphics{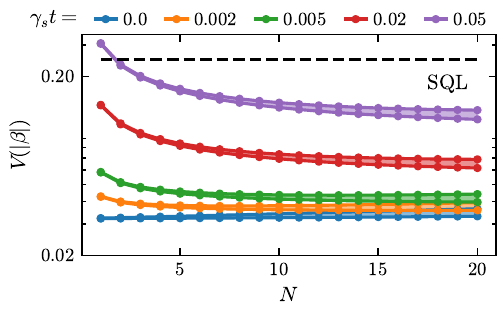}
    \caption{Classical Cram\'{e}r-Rao bound (CCRB) for displacement amplitude sensing with spin-dependent squeezed states exposed to spin dephasing. We use the time-reversal protocol of Fig.~\ref{fig:phase_insensitive_metrology}. Here, the displacement amplitude is $|\beta| = 0.02 \sqrt{N}$, and the spin-dependent squeezed state amplitude $z = 1.75$. The dephasing exponent is assumed to scale as $\gamma_s t \sqrt{N}$. The dependence on the angle of the displacement, $\phi$, is shown by the shading for each curve. Except for the $N =1, \gamma_s t = 0.05$ point, the CCRB is always below the SQL (black dashed line). For $\gamma_s t = 0.001$ the CCRB is approximately independent of $N$, while for $\gamma_s t = 0.005, 0.02$ and $0.05$ the CCRB decays uniformly with $N$. }
    \label{fig:noisyspin_phase_insensitive_metrology}
\end{figure}

In Fig.~\ref{fig:noisyspin_phase_insensitive_metrology}, we plot the CCRB on the estimation variance, $V(|\beta|) \geq 1/F_{|\beta|}[\rho(\beta)]$, as a function of $N$. The variances are obtained using the CFI in the presence of spin dephasing, cf. Eq.~\ref{eq:Noisy_CFI}. In the ideal noise-free case (blue line), the CCRB is only weakly dependent on $N$ and is minimized at $N=1$. The CCRB grows with $N$ since the displacement amplitude is amplified, moving the CFI away from the optimal point $|\beta| \rightarrow 0$. For a given $N$, as $\gamma_s t$ increases we observe that the CCRB grows since dephasing prevents distinguishing of the change in displacement signal from the effects of dephasing. Importantly, for a fixed value of $\gamma_s t$, the CCRB decreases with $N$. This is because the faster preparation time reduces effects from spin dephasing, and because the amplified displacement signal can be more easily distinguished from dephasing. The benefit of increasing $N$ is particularly apparent at larger spin dephasing rates: for the largest rate $\gamma_s t = 0.05$, our protocol surpasses the SQL only with $N > 1$. Therefore, provided the displacement amplitudes scales as $|\beta| \sqrt{N}$ and the spin dephasing exponent as $\gamma_s t \sqrt{N}$, there are parameter regimes where the metrological performance of our displacement amplitude sensing scheme improves with increasing numbers $N$ of spins$-1/2$. 

\subsection{Motional dephasing and heating}
Next, we consider sources of motional decoherence by numerically solving the Lindblad master equation, 
\begin{align}
    \dot{\rho} = -i[\hat{H}(t),\rho] + \sum_i \gamma_i \left(\hat{L}_i \rho \hat{L}_i - \frac{1}{2} \{\hat{L}_i^\dagger \hat{L}_i,\rho\} \right),
\end{align}
where $\hat{L}_i$ is the $i$th jump operator. As in the case of spin dephasing, we focus on motional decoherence occurring during the spin-dependent squeezing operation, and therefore assume that the GHZ state preparation and unknown displacement are noise-free. We realize spin-dependent squeezing using our protocol of Sec.~\ref{sec:SDSPreparation}, which uses the time-dependent spin-boson Hamiltonian of Eq.~\ref{eq:Hamiltonian}. 

We investigate the impact of motional dephasing which, in trapped-ion platforms, models fluctuations of the center-of-mass mode's frequency. The corresponding jump operator is,
\begin{align}
    \hat{L}_1 = \hat{n},
\end{align}
and occurs at a rate $\gamma_b$. We also consider motional heating, which is typically the most deleterious source of motional decoherence in trapped-ion platforms. Motional heating is described by the two jump operators, 
\begin{subequations}
    \begin{align}
        \hat{L}_1 &= \hat{a}, \\ 
        \hat{L}_2 &= \hat{a}^\dagger, 
    \end{align}
\end{subequations}
both occurring at approximately the same rate, $\kappa_b$. 

\begin{figure}
    \centering
    \includegraphics{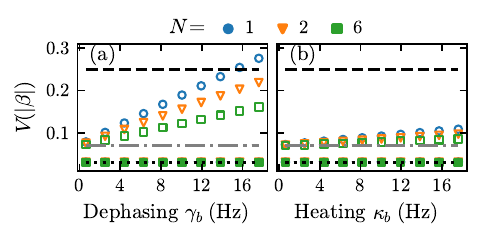}
    \caption{Metrological performance for displacement amplitude sensing with spin-dependent squeezed states under (a) motional dephasing and (b) motional heating. Metrological utility is measured through quantum (open markers) and classical (closed markers) Cram\'er-Rao bounds. The bounds are calculated for increasing ion numbers, with $N = 1$ (circles), $N = 2$ (triangles) and $N = 6$ (squares). We also show the SQL (black dashed line), and noise-free quantum (black dotted line) and classical (gray dash dot line) Cram\'er-Rao bounds.  
    We set the displacement $\beta = 0.02\sqrt{N}$ and the squeezing parameter $z = 1.5$. The spin-boson coupling strength is $g = 2\pi {\times} 30/\sqrt{N} \text{ kHz}$. Without noise, the spin-dependent squeezed state is prepared by our fast preparation protocol with fidelity $F > 0.99$. 
    }
    \label{fig:noisymotional_phase_insensitive_metrology}
\end{figure}

In Fig~\ref{fig:noisymotional_phase_insensitive_metrology}, we show the metrological utility of reference states obtained from spin-dependent squeezing subject to motional dephasing [panel (a)] and motional heating [panel (b)].
We first observe that the CCRB (open markers) obtained from the time-reversal protocol increases for increasing dephasing and heating rates. We also find that the QCRB (closed markers) minimally increases, although it is not visually apparent within the plotted range, suggesting a strong robustness of the QFI to experimentally relevant decoherence rates. 

We also investigate the metrological performance in the presence of motional decoherence for increasing numbers of ions, $N$ (see Fig.~\ref{fig:noisymotional_phase_insensitive_metrology}). For both motional heating and dephasing, we find that the CCRB for a given decay rate decreases with $N$, since the duration of the spin-dependent squeezing protocol decreases as $1/\sqrt{N}$. The Cram\'er-Rao bound under motional dephasing (panel (a)) is below the SQL for dephasing rates $\gamma_b \lesssim 15~\text{Hz}$ with $N=1$. This threshold increases with larger $N$, suggesting a feasible advantage in experiment given achievable parameters. The CCRB under motional heating (panel (b)) is below the SQL for all heating rates considered. 

\section{Conclusion and Outlook}
We have shown that the quantum Cram\'er-Rao bound of spin-dependent squeezed states saturates the Heisenberg limit for phase-insensitive estimation of a displacement's magnitude, and for simultaneous estimation of a displacement's real and imaginary components. We proposed explicit measurement protocols that follow Heisenberg scaling, and which can be readily implemented in current quantum hardware platforms. 
Moreover, we provided a protocol for preparing spin-dependent squeezed states in trapped-ion systems in a fast, scalable manner via dynamically modulated first-order sideband interactions. We showed that the attainable metrological utility is insensitive to spin dephasing and robust to motional heating and motional dephasing. Finally, for our explicit measurement protocols, we demonstrated that even in the presence of decoherence there are regimes where the metrological utility increases with the number of ions, realizing a collective enhancement. 

A natural extension of our work would be to develop robust measurement protocols. Moreover, robustness to decoherence occurring during state preparation could potentially be engineered by using quantum control~\cite{Leung2018, Milne2020, Matsos2024}. More broadly, we expect that our state preparation protocol can be generalized to generate other unitaries such as two-mode squeezing~\cite{MillicanEngineeringCVEntanglement2025} or spin-dependent tri-squeezing~\cite{bazavanSqueezingTrisqueezingQuadsqueezing2024,Sutherland2021PRA}, so as to expand the set of interactions that are available on hybrid many-spin many-boson systems~\cite{Bondinprep}.

One could also investigate the metrological performance of reference states prepared with alternative spin-boson entangling unitaries. For example, dispersive interactions have been considered for single-parameter displacement estimation~\cite{lewis-swanCavityQEDProtocolPrecise2020,barberenaAtomlightEntanglementPrecise2020}, but to the best of our knowledge, their suitability for phase-insensitive displacement sensing has not yet been studied. 

We also note that there is a wide variety of reference states that have been previously studied for quantum-enhanced displacement sensing. In the Supplemental Material~\cite{supp}, we provide a comparison of our spin-dependent squeezed states with other reference states. A detailed comparison of the metrological performance of various reference states in both ideal settings and in the presence of realistic experimental noise would be a valuable analysis that can guide experiments in choosing appropriate protocols tailored to their target hardware. 

Looking forward, our phase-insensitive sensing protocol naturally extends to large ion crystals for detecting weak forces or electric fields that induce small displacements. Scaling to $N$ ions both shortens the preparation time of spin-dependent squeezed states by $\sqrt{N}$ and amplifies the displacement signal by $
\sqrt{N}$. The phase insensitivity is beneficial for dark-matter searches, where the displacement phase is typically unknown, eliminating the need for phase synchronization between the signal and the sensor~\cite{gilmoreQuantumenhancedSensingDisplacements2021,budkerMillichargedDarkMatter2022}. The same approach applies to quantum-logic-enabled photon-recoil spectroscopy of molecular~\cite{Wolf2016, Chou2017} and highly charged ions~\cite{Kozlov2018, Micke2020}, which could benefit atomic clocks~\cite{King2022} and the search for new physics~\cite{Safronova2018}.

\begin{acknowledgments}
We thank Matteo Mazzanti and Rene Gerritsma for insightful discussions. 
A.S.N. is supported by the Dutch Research Council (NWO/OCW) as a part of the Quantum Software Consortium (project number 024.003.037), QDNL (Project No. NGF.1582.22.030) and ENWXL Grant (Project No. OCENW.XL21.XL21.122).
T. R. T. is supported by the Australian Research Council (FT220100359) and the U.S. Air Force Office of Scientific Research (FA2386-23-1-4062).
A.S. acknowledges support by the Department of Science and Technology, Govt. of India through the INSPIRE Faculty Award (DST/INSPIRE/04/2023/001486), by the Anusandhan National Research Foundation (ANRF), Govt. of India through the Prime Minister’s Early Career Research Grant (PMECRG) (ANRF/ECRG/2024/001160/PMS) and by IIT Madras through the New Faculty Initiation Grant (NFIG).
\end{acknowledgments}

\bibliography{library}

@article{Sutherland2021PRA,
  title = {Universal hybrid quantum computing in trapped ions},
  author = {Sutherland, R. T. and Srinivas, R.},
  journal = {Phys. Rev. A},
  volume = {104},
  issue = {3},
  pages = {032609},
  numpages = {8},
  year = {2021},
  month = {Sep},
  publisher = {American Physical Society},
  doi = {10.1103/PhysRevA.104.032609},
  url = {https://link.aps.org/doi/10.1103/PhysRevA.104.032609}
}

@article{Bondinprep,
  title = {},
  author = {Bond, L. J. and Valahu, C. H. and Tan, T. and Safavi-Naini, A.},
  journal = {{\it In preparation}},
  volume = {},
  issue = {},
  pages = {},
  numpages = {},
  year = {},
  month = {},
  publisher = {},
  doi = {},
  url = {}
}

@article{KramerCPC2018,
title = {QuantumOptics.jl: A Julia framework for simulating open quantum systems},
journal = {Comput. Phys. Commun.},
volume = {227},
pages = {109-116},
year = {2018},
issn = {0010-4655},
doi = {https://doi.org/10.1016/j.cpc.2018.02.004},
url = {https://www.sciencedirect.com/science/article/pii/S0010465518300328},
author = {Sebastian Krämer and David Plankensteiner and Laurin Ostermann and Helmut Ritsch}
}

@article{Julia-2017,
    title={Julia: A fresh approach to numerical computing},
    author={Bezanson, Jeff and Edelman, Alan and Karpinski, Stefan and Shah, Viral B},
    journal={SIAM {R}eview},
    volume={59},
    number={1},
    pages={65--98},
    year={2017},
    publisher={SIAM},
    doi={10.1137/141000671},
    url={https://epubs.siam.org/doi/10.1137/141000671}
}

@article{magnusExponentialSolutionDifferential1954,
  title = {On the Exponential Solution of Differential Equations for a Linear Operator},
  author = {Magnus, Wilhelm},
  year = {1954},
  month = nov,
  journal = {Comm. Pure Appl. Math.},
  volume = {7},
  number = {4},
  pages = {649--673},
  publisher = {Wiley},
  issn = {0010-3640, 1097-0312},
  doi = {10.1002/cpa.3160070404},
  urldate = {2025-07-17},
  copyright = {http://onlinelibrary.wiley.com/termsAndConditions\#vor},
  langid = {english}
}

@article{katzBodyInteractionsTrapped2022b,
  title = {N -{{Body Interactions}} between {{Trapped Ion Qubits}} via {{Spin-Dependent Squeezing}}},
  author = {Katz, Or and Cetina, Marko and Monroe, Christopher},
  year = {2022},
  month = aug,
  journal = {Phys. Rev. Lett.},
  volume = {129},
  number = {6},
  pages = {063603},
  issn = {0031-9007, 1079-7114},
  doi = {10.1103/PhysRevLett.129.063603},
  urldate = {2023-11-10},
  langid = {english},
}

@article{shapiraRobustTwoQubitGates2023,
  title = {Robust Two-Qubit Gates for Trapped Ions Using Spin-Dependent Squeezing},
  author = {Shapira, Yotam and Cohen, Sapir and Akerman, Nitzan and Stern, Ady and Ozeri, Roee},
  journal = {Phys. Rev. Lett.},
  volume = {130},
  issue = {3},
  pages = {030602},
  numpages = {7},
  year = {2023},
  month = {Jan},
  publisher = {American Physical Society},
  doi = {10.1103/PhysRevLett.130.030602},
  url = {https://link.aps.org/doi/10.1103/PhysRevLett.130.030602}
}

@article{Safronova2018,
  title = {Search for new physics with atoms and molecules},
  author = {Safronova, M. S. and Budker, D. and DeMille, D. and Kimball, Derek F. Jackson and Derevianko, A. and Clark, Charles W.},
  journal = {Rev. Mod. Phys.},
  volume = {90},
  issue = {2},
  pages = {025008},
  numpages = {106},
  year = {2018},
  month = {Jun},
  publisher = {American Physical Society},
  doi = {10.1103/RevModPhys.90.025008},
  url = {https://link.aps.org/doi/10.1103/RevModPhys.90.025008}
}

@article{King2022,
  title = {An optical atomic clock based on a highly charged ion},
  volume = {611},
  ISSN = {1476-4687},
  url = {http://dx.doi.org/10.1038/s41586-022-05245-4},
  DOI = {10.1038/s41586-022-05245-4},
  number = {7934},
  journal = {Nature},
  publisher = {Springer Science and Business Media LLC},
  author = {King,  Steven A. and Spieß,  Lukas J. and Micke,  Peter and Wilzewski,  Alexander and Leopold,  Tobias and Benkler,  Erik and Lange,  Richard and Huntemann,  Nils and Surzhykov,  Andrey and Yerokhin,  Vladimir A. and Crespo López-Urrutia,  José R. and Schmidt,  Piet O.},
  year = {2022},
  month = nov,
  pages = {43–47}
}

@article{Micke2020,
  title = {Coherent laser spectroscopy of highly charged ions using quantum logic},
  volume = {578},
  ISSN = {1476-4687},
  url = {http://dx.doi.org/10.1038/s41586-020-1959-8},
  DOI = {10.1038/s41586-020-1959-8},
  number = {7793},
  journal = {Nature},
  publisher = {Springer Science and Business Media LLC},
  author = {Micke,  P. and Leopold,  T. and King,  S. A. and Benkler,  E. and Spieß,  L. J. and Schm\"{o}ger,  L. and Schwarz,  M. and Crespo López-Urrutia,  J. R. and Schmidt,  P. O.},
  year = {2020},
  month = jan,
  pages = {60–65}
}

@article{Kozlov2018,
  title = {Highly charged ions: Optical clocks and applications in fundamental physics},
  author = {Kozlov, M. G. and Safronova, M. S. and Crespo L\'opez-Urrutia, J. R. and Schmidt, P. O.},
  journal = {Rev. Mod. Phys.},
  volume = {90},
  issue = {4},
  pages = {045005},
  numpages = {49},
  year = {2018},
  month = {Dec},
  publisher = {American Physical Society},
  doi = {10.1103/RevModPhys.90.045005},
  url = {https://link.aps.org/doi/10.1103/RevModPhys.90.045005}
}

@article{Chou2017,
  title = {Preparation and coherent manipulation of pure quantum states of a single molecular ion},
  volume = {545},
  ISSN = {1476-4687},
  url = {http://dx.doi.org/10.1038/nature22338},
  DOI = {10.1038/nature22338},
  number = {7653},
  journal = {Nature},
  publisher = {Springer Science and Business Media LLC},
  author = {Chou,  Chin-wen and Kurz,  Christoph and Hume,  David B. and Plessow,  Philipp N. and Leibrandt,  David R. and Leibfried,  Dietrich},
  year = {2017},
  month = may,
  pages = {203–207}
}

@article{Wolf2016,
  title = {Non-destructive state detection for quantum logic spectroscopy of molecular ions},
  volume = {530},
  ISSN = {1476-4687},
  url = {http://dx.doi.org/10.1038/nature16513},
  DOI = {10.1038/nature16513},
  number = {7591},
  journal = {Nature},
  publisher = {Springer Science and Business Media LLC},
  author = {Wolf,  Fabian and Wan,  Yong and Heip,  Jan C. and Gebert,  Florian and Shi,  Chunyan and Schmidt,  Piet O.},
  year = {2016},
  month = feb,
  pages = {457–460}
}

@article{Arrazola2020,
  title = {Hybrid Microwave-Radiation Patterns for High-Fidelity Quantum Gates with Trapped Ions},
  author = {Arrazola, I. and Plenio, M.B. and Solano, E. and Casanova, J.},
  journal = {Phys. Rev. Appl.},
  volume = {13},
  issue = {2},
  pages = {024068},
  numpages = {8},
  year = {2020},
  month = {Feb},
  publisher = {American Physical Society},
  doi = {10.1103/PhysRevApplied.13.024068},
  url = {https://link.aps.org/doi/10.1103/PhysRevApplied.13.024068}
}

@article{Kienzler2015,
  title = {Quantum harmonic oscillator state synthesis by reservoir engineering},
  volume = {347},
  ISSN = {1095-9203},
  url = {http://dx.doi.org/10.1126/science.1261033},
  DOI = {10.1126/science.1261033},
  number = {6217},
  journal = {Science},
  publisher = {American Association for the Advancement of Science (AAAS)},
  author = {Kienzler,  D. and Lo,  H.-Y. and Keitch,  B. and de Clercq,  L. and Leupold,  F. and Lindenfelser,  F. and Marinelli,  M. and Negnevitsky,  V. and Home,  J. P.},
  year = {2015},
  month = jan,
  pages = {53–56}
}

@article{Burd2019,
  title = {Quantum amplification of mechanical oscillator motion},
  volume = {364},
  ISSN = {1095-9203},
  url = {http://dx.doi.org/10.1126/science.aaw2884},
  DOI = {10.1126/science.aaw2884},
  number = {6446},
  journal = {Science},
  publisher = {American Association for the Advancement of Science (AAAS)},
  author = {Burd,  S. C. and Srinivas,  R. and Bollinger,  J. J. and Wilson,  A. C. and Wineland,  D. J. and Leibfried,  D. and Slichter,  D. H. and Allcock,  D. T. C.},
  year = {2019},
  month = jun,
  pages = {1163–1165}
}

@article{Matsos2024,
  title = {Robust and Deterministic Preparation of Bosonic Logical States in a Trapped Ion},
  author = {Matsos, V. G. and Valahu, C. H. and Navickas, T. and Rao, A. D. and Millican, M. J. and Kolesnikow, X. C. and Biercuk, M. J. and Tan, T. R.},
  journal = {Phys. Rev. Lett.},
  volume = {133},
  issue = {5},
  pages = {050602},
  numpages = {7},
  year = {2024},
  month = {Jul},
  publisher = {American Physical Society},
  doi = {10.1103/PhysRevLett.133.050602},
  url = {https://link.aps.org/doi/10.1103/PhysRevLett.133.050602}
}

@article{katzDemonstrationThreeFourbody2023,
  title = {Demonstration of Three- and Four-Body Interactions between Trapped-Ion Spins},
  author = {Katz, Or and Feng, Lei and Risinger, Andrew and Monroe, Christopher and Cetina, Marko},
  year = {2023},
  month = oct,
  journal = {Nat. Phys.},
  volume = {19},
  number = {10},
  pages = {1452--1458},
  publisher = {{Springer Science and Business Media LLC}},
  issn = {1745-2473, 1745-2481},
  doi = {10.1038/s41567-023-02102-7},
  urldate = {2025-08-01},
  copyright = {https://www.springernature.com/gp/researchers/text-and-data-mining},
  langid = {english},
}

@article{bazavanSqueezingTrisqueezingQuadsqueezing2024,
  title = {Squeezing, Trisqueezing and Quadsqueezing in a Hybrid Oscillator--Spin System},
  author = {B{\u a}z{\u a}van, O. and Saner, S. and Webb, D. J. and Ainley, E. M. and Drmota, P. and Nadlinger, D. P. and Araneda, G. and Lucas, D. M. and Ballance, C. J. and Srinivas, R.},
  year = 2026,
  month = may,
  journal = {Nat. PHys.},
  volume = {22},
  number = {5},
  pages = {757--762},
  issn = {1745-2473, 1745-2481},
  doi = {10.1038/s41567-026-03222-6},
  urldate = {2026-05-24},
  langid = {english}
}

@misc{sanerGeneratingArbitrarySuperpositions2024,
  title = {Generating Arbitrary Superpositions of Nonclassical Quantum Harmonic Oscillator States},
  author = {Saner, S. and B{\u a}z{\u a}van, O. and Webb, D. J. and Araneda, G. and Lucas, D. M. and Ballance, C. J. and Srinivas, R.},
  year = {2024},
  publisher = {arXiv},
  doi = {10.48550/ARXIV.2409.03482},
  urldate = {2025-09-03},
  copyright = {arXiv.org perpetual, non-exclusive license}
}

@article{carolloQuantumnessMultiparameterQuantum2019,
  title = {On Quantumness in Multi-Parameter Quantum Estimation},
  author = {Carollo, Angelo and Spagnolo, Bernardo and Dubkov, Alexander A and Valenti, Davide},
  year = {2019},
  month = sep,
  journal = {J. Stat. Mech. },
  volume = {2019},
  number = {9},
  pages = {094010},
  issn = {1742-5468},
  doi = {10.1088/1742-5468/ab3ccb},
  urldate = {2025-08-05},
}

@article{gilmoreQuantumenhancedSensingDisplacements2021,
  title = {Quantum-Enhanced Sensing of Displacements and Electric Fields with Two-Dimensional Trapped-Ion Crystals},
  author = {Gilmore, Kevin A. and Affolter, Matthew and {Lewis-Swan}, Robert J. and Barberena, Diego and Jordan, Elena and Rey, Ana Maria and Bollinger, John J.},
  year = {2021},
  month = aug,
  journal = {Science},
  volume = {373},
  number = {6555},
  pages = {673--678},
  issn = {0036-8075, 1095-9203},
  doi = {10.1126/science.abi5226},
  urldate = {2023-05-17},
  langid = {english},
}

@article{Gottesman2001,
  title = {Encoding a qubit in an oscillator},
  author = {Gottesman, Daniel and Kitaev, Alexei and Preskill, John},
  journal = {Phys. Rev. A},
  volume = {64},
  issue = {1},
  pages = {012310},
  numpages = {21},
  year = {2001},
  month = {Jun},
  publisher = {American Physical Society},
  doi = {10.1103/PhysRevA.64.012310},
  url = {https://link.aps.org/doi/10.1103/PhysRevA.64.012310}
}

@article{goreckiQuantumMetrologyNoisy2022,
  title = {Quantum {{Metrology}} of {{Noisy Spreading Channels}}},
  author = {G{\'o}recki, Wojciech and Riccardi, Alberto and Maccone, Lorenzo},
  year = {2022},
  month = dec,
  journal = {Phys. Rev. Lett.},
  volume = {129},
  number = {24},
  pages = {240503},
  issn = {0031-9007, 1079-7114},
  doi = {10.1103/PhysRevLett.129.240503},
  urldate = {2025-09-03},
  langid = {english},
}

@article{genoniOptimalEstimationJoint2013,
  title = {Optimal Estimation of Joint Parameters in Phase Space},
  author = {Genoni, M. G. and Paris, M. G. A. and Adesso, G. and Nha, H. and Knight, P. L. and Kim, M. S.},
  year = {2013},
  month = jan,
  journal = {Phys. Rev. A},
  volume = {87},
  number = {1},
  pages = {012107},
  issn = {1050-2947, 1094-1622},
  doi = {10.1103/PhysRevA.87.012107},
  urldate = {2025-05-08},
  copyright = {http://link.aps.org/licenses/aps-default-license},
  langid = {english}
}

@article{bressaniniMultiparameterQuantumEstimation2024,
doi = {10.1088/1751-8121/ad6364},
url = {https://doi.org/10.1088/1751-8121/ad6364},
year = {2024},
month = {jul},
publisher = {IOP Publishing},
volume = {57},
number = {31},
pages = {315305},
author = {Bressanini, Gabriele and Genoni, Marco G and Kim, M S and Paris, Matteo G A},
title = {Multi-parameter quantum estimation of single- and two-mode pure Gaussian states},
journal = {. Phys. A: Math. Theor.}
}

@article{duivenvoordenSingleModeDisplacementSensor2017,
  title = {Single-mode displacement sensor},
  author = {Duivenvoorden, Kasper and Terhal, Barbara M. and Weigand, Daniel},
  journal = {Phys. Rev. A},
  volume = {95},
  issue = {1},
  pages = {012305},
  numpages = {15},
  year = {2017},
  month = {Jan},
  publisher = {American Physical Society},
  doi = {10.1103/PhysRevA.95.012305},
  url = {https://link.aps.org/doi/10.1103/PhysRevA.95.012305}
}

@article{valahuQuantumenhancedMultiparameterSensing2025,
  title = {Quantum-Enhanced Multiparameter Sensing in a Single Mode},
  author = {Valahu, Christophe H. and Stafford, Matthew P. and Huang, Zixin and Matsos, Vassili G. and Millican, Maverick J. and Chalermpusitarak, Teerawat and Menicucci, Nicolas C. and Combes, Joshua and Baragiola, Ben Q. and Tan, Ting Rei},
  year = 2025,
  month = sep,
  journal = {Sci. Adv.},
  volume = {11},
  number = {39},
  pages = {eadw9757},
  issn = {2375-2548},
  doi = {10.1126/sciadv.adw9757},
  urldate = {2025-10-29},
  langid = {english}
}

@article{dengQuantumenhancedMetrologyLarge2024,
  title = {Quantum-Enhanced Metrology with Large {{Fock}} States},
  author = {Deng, Xiaowei and Li, Sai and Chen, Zi-Jie and Ni, Zhongchu and Cai, Yanyan and Mai, Jiasheng and Zhang, Libo and Zheng, Pan and Yu, Haifeng and Zou, Chang-Ling and Liu, Song and Yan, Fei and Xu, Yuan and Yu, Dapeng},
  year = {2024},
  month = dec,
  journal = {Nat. Phys.},
  volume = {20},
  number = {12},
  pages = {1874--1880},
  issn = {1745-2473, 1745-2481},
  doi = {10.1038/s41567-024-02619-5},
  urldate = {2025-05-08},
  langid = {english}
}

@article{wolfMotionalFockStates2019a,
  title = {Motional {{Fock}} States for Quantum-Enhanced Amplitude and Phase Measurements with Trapped Ions},
  author = {Wolf, Fabian and Shi, Chunyan and Heip, Jan C. and Gessner, Manuel and Pezz{\`e}, Luca and Smerzi, Augusto and Schulte, Marius and Hammerer, Klemens and Schmidt, Piet O.},
  year = {2019},
  month = jul,
  journal = {Nat Commun},
  volume = {10},
  number = {1},
  pages = {2929},
  issn = {2041-1723},
  doi = {10.1038/s41467-019-10576-4},
  urldate = {2025-05-08},
  langid = {english}
}

@misc{liMultiparameterQuantumMetrology2023,
  title = {Multi-Parameter Quantum Metrology with Stabilized Multi-Mode Squeezed State},
  author = {Li, Yue and Cheng, Xu and Wang, Lingna and Zhao, Xingyu and Hou, Waner and Li, Yi and Rehan, Kamran and Zhu, Mingdong and Yan, Lin and Qin, Xi and Peng, Xinhua and Yuan, Haidong and Lin, Yiheng and Du, Jiangfeng},
  year = {2023},
  month = dec,
  number = {arXiv:2312.10379},
  eprint = {2312.10379},
  publisher = {arXiv},
  doi = {10.48550/arXiv.2312.10379},
  urldate = {2025-07-24},
  archiveprefix = {arXiv}
}

@article{Milne2020,
  title = {Phase-Modulated Entangling Gates Robust to Static and Time-Varying Errors},
  author = {Milne, Alistair R. and Edmunds, Claire L. and Hempel, Cornelius and Roy, Federico and Mavadia, Sandeep and Biercuk, Michael J.},
  journal = {Phys. Rev. Appl.},
  volume = {13},
  issue = {2},
  pages = {024022},
  numpages = {14},
  year = {2020},
  month = {Feb},
  publisher = {American Physical Society},
  doi = {10.1103/PhysRevApplied.13.024022},
  url = {https://link.aps.org/doi/10.1103/PhysRevApplied.13.024022}
}

@article{Leung2018,
  title = {Robust 2-Qubit Gates in a Linear Ion Crystal Using a Frequency-Modulated Driving Force},
  author = {Leung, Pak Hong and Landsman, Kevin A. and Figgatt, Caroline and Linke, Norbert M. and Monroe, Christopher and Brown, Kenneth R.},
  journal = {Phys. Rev. Lett.},
  volume = {120},
  issue = {2},
  pages = {020501},
  numpages = {4},
  year = {2018},
  month = {Jan},
  publisher = {American Physical Society},
  doi = {10.1103/PhysRevLett.120.020501},
  url = {https://link.aps.org/doi/10.1103/PhysRevLett.120.020501}
}

@article{Wineland1998,
  title = {Experimental issues in coherent quantum-state manipulation of trapped atomic ions},
  volume = {103},
  ISSN = {1044-677X},
  url = {http://dx.doi.org/10.6028/jres.103.019},
  DOI = {10.6028/jres.103.019},
  number = {3},
  journal = {J. Res. Natl. Inst. Stand. Technol.},
  publisher = {National Institute of Standards and Technology (NIST)},
  author = {Wineland,  D.J. and Monroe,  C. and Itano,  W.M. and Leibfried,  D. and King,  B.E. and Meekhof,  D.M.},
  year = {1998},
  month = may,
  pages = {259}
}

@article{toscanoSubPlanckPhasespaceStructures2006a,
  title = {Sub-{{Planck}} Phase-Space Structures and {{Heisenberg-limited}} Measurements},
  author = {Toscano, F. and Dalvit, D. A. R. and Davidovich, L. and Zurek, W. H.},
  year = {2006},
  month = feb,
  journal = {Phys. Rev. A},
  volume = {73},
  number = {2},
  pages = {023803},
  issn = {1050-2947, 1094-1622},
  doi = {10.1103/PhysRevA.73.023803},
  urldate = {2025-09-03},
  copyright = {http://link.aps.org/licenses/aps-default-license},
  langid = {english}
}

@article{degenQuantumSensing2017,
  title = {Quantum sensing},
  author = {Degen, C. L. and Reinhard, F. and Cappellaro, P.},
  journal = {Rev. Mod. Phys.},
  volume = {89},
  issue = {3},
  pages = {035002},
  numpages = {39},
  year = {2017},
  month = {Jul},
  publisher = {American Physical Society},
  doi = {10.1103/RevModPhys.89.035002},
  url = {https://link.aps.org/doi/10.1103/RevModPhys.89.035002}
}

@article{duSearchInvisibleAxion2018,
  title = {Search for {{Invisible Axion Dark Matter}} with the {{Axion Dark Matter Experiment}}},
  author = {Du, N. and Force, N. and Khatiwada, R. and Lentz, E. and Ottens, R. and Rosenberg, L. J and Rybka, G. and Carosi, G. and Woollett, N. and Bowring, D. and Chou, A. S. and Sonnenschein, A. and Wester, W. and Boutan, C. and Oblath, N. S. and Bradley, R. and Daw, E. J. and Dixit, A. V. and Clarke, J. and O'Kelley, S. R. and Crisosto, N. and Gleason, J. R. and Jois, S. and Sikivie, P. and Stern, I. and Sullivan, N. S. and Tanner, D. B and Hilton, G. C. and {ADMX Collaboration}},
  year = {2018},
  month = apr,
  journal = {Phys. Rev. Lett.},
  volume = {120},
  number = {15},
  pages = {151301},
  issn = {0031-9007, 1079-7114},
  doi = {10.1103/PhysRevLett.120.151301},
  urldate = {2025-09-03},
  langid = {english}
}

@article{backesQuantumEnhancedSearch2021,
  title = {A Quantum Enhanced Search for Dark Matter Axions},
  author = {Backes, K. M. and Palken, D. A. and Kenany, S. Al and Brubaker, B. M. and Cahn, S. B. and Droster, A. and Hilton, Gene C. and Ghosh, Sumita and Jackson, H. and Lamoreaux, S. K. and Leder, A. F. and Lehnert, K. W. and Lewis, S. M. and Malnou, M. and Maruyama, R. H. and Rapidis, N. M. and Simanovskaia, M. and Singh, Sukhman and Speller, D. H. and Urdinaran, I. and Vale, Leila R. and Van Assendelft, E. C. and Van Bibber, K. and Wang, H.},
  year = {2021},
  month = feb,
  journal = {Nature},
  volume = {590},
  number = {7845},
  pages = {238--242},
  issn = {0028-0836, 1476-4687},
  doi = {10.1038/s41586-021-03226-7},
  urldate = {2025-09-03},
  langid = {english}
}

@article{zhengQuantumenhancedDarkMatter2025,
  title = {Quantum-Enhanced Dark Matter Search Using Cat States},
  author = {Zheng, Pan and Cai, Yanyan and Xu, Bin and Wen, Shengcheng and Zhang, Libo and Ni, Zhongchu and Mai, Jiasheng and Zeng, Yanjie and Lin, Lin and Hu, Ling and Deng, Xiaowei and Liu, Song and Shu, Jing and Xu, Yuan and Yu, Dapeng},
  journal = {Phys. Rev. Lett.},
  volume = {136},
  issue = {17},
  pages = {171002},
  numpages = {8},
  year = {2026},
  month = {Apr},
  publisher = {American Physical Society},
  doi = {10.1103/wbhn-v1sw},
  url = {https://link.aps.org/doi/10.1103/wbhn-v1sw}
}

@article{braggioQuantumEnhancedSensingAxion2025,
  title = {Quantum-{{Enhanced Sensing}} of {{Axion Dark Matter}} with a {{Transmon-Based Single Microwave Photon Counter}}},
  author = {Braggio, C. and Balembois, L. and Di Vora, R. and Wang, Z. and Travesedo, J. and Pallegoix, L. and Carugno, G. and Ortolan, A. and Ruoso, G. and Gambardella, U. and D'Agostino, D. and Bertet, P. and Flurin, E.},
  year = {2025},
  month = apr,
  journal = {Phys. Rev. X},
  volume = {15},
  number = {2},
  pages = {021031},
  issn = {2160-3308},
  doi = {10.1103/PhysRevX.15.021031},
  urldate = {2025-09-03},
  langid = {english}
}

@article{drechslerStateDependentMotional2020,
  title = {State-dependent motional squeezing of a trapped ion: Proposed method and applications},
  author = {Drechsler, Mart\'{\i}n and Bel\'en Far\'{\i}as, M. and Freitas, Nahuel and Schmiegelow, Christian T. and Paz, Juan Pablo},
  journal = {Phys. Rev. A},
  volume = {101},
  issue = {5},
  pages = {052331},
  numpages = {4},
  year = {2020},
  month = {May},
  publisher = {American Physical Society},
  doi = {10.1103/PhysRevA.101.052331},
  url = {https://link.aps.org/doi/10.1103/PhysRevA.101.052331}
}

@article{delgrossoControlledsqueezeGateSuperconducting2025,
  title = {Controlled-Squeeze Gate in Superconducting Quantum Circuits},
  author = {Del Grosso, Nicol{\'a}s F. and Corti{\~n}as, Rodrigo G. and Villar, Paula I. and Lombardo, Fernando C. and Paz, Juan Pablo},
  year = {2025},
  month = apr,
  journal = {Phys. Rev. A},
  volume = {111},
  number = {4},
  pages = {042606},
  issn = {2469-9926, 2469-9934},
  doi = {10.1103/PhysRevA.111.042606},
  urldate = {2025-09-03},
  langid = {english}
}

@article{cardosoSuperpositionTwomodeSqueezed2021,
  title = {Superposition of Two-Mode Squeezed States for Quantum Information Processing and Quantum Sensing},
  author = {Cardoso, Fernando R. and Rossatto, Daniel Z. and Fernandes, Gabriel P. L. M. and Higgins, Gerard and {Villas-Boas}, Celso J.},
  year = {2021},
  month = jun,
  journal = {Phys. Rev. A},
  volume = {103},
  number = {6},
  pages = {062405},
  issn = {2469-9926, 2469-9934},
  doi = {10.1103/PhysRevA.103.062405},
  urldate = {2025-09-03},
  langid = {english}
}

@article{armanGeneratingOverlapCompass2024,
  title = {Generating Overlap between Compass States and Squeezed, Displaced, or {{Fock}} States},
  author = {{Arman} and Panigrahi, Prasanta K.},
  year = {2024},
  month = mar,
  journal = {Phys. Rev. A},
  volume = {109},
  number = {3},
  pages = {033724},
  issn = {2469-9926, 2469-9934},
  doi = {10.1103/PhysRevA.109.033724},
  urldate = {2025-09-03},
  langid = {english}
}

@article{sandersSuperpositionTwoSqueezed1989,
  title = {Superposition of Two Squeezed Vacuum States and Interference Effects},
  author = {Sanders, Barry C.},
  year = {1989},
  month = apr,
  journal = {Phys. Rev. A},
  volume = {39},
  number = {8},
  pages = {4284--4287},
  issn = {0556-2791},
  doi = {10.1103/PhysRevA.39.4284},
  urldate = {2025-09-03},
  copyright = {http://link.aps.org/licenses/aps-default-license},
  langid = {english}
}

@article{dixitSearchingDarkMatter2021,
  title = {Searching for {{Dark Matter}} with a {{Superconducting Qubit}}},
  author = {Dixit, Akash V. and Chakram, Srivatsan and He, Kevin and Agrawal, Ankur and Naik, Ravi K. and Schuster, David I. and Chou, Aaron},
  year = {2021},
  month = apr,
  journal = {Phys. Rev. Lett.},
  volume = {126},
  number = {14},
  pages = {141302},
  issn = {0031-9007, 1079-7114},
  doi = {10.1103/PhysRevLett.126.141302},
  urldate = {2025-09-03},
  langid = {english}
}

@article{liuQuantumFisherInformation2020,
doi = {10.1088/1751-8121/ab5d4d},
url = {https://doi.org/10.1088/1751-8121/ab5d4d},
year = {2019},
month = {dec},
publisher = {IOP Publishing},
volume = {53},
number = {2},
pages = {023001},
author = {Liu, Jing and Yuan, Haidong and Lu, Xiao-Ming and Wang, Xiaoguang},
title = {Quantum Fisher information matrix and multiparameter estimation},
journal = {J. Phys. A: Math. Theor.}
}

@article{barberenaAtomlightEntanglementPrecise2020,
  title = {Atom-light entanglement for precise field sensing in the optical domain},
  author = {Barberena, D. and Lewis-Swan, R. J. and Thompson, J. K. and Rey, A. M.},
  journal = {Phys. Rev. A},
  volume = {102},
  issue = {5},
  pages = {052615},
  numpages = {23},
  year = {2020},
  month = {Nov},
  publisher = {American Physical Society},
  doi = {10.1103/PhysRevA.102.052615},
  url = {https://link.aps.org/doi/10.1103/PhysRevA.102.052615}
}

@article{lewis-swanCavityQEDProtocolPrecise2020,
  title = {Protocol for Precise Field Sensing in the Optical Domain with Cold Atoms in a Cavity},
  author = {Lewis-Swan, Robert J. and Barberena, Diego and Muniz, Juan A. and Cline, Julia R. K. and Young, Dylan and Thompson, James K. and Rey, Ana Maria},
  journal = {Phys. Rev. Lett.},
  volume = {124},
  issue = {19},
  pages = {193602},
  numpages = {6},
  year = {2020},
  month = {May},
  publisher = {American Physical Society},
  doi = {10.1103/PhysRevLett.124.193602},
  url = {https://link.aps.org/doi/10.1103/PhysRevLett.124.193602}
}

@misc{pezzeAdvancesMultiparameterQuantum2025,
  title = {Advances in Multiparameter Quantum Sensing and Metrology},
  author = {Pezz{\`e}, Luca and Smerzi, Augusto},
  year = {2025},
  month = feb,
  number = {arXiv:2502.17396},
  eprint = {2502.17396},
  publisher = {arXiv},
  doi = {10.48550/arXiv.2502.17396},
  urldate = {2025-07-24},
  archiveprefix = {arXiv}
}

@article{ayyashDrivenMultiphotonQubitresonator2024,
  title = {Driven Multiphoton Qubit-Resonator Interactions},
  author = {Ayyash, Mohammad and Xu, Xicheng and Ashhab, Sahel and Mariantoni, M.},
  year = {2024},
  month = nov,
  journal = {Phys. Rev. A},
  volume = {110},
  number = {5},
  pages = {053711},
  issn = {2469-9926, 2469-9934},
  doi = {10.1103/PhysRevA.110.053711},
  urldate = {2025-09-03},
  langid = {english}
}

@article{liuHybridOscillatorQubitQuantum2025,
  title = {Hybrid Oscillator-Qubit Quantum Processors: Instruction Set Architectures, Abstract Machine Models, and Applications},
  author = {Liu, Yuan and Singh, Shraddha and Smith, Kevin C. and Crane, Eleanor and Martyn, John M. and Eickbusch, Alec and Schuckert, Alexander and Li, Richard D. and Sinanan-Singh, Jasmine and Soley, Micheline B. and Tsunoda, Takahiro and Chuang, Isaac L. and Wiebe, Nathan and Girvin, Steven M.},
  journal = {PRX Quantum},
  volume = {7},
  issue = {1},
  pages = {010201},
  numpages = {166},
  year = {2026},
  month = {Jan},
  publisher = {American Physical Society},
  doi = {10.1103/4rf7-9tfx},
  url = {https://link.aps.org/doi/10.1103/4rf7-9tfx}
}

@article{hopePreparationConditionallysqueezedStates2025,
  title = {Preparation of conditionally squeezed states in qubit-oscillator systems},
  author = {Hope, Marius K. and Lidal, Jonas and Massel, Francesco},
  journal = {Phys. Rev. Res.},
  volume = {8},
  issue = {1},
  pages = {L012046},
  numpages = {6},
  year = {2026},
  month = {Feb},
  publisher = {American Physical Society},
  doi = {10.1103/jrrb-hymw},
  url = {https://link.aps.org/doi/10.1103/jrrb-hymw}
}

@article{grochowskiOptimalPhaseinsensitiveForce2025,
  title = {Optimal Phase-Insensitive Force Sensing with Non-Gaussian States},
  author = {Grochowski, Piotr T. and Filip, Radim},
  journal = {Phys. Rev. Lett.},
  volume = {135},
  issue = {23},
  pages = {230802},
  numpages = {11},
  year = {2025},
  month = {Dec},
  publisher = {American Physical Society},
  doi = {10.1103/7pyw-tgjd},
  url = {https://link.aps.org/doi/10.1103/7pyw-tgjd}
}

@article{munroWeakforceDetectionSuperposed2002,
  title = {Weak-Force Detection with Superposed Coherent States},
  author = {Munro, W. J. and Nemoto, K. and Milburn, G. J. and Braunstein, S. L.},
  year = {2002},
  month = aug,
  journal = {Phys. Rev. A},
  volume = {66},
  number = {2},
  pages = {023819},
  issn = {1050-2947, 1094-1622},
  doi = {10.1103/PhysRevA.66.023819},
  urldate = {2025-09-16},
  copyright = {http://link.aps.org/licenses/aps-default-license},
  langid = {english}
}

@article{burdExperimentalSpeedupQuantum2024,
  title = {Experimental {{Speedup}} of {{Quantum Dynamics}} through {{Squeezing}}},
  author = {Burd, S. C. and Knaack, H. M. and Srinivas, R. and Arenz, C. and Collopy, A. L. and Stephenson, L. J. and Wilson, A. C. and Wineland, D. J. and Leibfried, D. and Bollinger, J. J. and Allcock, D. T. C. and Slichter, D. H.},
  year = {2024},
  month = apr,
  journal = {PRX Quantum},
  volume = {5},
  number = {2},
  pages = {020314},
  issn = {2691-3399},
  doi = {10.1103/PRXQuantum.5.020314},
  urldate = {2025-05-08},
  langid = {english}
}

@article{maQuantumSpinSqueezing2011,
title = {Quantum spin squeezing},
journal = {Phys. Rep.},
volume = {509},
number = {2},
pages = {89-165},
year = {2011},
issn = {0370-1573},
doi = {https://doi.org/10.1016/j.physrep.2011.08.003},
url = {https://www.sciencedirect.com/science/article/pii/S0370157311002201},
author = {Jian Ma and Xiaoguang Wang and C.P. Sun and Franco Nori}
}

@article{demkowicz-dobrzanskiMultiparameterEstimationQuantum2020,
doi = {10.1088/1751-8121/ab8ef3},
url = {https://doi.org/10.1088/1751-8121/ab8ef3},
year = {2020},
month = {aug},
publisher = {IOP Publishing},
volume = {53},
number = {36},
pages = {363001},
author = {Demkowicz-Dobrzański, Rafał and Górecki, Wojciech and Guţă, Mădălin},
title = {Multi-parameter estimation beyond quantum Fisher information},
journal = {J. Phys. A: Math. Theor.}
}

@article{TURNER199067,
  title = {Windows on the Axion},
  author = {Turner, Michael S.},
  year = 1990,
  journal = {Phys. Rep.},
  volume = {197},
  number = {2},
  pages = {67--97},
  issn = {0370-1573},
  doi = {10.1016/0370-1573(90)90172-X}
}

@article{budkerMillichargedDarkMatter2022,
  title = {Millicharged Dark Matter Detection with Ion Traps},
  author = {Budker, Dmitry and Graham, Peter W. and Ramani, Harikrishnan and Schmidt-Kaler, Ferdinand and Smorra, Christian and Ulmer, Stefan},
  journal = {PRX Quantum},
  volume = {3},
  issue = {1},
  pages = {010330},
  numpages = {18},
  year = {2022},
  month = {Feb},
  publisher = {American Physical Society},
  doi = {10.1103/PRXQuantum.3.010330},
  url = {https://link.aps.org/doi/10.1103/PRXQuantum.3.010330}
}

@article{labarcaQuantumSensingDisplacements2025,
  title = {Quantum Sensing of Displacements with Stabilized Gottesman-Kitaev-Preskill States},
  author = {Labarca, Lautaro and Turcotte, Sara and Blais, Alexandre and Royer, Baptiste},
  journal = {PRX Quantum},
  volume = {7},
  issue = {2},
  pages = {020301},
  numpages = {24},
  year = {2026},
  month = {Apr},
  publisher = {American Physical Society},
  doi = {10.1103/qmss-lc5x},
  url = {https://link.aps.org/doi/10.1103/qmss-lc5x}
}

@article{gilmoreAmplitudeSensingZeroPoint2017,
  title = {Amplitude {{Sensing}} below the {{Zero-Point Fluctuations}} with a {{Two-Dimensional Trapped-Ion Mechanical Oscillator}}},
  author = {Gilmore, K. A. and Bohnet, J. G. and Sawyer, B. C. and Britton, J. W. and Bollinger, J. J.},
  year = 2017,
  month = jun,
  journal = {Phys. Rev. Lett.},
  volume = {118},
  number = {26},
  pages = {263602},
  issn = {0031-9007, 1079-7114},
  doi = {10.1103/PhysRevLett.118.263602},
  urldate = {2023-04-13},
  langid = {english}
}

@article{delaneyMeasurementMotionQuantum2019,
  title = {Measurement of {{Motion}} beyond the {{Quantum Limit}} by {{Transient Amplification}}},
  author = {Delaney, R. D. and Reed, A. P. and Andrews, R. W. and Lehnert, K. W.},
  year = 2019,
  month = oct,
  journal = {Phys. Rev. Lett.},
  volume = {123},
  number = {18},
  pages = {183603},
  issn = {0031-9007, 1079-7114},
  doi = {10.1103/PhysRevLett.123.183603},
  urldate = {2025-10-21},
  langid = {english}
}

@article{kolkowitzCoherentSensingMechanical2012,
  title = {Coherent {{Sensing}} of a {{Mechanical Resonator}} with a {{Single-Spin Qubit}}},
  author = {Kolkowitz, Shimon and Bleszynski Jayich, Ania C. and Unterreithmeier, Quirin P. and Bennett, Steven D. and Rabl, Peter and Harris, J. G. E. and Lukin, Mikhail D.},
  year = 2012,
  month = mar,
  journal = {Science},
  volume = {335},
  number = {6076},
  pages = {1603--1606},
  issn = {0036-8075, 1095-9203},
  doi = {10.1126/science.1216821},
  urldate = {2025-10-20},
  langid = {english}
}

@article{schrepplerOpticallyMeasuringForce2014,
  title = {Optically Measuring Force near the Standard Quantum Limit},
  author = {Schreppler, Sydney and Spethmann, Nicolas and Brahms, Nathan and Botter, Thierry and Barrios, Maryrose and {Stamper-Kurn}, Dan M.},
  year = 2014,
  month = jun,
  journal = {Science},
  volume = {344},
  number = {6191},
  pages = {1486--1489},
  issn = {0036-8075, 1095-9203},
  doi = {10.1126/science.1249850},
  urldate = {2025-10-21},
  langid = {english}
}

@article{ayyashDispersiveRegimeMultiphoton2025,
  title = {Dispersive Regime of Multiphoton Qubit-Oscillator Interactions},
  author = {Ayyash, Mohammad and Ashhab, Sahel},
  year = 2025,
  month = aug,
  journal = {Phys. Rev. A},
  volume = {112},
  number = {2},
  pages = {023713},
  issn = {2469-9926, 2469-9934},
  doi = {10.1103/hrc2-7nqg},
  urldate = {2025-10-21},
  langid = {english}
}

@article{ludlowOpticalAtomicClocks2015,
  title = {Optical Atomic Clocks},
  author = {Ludlow, Andrew D. and Boyd, Martin M. and Ye, Jun and Peik, E. and Schmidt, P. O.},
  year = 2015,
  month = jun,
  journal = {Rev. Mod. Phys.},
  volume = {87},
  number = {2},
  pages = {637--701},
  issn = {0034-6861, 1539-0756},
  doi = {10.1103/RevModPhys.87.637},
  urldate = {2025-10-20},
  copyright = {http://link.aps.org/licenses/aps-default-license},
  langid = {english}
}

@article{aasiEnhancedSensitivityLIGO2013,
  title = {Enhanced Sensitivity of the {{LIGO}} Gravitational Wave Detector by Using Squeezed States of Light},
  author = {Aasi, J. and Abadie, J. and Abbott, B. P. and others},
  year = 2013,
  month = aug,
  journal = {Nature Photon},
  volume = {7},
  number = {8},
  pages = {613--619},
  issn = {1749-4893},
  doi = {10.1038/nphoton.2013.177}
}

@article{AbbottObservation,
  title = {Observation of Gravitational Waves from a Binary Black Hole Merger},
  author = {Abbott, B. P. and Abbott, R. and Abbott, T. D. and others},
  collaboration = {LIGO Scientific Collaboration and Virgo Collaboration},
  journal = {Phys. Rev. Lett.},
  volume = {116},
  issue = {6},
  pages = {061102},
  numpages = {16},
  year = {2016},
  month = {Feb},
  publisher = {American Physical Society},
  doi = {10.1103/PhysRevLett.116.061102},
  url = {https://link.aps.org/doi/10.1103/PhysRevLett.116.061102}
}

@article{aeppliClock8102024,
  title = {Clock with $8\ifmmode\times\else\texttimes\fi{}{10}^{\ensuremath{-}19}$ Systematic Uncertainty},
  author = {Aeppli, Alexander and Kim, Kyungtae and Warfield, William and Safronova, Marianna S. and Ye, Jun},
  year = 2024,
  month = jul,
  journal = {Phys. Rev. Lett.},
  volume = {133},
  number = {2},
  pages = {023401},
  issn = {0031-9007, 1079-7114},
  doi = {10.1103/PhysRevLett.133.023401},
  urldate = {2025-10-21},
  langid = {english}
}

@article{shawErasureCoolingControl2025,
  title = {Erasure Cooling, Control, and Hyperentanglement of Motion in Optical Tweezers},
  author = {Shaw, Adam L. and Scholl, Pascal and Finkelstein, Ran and Tsai, Richard Bing-Shiun and Choi, Joonhee and Endres, Manuel},
  year = 2025,
  month = may,
  journal = {Science},
  volume = {388},
  number = {6749},
  pages = {845--849},
  issn = {0036-8075, 1095-9203},
  doi = {10.1126/science.adn2618},
  urldate = {2025-10-24},
  langid = {english}
}

@article{fluhmannEncodingQubitTrappedion2019,
  title = {Encoding a Qubit in a Trapped-Ion Mechanical Oscillator},
  author = {Fl{\"u}hmann, C. and Nguyen, T. L. and Marinelli, M. and Negnevitsky, V. and Mehta, K. and Home, J. P.},
  year = 2019,
  month = feb,
  journal = {Nature},
  volume = {566},
  number = {7745},
  pages = {513--517},
  issn = {0028-0836, 1476-4687},
  doi = {10.1038/s41586-019-0960-6},
  urldate = {2025-10-24},
  langid = {english}
}

@article{campagne-ibarcqQuantumErrorCorrection2020,
  title = {Quantum Error Correction of a Qubit Encoded in Grid States of an Oscillator},
  author = {{Campagne-Ibarcq}, P. and Eickbusch, A. and Touzard, S. and {Zalys-Geller}, E. and Frattini, N. E. and Sivak, V. V. and Reinhold, P. and Puri, S. and Shankar, S. and Schoelkopf, R. J. and Frunzio, L. and Mirrahimi, M. and Devoret, M. H.},
  year = 2020,
  month = aug,
  journal = {Nature},
  volume = {584},
  number = {7821},
  pages = {368--372},
  issn = {0028-0836, 1476-4687},
  doi = {10.1038/s41586-020-2603-3},
  urldate = {2025-10-24},
  langid = {english}
}

@article{penasaMeasurementMicrowaveField2016,
  title = {Measurement of a Microwave Field Amplitude beyond the Standard Quantum Limit},
  author = {Penasa, M. and Gerlich, S. and Rybarczyk, T. and M{\'e}tillon, V. and Brune, M. and Raimond, J. M. and Haroche, S. and Davidovich, L. and Dotsenko, I.},
  year = 2016,
  month = aug,
  journal = {Phys. Rev. A},
  volume = {94},
  number = {2},
  pages = {022313},
  issn = {2469-9926, 2469-9934},
  doi = {10.1103/PhysRevA.94.022313},
  urldate = {2025-10-24},
  copyright = {http://link.aps.org/licenses/aps-default-license},
  langid = {english}
}

@incollection{greenbergerGoingBellsTheorem1989,
  title = {Going {{Beyond Bell}}'s {{Theorem}}},
  booktitle = {Bell's {{Theorem}}, {{Quantum Theory}} and {{Conceptions}} of the {{Universe}}},
  author = {Greenberger, Daniel M. and Horne, Michael A. and Zeilinger, Anton},
  editor = {Kafatos, Menas},
  year = 1989,
  pages = {69--72},
  publisher = {Springer Netherlands},
  address = {Dordrecht},
  doi = {10.1007/978-94-017-0849-4_10},
  urldate = {2025-10-29},
  isbn = {978-90-481-4058-9 978-94-017-0849-4},
  langid = {english}
}

@article{kotibhaskarProgrammableXYtypeCouplings2024,
  title = {Programmable {{XY-type}} Couplings through Parallel Spin-Dependent Forces on the Same Trapped Ion Motional Modes},
  author = {Kotibhaskar, Nikhil and Shih, Chung-You and Motlakunta, Sainath and Vogliano, Anthony and Hahn, Lewis and Chen, Yu-Ting and Islam, Rajibul},
  year = 2024,
  month = jul,
  journal = {Phys. Rev. Res.},
  volume = {6},
  number = {3},
  pages = {033038},
  issn = {2643-1564},
  doi = {10.1103/PhysRevResearch.6.033038},
  urldate = {2025-10-29},
  langid = {english}
}

@article{KoppenhoferRevisitingImpactofDissipation2023,
  title = {Revisiting the impact of dissipation on time-reversed one-axis-twist quantum-sensing protocols},
  author = {Koppenh\"ofer, Martin and Clerk, A. A.},
  journal = {Phys. Rev. Res.},
  volume = {5},
  issue = {4},
  pages = {043279},
  numpages = {16},
  year = {2023},
  month = {Dec},
  publisher = {American Physical Society},
  doi = {10.1103/PhysRevResearch.5.043279},
  url = {https://link.aps.org/doi/10.1103/PhysRevResearch.5.043279}
}

@article{MillicanEngineeringCVEntanglement2025,
  title = {Engineering Continuous-Variable Entanglement in Mechanical Oscillators with Optimal Control},
  author = {Millican, Maverick J. and Matsos, Vassili G. and Valahu, Christophe H. and Navickas, Tomas and Bond, Liam J. and Tan, Ting Rei},
  journal = {Phys. Rev. Lett.},
  volume = {135},
  issue = {23},
  pages = {233604},
  numpages = {6},
  year = {2025},
  month = {Dec},
  publisher = {American Physical Society},
  doi = {10.1103/2ntn-qh3t},
  url = {https://link.aps.org/doi/10.1103/2ntn-qh3t}
}

@article{Fadel_2025,
doi = {10.1088/1361-6633/ae00d8},
url = {https://doi.org/10.1088/1361-6633/ae00d8},
year = {2025},
month = {oct},
publisher = {IOP Publishing},
volume = {88},
number = {10},
pages = {106001},
author = {Fadel, Matteo and Roux, Noah and Gessner, Manuel},
title = {Quantum metrology with a continuous-variable system},
journal = {Reports on Progress in Physics}
}

@article{YinFastAccurateGHZ,
  title = {Fast and Accurate Greenberger-Horne-Zeilinger Encoding Using All-to-All Interactions},
  author = {Yin, Chao},
  journal = {Phys. Rev. Lett.},
  volume = {134},
  issue = {13},
  pages = {130604},
  numpages = {8},
  year = {2025},
  month = {Apr},
  publisher = {American Physical Society},
  doi = {10.1103/PhysRevLett.134.130604},
  url = {https://link.aps.org/doi/10.1103/PhysRevLett.134.130604}
}

@misc{supp,
  note = {See Supplemental Material at [URL will be inserted by publisher] for (i) a comparison to other reference states, (ii) review of quantum multi-parameter estimation, standard quantum limit and Heisenberg limit, (iii) derivations of the metrological performance of spin-dependent squeezed states, (iv) derivations of the measurement protocols, (v) additional details on the state preparation protocol and (vi) derivations of the metrological performance in the presence of noise, which include Refs.~\cite{armanGeneratingOverlapCompass2024,bressaniniMultiparameterQuantumEstimation2024,demkowicz-dobrzanskiMultiparameterEstimationQuantum2020,Fadel_2025,Gottesman2001,sandersSuperpositionTwoSqueezed1989,toscanoSubPlanckPhasespaceStructures2006a}
}
}


\clearpage
\newpage
\mbox{~}

\setcounter{section}{0}
\setcounter{equation}{0}
\setcounter{figure}{0}
\setcounter{table}{0}
\setcounter{page}{1}

\renewcommand\thefigure{\arabic{figure}}

\let\thesectionWithoutS\thesection 
\renewcommand\thesection{S\thesectionWithoutS}

\let\theequationWithoutS\theequation 
\renewcommand\theequation{S\theequationWithoutS}
\let\thefigureWithoutS\thefigure 
\renewcommand\thefigure{S\thefigureWithoutS}

\title{Supplemental Material: \mytitle}

\maketitle

\onecolumngrid

\section{Comparison to other reference states}\label{sec:Comparison}

To contextualize our results, we provide in Table~\ref{tab:MetrologyComparison} a non-exhaustive comparison of our spin-dependent squeezed states with other reference states that are metrologically useful for displacement sensing. 

\NiceMatrixOptions{cell-space-limits=5pt}
\setlength{\tabcolsep}{5pt} 

\begin{table*}
\centering
\begin{NiceTabular}{
    *{7}{c}
}[hvlines]
\toprule[0.9pt]
Reference State & Definition & Resources & ${V(\beta_{\rm re})}$ & ${V(|\beta|)}$ & ${V(\beta_{\rm re}) + V(\beta_{\rm im})}$ & Measurement \\
\toprule[0.9pt]

Standard quantum limit (SQL) &
$\hat{D}(\alpha)\ket{0}_b$ &
1b &
$\dfrac{1}{4}$ &
$\dfrac{1}{4}$ &
$\dfrac{1}{2}$ & 
$\hat{x}$, $\hat{p}$ \\ 

Heisenberg limit &
- &
1b &
$\dfrac{1}{16\langle \hat{n} \rangle + 4}$ &
$\dfrac{1}{8 \langle \hat{n} \rangle + 4}$ &
$\dfrac{1}{4 \langle \hat{n} \rangle + 2}$ & 
- \\ 

\toprule[0.9pt]

Squeezed &
$\hat{S}(\zeta) \ket{0}_b$ &
1b &
\cellcolor{green!15}$\dfrac{1}{16\langle \hat{n} \rangle + 4}$ &
\cellcolor{green!15}$\dfrac{1}{8\langle \hat{n} \rangle + 4}^{*}$  &
\cellcolor{red!15}$\langle \hat{n} \rangle + \dfrac{1}{2}$ & 
$\ketbra{n}$, $\hat{\Pi}$ \\

Spin-dependent displaced~\cite{gilmoreQuantumenhancedSensingDisplacements2021} &
$\hat{D}(\alpha \hat{J}_z)\ket{0}_b \ket{\psi}_s$ &
1b, Ns &
\cellcolor{green!15}{$\dfrac{1}{16 \langle \hat{n} \rangle + 4}$} &
- &
\cellcolor{red!15}{$\dfrac{1}{16 \langle \hat{n} \rangle + 4} + \dfrac{1}{4}$} & 
$\hat{J}_\alpha$ \\

Single-mode Fock~\cite{wolfMotionalFockStates2019a,dengQuantumenhancedMetrologyLarge2024} &
$(1/\sqrt{n!}) \hat{a}^{\dagger n} \ket{0}_b$ &
1b &
\cellcolor{green!15}$\dfrac{1}{8\langle \hat{n} \rangle + 4}$ &
\cellcolor{green!15}$\dfrac{1}{8\langle \hat{n} \rangle + 4}$ &
\cellcolor{green!15}$\dfrac{1}{4\langle \hat{n} \rangle + 2}$ & 
$\hat{\Pi}$ \\

Single-mode grid~\cite{Gottesman2001, duivenvoordenSingleModeDisplacementSensor2017,valahuQuantumenhancedMultiparameterSensing2025} & 
$^{**}$ &
1b &
\cellcolor{green!15}$\dfrac{1}{8\langle \hat{n} \rangle + 4}$ &
\cellcolor{green!15}$\dfrac{1}{8\langle \hat{n} \rangle + 4}$ &
\cellcolor{green!15}$\dfrac{1}{4\langle \hat{n} \rangle + 2}$ & 
$e^{i l \hat{x}}, e^{i l \hat{p}}$ \\

Two-mode squeezed~\cite{genoniOptimalEstimationJoint2013,bressaniniMultiparameterQuantumEstimation2024} &
$\hat{S}_2 \ket{0}_b \ket{0}_b$ &
2b &
\cellcolor{green!15}$\dfrac{1}{8\langle \hat{n} \rangle + 4}$ &
\cellcolor{green!15}$\dfrac{1}{8\langle \hat{n} \rangle + 4}$ &
\cellcolor{green!15}$\dfrac{1}{4\langle \hat{n} \rangle + 2}$ & 
$\hat{x}_1$, $\hat{p}_2$ \\

\makecell{Spin-dependent squeezed \\ (this work)} &
$\hat{S}(\zeta \hat{J}_z) \ket{0}_b \ket{\psi}_s $ &
1b, Ns &
\cellcolor{green!15}$\dfrac{1}{8\langle \hat{n} \rangle + 4}$ &
\cellcolor{green!15}$\dfrac{1}{8\langle \hat{n} \rangle + 4}$ &
\cellcolor{green!15}$\dfrac{1}{4\langle \hat{n} \rangle + 2}$ & 
$^{***}$ \\
\bottomrule
\end{NiceTabular}
\caption{
\textbf{Comparison of reference states for displacement sensing.} In the first row we report the standard quantum limit (SQL), which is defined by the QCRB of single-mode coherent states. In the second row we report the Heisenberg limit, which is the QCRB minimized over all possible single-mode bosonic states. For more details of both of these definitions, see Sec.~\ref{sec:SQL_HS}. 
Below, for each reference state, column-by-column we report: (i) the definition of the state, where $\hat{D}(\alpha) = e^{\alpha \hat{a}^\dagger - \alpha^* \hat{a}}$ is the displacement operator, $\hat{S}(\zeta) = e^{(\zeta \hat{a}^{\dagger 2} - \zeta^* \hat{a}^2)/2}$ is the squeeze operator and $\hat{S}_2(\zeta) = e^{\zeta^* \hat{a} \hat{b} - \zeta \hat{a}^\dagger \hat{b}^\dagger}$ is the two-mode squeeze operator. (ii) The required physical resources, using the notation ``b'' for a bosonic mode and ``s'' for a spin (e.g. ``1b, Ns'' denotes 1 bosonic mode and N spins). (iii) The QCRB for phase-aligned single-parameter estimation of $\beta_{\rm re}$ (or equivalently, $\beta_{\rm im}$). (iv) The QCRB for phase-insensitive estimation of a displacement's magnitude, $|\beta|$. (v) The QCRB for joint-estimation of a displacement's real ($\beta_{\rm re}$) and imaginary ($\beta_{\rm im}$) components. (vi) Possible choices of measurement operators whose CCRB follow the corresponding QCRB. 
Green cell shading indicates that a reference state's QCRB follows Heisenberg scaling, $\propto 1/\langle \hat{n} \rangle$, while red indicates that it does not. 
($^{*}$) The variance $V(|\beta|)$ for squeezed states distinguishes that this result is for a phase-twirled squeezed state, see main text for details. 
($^{**}$) An exact definition for general single-mode grid states can be found in Ref.~\cite{Gottesman2001} with their application to displacement sensing in Refs.~\cite{duivenvoordenSingleModeDisplacementSensor2017, valahuQuantumenhancedMultiparameterSensing2025}.
($^{***}$) The measurements associated with the spin-dependent squeezed state are detailed in this work (see main text).
For the measurement operators, $\ketbra{n}{n}$ is the projection onto the $n$th Fock state, $\hat{\Pi}$ is the bosonic parity operator, and $e^{i l \hat{x}}$ and $e^{i l \hat{p}}$ are modular operators where $l$ is some modular length. The reported measurement operators are indicative only, and other choices may be possible. The list of reference states is non-exhaustive, and this table should serve as an exemplary comparison only.
}
\label{tab:MetrologyComparison}
\end{table*}

For the phase-aligned sensing of either $\beta_{\rm re}$ or $\beta_{\rm im}$, the Heisenberg limit (HL) is $1/(16\langle \hat{n} \rangle + 4)$, see Sec.~\ref{sec:SQL_HS}. The QCRB for squeezed states and spin-dependent squeezed states saturate the HL, while the QCRB for single-mode Fock, single-mode grid, two-mode squeezed and spin-dependent squeezed states is a factor of two larger, and therefore follows but does not saturate the HL. 

For phase-insensitive sensing of $|\beta|$, pure single-mode bosonic squeezed states cannot be used because their QCRB is phase-dependent, $Q_{|\beta|} = \langle \hat{n} \rangle/[\rm{c}(\varphi)^2 + 16\langle \hat{n} \rangle^2 \rm{s}(\varphi)^2]$. Instead, in Table~\ref{tab:MetrologyComparison} we report the QCRB for phase-twirled squeezed states~\cite{goreckiQuantumMetrologyNoisy2022}. Because the phase-twirled state is a mixed state, we use a double asterisk to distinguish it from all of the other (pure) reference states. The QCRB for spin-dependent displaced states is similarly phase-dependent. We leave the calculation of the performance of phase-twirled spin-dependent displaced states to future work. The HL for the estimation of $|\beta|$ is $1/(8 \langle \hat{n} \rangle + 4)$, and therefore phase-twirled squeezed, single-mode Fock, single-mode grid, two-mode squeezed and spin-dependent squeezed states all saturate the HL. 

For phase-aligned joint estimation of $\beta_{\rm re}$ and $\beta_{\rm im}$, the SQL is $1/2$. As such, the QCRB for squeezed states never falls below the SQL, while the QCRB for spin-dependent displaced states is never less than a factor of two smaller. The HL is $1/(4 \langle \hat{n} \rangle + 2)$, and therefore single-mode Fock, single-mode grid, two-mode squeezed and spin-dependent squeezed states all saturate the HL. 

We emphasize that single-mode Fock, single-mode grid, two-mode squeezed and spin-dependent squeezed states all exhibit identical metrological performance for displacement sensing. The differences between these states lie in the required experimental resources, their preparation protocols, and the observable that should be measured to access the metrological advantage. Spin-dependent squeezed states have the benefit that they (1) require a single bosonic mode and a minimum of one spin, with the capability to scale to $N$ spins, (2) can be prepared in many-ion crystals in a fast and scalable manner, as we showed in Sec.~\ref{sec:SDSPreparation} of the main text, and (3) have spin-measurement sequences that follow Heisenberg scaling, as we proposed in Sec.~\ref{sec:Sensing} of the main text.

\section{Review of quantum multi-parameter estimation}
In this section, we review quantum multi-parameter estimation. For a more complete review, see e.g. Refs.~\cite{liuQuantumFisherInformation2020,demkowicz-dobrzanskiMultiparameterEstimationQuantum2020}. 
Consider a quantum state $\rho(\bm{\theta})$ that encodes a set of $d$ real parameters, $\bm{\theta} = \{\theta_1,\dots,\theta_d\}$. The goal is to estimate $\bm{\theta}$ from a set of $M$ independent measurement outcomes $\bm{x} = \{x_1,\dots,x_M\}$, using a locally unbiased estimator $\tilde{\bm{\theta}}(\bm{x})$, which returns an estimate for $\bm{\theta}$ given $\bm{x}$. The measurement outcomes $\bm{x}$ are described by a positive operator-valued measurement (POVM), $\bm{\Pi} = (\Pi_1,\Pi_2,\dots)$. The probability of a particular sequence of measurement outcomes is given by $p(\bm{x}|\bm{\theta}) = \prod_k p(x_k|\bm{\theta})$, where $p(x_k|\bm{\theta}) = \Tr(\Pi_k {\rho}(\bm{\theta}))$ is the probability of obtaining the measurement outcome $x_k$, with $\Pi_k$ the associated measurement operator. 
Note that in the case of single-parameter estimation, all of the matrix expressions introduced below reduce to scalar quantities. 

For a given quantum state ${\rho}(\bm{\theta})$, unbiased estimator $\tilde{\bm{\theta}}$ and POVM $\bm{\Pi}$, the uncertainty of the estimation is quantified by the covariance matrix,
\begin{align}
    \bm{V}(\bm{\theta}) = \sum_{\bm{x}} p(\bm{x}|\bm{\theta})[\bm{\theta} - \tilde{\bm{\theta}}(\bm{x})][\bm{\theta} - \tilde{\bm{\theta}}(\bm{x})]^T,
\end{align}
where the summation runs over all possible measurements. Minimizing over all possible unbiased estimators yields a lower bound for the variance, known as the classical multi-parameter Cram\'er-Rao bound, 
\begin{align}
    \textbf{V} \geq \frac{1}{M} \bm{F}^{-1}, \label{eq:QMP-CCRB}
\end{align}
where $\bm{F}$ is the classical Fisher information matrix, whose elements are defined as, 
\begin{align}
    {F}_{ij} = \sum_k \frac{1}{p(x_k|\bm{\theta})} \partial_i p(x_k|\bm{\theta}) \partial_j p(x_k|\bm{\theta}),
\end{align}
where $\partial_i = \partial/\partial \theta_i$. Intuitively, the diagonal elements $F_{ii}$ quantify the amount of information about $\theta_i$ that is contained in the measurement outcomes $\bm{x}$. The off-diagonal elements $F_{ij}$ quantify the degree of correlation between $\theta_i$ and $\theta_j$. The lower bound of Eq.~\ref{eq:QMP-CCRB} is minimized when the diagonal elements of $\bm{F}$ are maximized while the off-diagonal elements are minimized. It is often more convenient to work with a scalar quantity, so taking the trace of Eq.~\ref{eq:QMP-CCRB} lower bounds the sum of variances according to, 
\begin{align}
    \Tr(\bm{V}) \geq \frac{1}{M} \Tr(\bm{F}^{-1}). 
\end{align}
After setting $M = 1$, we refer to this as the classical Cram\'er-Rao bound (CCRB). 

Additionally minimizing over all possible POVMs yields a lower bound on the variance known as the quantum multi-parameter Cram\'er-Rao bound, 
\begin{align}
    \bm{V} \geq \frac{1}{M} \bm{Q}^{-1},
    \label{eq:QMP-QCRB}
\end{align}
where $\bm{Q}$ is the quantum Fisher information matrix (QFIM), whose elements are defined as,
\begin{align}
    Q_{ij} &= \frac{1}{2}{\rm Tr}[ \{\hat{L}_i,\hat{L}_j\} {\rho}(\bm{\theta})].
    \label{eq:QFIM}
\end{align}
The quantum multi-parameter Cram\'er-Rao bound depends only on the choice of reference state, $\ket{\psi_{\rm ref}}$. Here, $\hat{L}_i$ is the symmetric logarithmic derivative (SLD), defined implicitly via $\partial_i {\rho}(\bm{\theta}) = ({\rho}(\bm{\theta}) \hat{L}_i + \hat{L}_i {\rho}(\bm{\theta}))/2$. 
Similarly to the multi-parameter classical Cram\'er-Rao bound, the quantum multi-parameter Cram\'er-Rao bound is minimized when the diagonal elements of $\bm{Q}$ are maximized while the off-diagonal elements are minimized. Taking the trace of Eq.~\ref{eq:QMP-QCRB} lower bounds the sum of variances according to,
\begin{align}
    \Tr(\bm{V}) \geq \frac{1}{M} \Tr(\bm{Q}^{-1}). \label{eq:sm:QCRB}
\end{align}
After setting $M = 1$, we refer to this as the quantum Cram\'er-Rao bound (QCRB). The QCRB is the main quantifier of a reference state's metrological utility. 

For multi-parameter estimation, it is not always possible to attain the lower bound of Eq.~\ref{eq:sm:QCRB}. This is because the optimal POVM for the parameter $\theta_i$ may be different to the optimal POVM for the parameter $\theta_j$, and there is no guarantee that the two different optimal POVMs are compatible. 
A necessary and sufficient condition for the attainability of Eq.~\ref{eq:sm:QCRB} is that the Uhlmann curvature, $\mathbf{D}$, vanishes. That is, for all $i,j$, 
\begin{align}
    D_{ij} = {\rm Tr}([\hat{L}_i,\hat{L}_j] {\rho}(\bm{\theta})) = 0. \label{eq:QCRB_AttainCondition}
\end{align}
The diagonal elements $D_{ii}$ are zero by definition, while the off-diagonal elements $D_{ij}$ reflect the degree of non-commutativity between a pair of SLDs, $\hat{L}_i$ and $\hat{L}_j$. Intuitively, Eq.~\ref{eq:QCRB_AttainCondition} is required for the attainability of Eq.~\ref{eq:sm:QCRB} because if $\mathbf{D}$ vanishes, then there exists a pair of SLDs that commute and whose simultaneous eigenstates can be used to construct a jointly optimal measurement. 
The deviation from Eq.~\ref{eq:sm:QCRB} is quantified by the so-called asymptotic incompatibility~\cite{carolloQuantumnessMultiparameterQuantum2019},
\begin{align}
    R = ||i \bm{Q}^{-1} \bm{D} ||_\infty, \label{eq:AsymptoticAttainability}
\end{align}
where $||.||_\infty$ denotes the largest eigenvalue of the matrix. It is bounded as $0 \leq R \leq 1$. Provided that there are asymptotically many copies of the reference state, one obtains the following inequality,
\begin{align}
    {\rm QCRB} \leq {\rm HCRB} \leq (1+R) {\rm QCRB}, 
\end{align}
where the HCRB is the Holevo-Cram\'{e}r-Rao bound~\cite{demkowicz-dobrzanskiMultiparameterEstimationQuantum2020}, whose computation for spin-dependent squeezed states we leave to future work. The QCRB of Eq.~\ref{eq:sm:QCRB} is therefore attainable within a factor of at most two. 

\section{Standard Quantum Limit and the Heisenberg Limit}\label{sec:SQL_HS}
In this section we derive expressions for the standard quantum limit (SQL) and the Heisenberg limit (HL) for displacement sensing with a single bosonic mode. Consider a pure reference state $\ket{\psi_{\rm ref}}$ that interacts linearly with a signal generated by $\hat{\mathcal{O}}$ according to $\hat{U}(\theta) = \exp\mathopen{(}-i \theta \hat{O}\mathclose{)}$. For example, in the case of displacement sensing of an unknown displacement $\beta = \beta_\mathrm{re} + i \beta_\mathrm{im}$, the generator $\hat{\mathcal{O}}$ is a linear combination of the displacement $\hat{x} = \hat{a}^\dagger + \hat{a}$ and momentum $\hat{p} = i (\hat{a}^\dagger - \hat{a})$ operators. 

\subsection{Standard quantum limit}
The standard quantum limit (SQL) for displacement sensing is defined as the QCRB obtained when using coherent states as reference states, i.e. $\ket{\psi_{\rm ref}} = \hat{D}(\alpha)\ket{0}_b$, where $\alpha \in \mathbb{C}$ is the displacement amplitude. 
If the relative phase between the reference state and the displacement is fixed, then with no loss of generality we may assume that either $\beta \in \mathbb{R}$, so the generator is $\hat{\mathcal{O}} = \hat{p}$, or $\beta \in i \mathbb{R}$, so the generator is $\hat{\mathcal{O}} = \hat{x}$. The variance for phase-aligned single-parameter estimation is then lower bounded according to, 
\begin{align}
    V(\beta_{\rm re}) \geq \frac{1}{4}, \qquad  
    V(\beta_{\rm im}) \geq \frac{1}{4}, 
\end{align}
where we used the definition $Q = 4 (\Delta \hat{\mathcal{O}})^2_{\psi}$ for the single-parameter quantum Fisher information using pure reference states~\cite{liuQuantumFisherInformation2020}, and calculated the variances $(\Delta \hat{x})^2_{\psi_{\rm ref}} = (\Delta \hat{p})^2_{\psi_{\rm ref}} = 1$ for coherent states. For the joint estimation of $\beta_{\rm re}$ and $\beta_{\rm im}$, the QFIM is $\mathbf{Q} = 4 \mathbb{1}$, and therefore the sum of variances is lower bounded according to the QCRB as, 
\begin{align}
      V(\beta_{\rm re}) + V(\beta_{\rm im}) &\geq \frac{1}{2}.
\end{align}
Finally, for phase-insensitive estimation of the displacement's magnitude $|\beta|$, calculating the single-parameter QCRB either by a change of basis of the QFIM or by phase-twirling coherent states, we obtain, 
\begin{align}
    V(|\beta|) \geq \frac{1}{4}. 
\end{align}
We refer to all of these results as the standard quantum limit (SQL). 

\subsection{Heisenberg Limit}
The Heisenberg limit is defined as the QCRB minimized over all possible single-mode bosonic reference states. If the relative phase between the reference state and the displacement is fixed, the QFI for the estimation of $\beta_{\rm re}$ or $\beta_{\rm im}$ is upper bounded as (for a derivation see e.g. Sec~4.1.2 of Ref.~\cite{Fadel_2025}),
\begin{align}
    Q \leq 4 \left[ 2 \langle \hat{n} \rangle + 1 + 2 \sqrt{\langle \hat{n} \rangle (\langle \hat{n} \rangle - 1)} \right],
\end{align}
The QCRBs of single-parameter displacement estimation, minimized over all possible reference states, are therefore lower bounded by, 
\begin{align}
    V(\beta_{\rm re}) \geq \frac{1}{8 \langle \hat{n} \rangle + 4 + 8 \sqrt{\langle \hat{n} \rangle(\langle \hat{n} \rangle - 1)}}, \qquad V(\beta_{\rm im}) \geq \frac{1}{8 \langle \hat{n} \rangle + 4 + 8 \sqrt{\langle \hat{n} \rangle(\langle \hat{n} \rangle - 1)}}. 
\end{align}
For reference states with large mode occupations, $\langle \hat{n} \rangle \gg 1$, the minimized QCRBs are approximately, 
\begin{align}
    V(\beta_{\rm re}) \geq \frac{1}{16 \langle \hat{n} \rangle + 4}, \qquad V(\beta_{\rm im}) \geq \frac{1}{16 \langle \hat{n} \rangle + 4}, 
\end{align}
which we refer to as the Heisenberg limit for the single-parameter estimation of either $\beta_{\rm re}$ or $\beta_{\rm im}$. 

For the joint estimation of $\beta_{\rm re}$ and $\beta_{\rm im}$, we note that the diagonal elements of the QFIM, $Q_{11} = 4(\Delta \hat{x})^2$ and $Q_{22} = 4(\Delta \hat{p})^2$, are maximized when the reference state is centered at the phase space origin, i.e. when $\langle \hat{x} \rangle = \langle \hat{p} \rangle = 0$. Then, the sum of variances is upper bounded using the fact that $\hat{x}^2 + \hat{p}^2 = 4 \hat{n} + 2$, i.e. $\tr(V) \leq 4 \langle \hat{n} \rangle + 2$. Using the fact that $\tr(V^{-1})$ is minimized when $V$ is proportional to the identify, the minimized multi-parameter QCRB is, 
\begin{align}
    V(\beta_{\rm re}) + V(\beta_{\rm im}) \geq \frac{1}{4 \langle \hat{n} \rangle + 2}. \label{eq:HS_MultiParameter}
\end{align}
which we refer to as the Heisenberg limit for the joint estimation of $\beta_{\rm re}$ and $\beta_{\rm im}$. 

Finally, for the estimation of the displacement magnitude $|\beta|$, a similar calculation using phase-averaged states, as performed in Ref.~\cite{goreckiQuantumMetrologyNoisy2022}, yields, 
\begin{align}
    V(|\beta|) \geq  \frac{1}{8 \langle \hat{n} \rangle + 4}, \label{eq:HS_Amplitude}
\end{align}
which we refer to as the Heisenberg limit for phase-insensitive estimation of the displacement's magnitude, $|\beta|$.

\section{Metrological performance of spin-dependent squeezed-displaced states}
We derive analytical expressions for the quantum Fisher information matrix (QFIM) for a reference state obtained by applying spin-dependent squeezing and a spin-dependent displacement to the bosonic vacuum and any collective spin state. Such a reference state can be written in the Dicke basis as, 
\begin{align}
    |\psi(\alpha\hat{J_z},\zeta \hat{J}_z)\rangle = \sum_m c_m \hat{D}(\alpha m) \hat{S}(\zeta m) \ket{0}_b \ket{m}_s, 
\end{align}
with Dicke basis coefficients $c_m \in \mathbb{C}$. The spin-dependent displacement is controlled by $\alpha \in \mathbb{C}$, while without loss of generality we assume that the spin-dependent squeezing parameter is $\zeta \in \mathbb{R}$. The reference state is then subjected to a spin-independent displacement $\hat{D}(\beta)$ with unknown $\beta \in \mathbb{C}$, producing, 
\begin{align}
    \ket{\psi(\bm{\theta})} = \hat{D}(\beta) |\psi(\alpha \hat{J}_z, \zeta \hat{J}_z)\rangle,
\end{align}
where $\bm{\theta}$ is either $\bm{\theta} = |\beta|$ for phase-insensitive displacement amplitude sensing, or $\bm{\theta} = (\beta_{\rm re},\beta_{\rm im})$
The reference state's metrological utility is quantified by the QFIM of Eq.~\ref{eq:QFIM}, which is obtained by computing the symmetric log derivative (SLD). Because $\ket{\psi(\bm{\theta})}$ is a pure state, the SLD is defined as,
\begin{align}
    L_i = 2 \partial_i \rho = 2(\ketbra{\psi(\bm{\theta})}{\partial_i \psi(\bm{\theta})} + \ketbra{\partial_i \psi(\bm{\theta})}{\psi(\bm{\theta})}). 
\end{align}

\subsection{Multi-parameter displacement sensing}
For joint estimation where $\bm{\theta} = \{\beta_{\rm re}, \beta_{\rm im}\}$, computing the SLDs yields, 
\begin{subequations}
\begin{align}
    \partial_{\theta_1} \ket{\psi} &= \sum_m c_m \left[ - i \Im(2 \alpha m + \beta) \ket{e_{0,m}} + e^{\zeta m} \ket{e_{1,m}} \right] \ket{m}, \\ 
    \partial_{\theta_2} \ket{\psi} &= \sum_m c_m \left[ i \Re(2 \alpha m + \beta) \ket{e_{0,m}} + i e^{-\zeta m} \ket{e_{1,m}} \right] \ket{m}, 
\end{align}
\end{subequations}
where we introduced the bosonic basis states $\ket{e_{n,m}} \equiv \hat{D}(\beta) \hat{D}(\alpha m) \hat{S}(\zeta m) \hat{a}^{\dagger n}\ket{0}_b/\sqrt{n!}$. After some calculations, we arrive at the following expressions for the diagonal elements of the QFIM, 
\begin{subequations}
    \begin{align}
        \mathbf{Q}_{11} &= 4\left[\sum_m |{c}_m|^2 (\Im(\gamma_m)^2+e^{2 \zeta m})\right] - 4\left[\sum_m |{c}_m|^2 \Im(\gamma_m)\right]^2 = 4\mathbf{K}_{\Im(\gamma_m),\Im(\gamma_m),\vec{w}} + 4\left[\sum_m |c_m|^2 e^{2 \zeta m}\right],  \\ 
        \mathbf{Q}_{22} &= 4\left[\sum_m |{c}_m|^2 (\Re(\gamma_m)^2+e^{-2 \zeta m})\right] - 4\left[\sum_m |{c}_m|^2 \Re(\gamma_m)\right]^2 = 4\mathbf{K}_{\Re(\gamma_m),\Re(\gamma_m),\vec{w}} + 4\left[\sum_m |c_m|^2 e^{-2 \zeta m}\right]. 
    \end{align}
\end{subequations}
For the off-diagonal elements, which are symmetric, we obtain, 
\begin{align}
    \mathbf{Q}_{12} = \mathbf{Q}_{21} = 4\left[\sum_m |{c}_m|^2 \Im(\gamma_m)\right]\left[\sum_m |{c}_m|^2 \Re(\gamma_m)\right] -4\left[\sum_m |{c}_m|^2 \Im(\gamma_m)\Re(\gamma_m)\right] = -4\mathbf{K}_{\Im(\gamma_m),\Re(\gamma_m),\vec{w}}. 
\end{align}
In the above expressions we introduced $\gamma_m \equiv 2 \alpha m + \beta$, and used the weighted covariance matrix $\mathbf{K}$ of $\mathbf{X} = (X_1,X_2,\dots,X_n)^T$ with corresponding non-negative weights $\vec{w}$, whose elements are defined as, 
\begin{align}
    \mathbf{K}_{X_j,X_k,\vec{w}} = \sum_{i} \omega_i (x_{ij} - \bar{x}_j)(x_{ik} - \bar{x}_k) = \left(\sum_i \omega_i x_{ij} x_{ik}\right) - \left(\sum_i w_i x_{ik} \right) \left( \sum_i w_i x_{ij} \right),
\end{align}
with $\bar{x}_j = \sum_i w_i x_{ij}$ the weighted mean. For our calculations, we substituted $\mathbf{X} = (\Im(\bm{\gamma}),\Re(\bm{\gamma}))$ and $w_i = |c_i|^2$, where $\bm{\gamma} = (\gamma_{-N/2},\gamma_{-N/2+1},\dots,\gamma_{N/2})^T$. We can finally obtain a compact expression for the QFIM as a function of $\mathbf{K}$,
\begin{align}
        \mathbf{Q} = 4\mathbf{P}\mathbf{K}\mathbf{P} + 4\begin{pmatrix}
        \sum_m |c_m|^2 e^{2 \zeta m} & 0 \\ 0 & \sum_m |c_m|^2 e^{-2 \zeta m}
    \end{pmatrix},
\end{align}
where we introduced the matrix $\mathbf{P} = \rm{diag}(1,-1)$ to make the off-diagonal elements negative. 

\subsection{Limiting cases}
Here, we consider three limiting cases of the general spin-dependent squeezed, spin-dependent displaced spin-boson reference state. 

\subsubsection{Spin-dependent squeezed states}
When the reference state is spin-dependent squeezing only, i.e. when $\alpha = 0$, then $\gamma_m = \beta$ and therefore $\mathbf{K} = 0$. The QFIM simplifies to,
\begin{align}
    \mathbf{Q} = 4\begin{pmatrix} \sum_m |{c}_m|^2 e^{2 \zeta m} & 0 \\ 0 & \sum_m |{c}_m|^2 e^{-2 \zeta m}    \end{pmatrix}. 
\end{align}
We observe that the off-diagonal elements vanish, and that the diagonal elements grow exponentially with $\zeta$. 

A special case is obtained when the initial spin state has symmetric Dicke basis weights, i.e. if $|c_{-m}| = |c_m|$. Spin states that satisfy this property include GHZ states and any spin coherent state in the $x-y$ plane of the collective Bloch sphere. Using the fact that $e^{x} + e^{- x} = 2 {\cosh(x)}$, the QFIM for odd $N$ (a similar expression holds for $N$ even) simplifies to, 
\begin{align}
    \mathbf{Q} = 8  \sum_{m>0} \left[|{c}_m|^2 \cosh(2\zeta m) \right] \mathbb{1}  = (8 \langle \hat{n} \rangle + 4)\mathbb{1}. \label{eq:QFIM_SDS}
\end{align}
To obtain the last equality, we used the fact for spin-dependent squeezing acting on a spin state with symmetric Dicke weights, the average bosonic mode occupation is
\begin{align}
    \langle \psi(0,\zeta \hat{J}_z) | \hat{n} | \psi(0,\zeta \hat{J}_z)\rangle = \sum_m |c_m|^2 \langle \zeta m|\hat{n} |\zeta m \rangle_b = 2\sum_{m>0} |c_m|^2 \sinh^2(\zeta m),
\end{align}
where we used the identity $\sum_{m>0} |c_m|^2 = 1/2$ which arises from normalization, and the fact that the average bosonic occupation of a squeezed state with squeezing amplitude $\zeta m$ is $\langle \hat{n}\rangle = {\sinh^2(|\zeta m|)}$. From the QFIM, the multi-parameter QCRB for the sum of the variances is, 
\begin{align}
    V(\beta_{\rm re}) + V(\beta_{\rm im}) \geq \frac{1}{4\langle \hat{n} \rangle + 2}, 
\end{align}
which we report in Eq.~\ref{eq:QCRB_MP_SDS} of the main text. We emphasize that the QCRB saturates the multi-parameter Heisenberg limit of Eq.~\ref{eq:HS_MultiParameter}. Therefore, the reference state obtained by applying spin-dependent squeezing to any collective spin state with symmetric Dicke weights is optimal for the joint estimation of $\beta_{\rm re}$ and $\beta_{\rm im}$. 

The attainability of the QCRB is determined by the Uhlmann curvature of Eq.~\ref{eq:QCRB_AttainCondition}. For the symmetric Dicke basis spin-dependent squeezed state, 
\begin{align}
    \mathbf{D} = \begin{pmatrix}
        0 & 4 \\ -4 & 0
    \end{pmatrix}. \label{eq:Uhlmann}
\end{align}
Because $\mathbf{D}$ does not vanish, the QCRB is, in general, not attainable. The deviation from the QCRB is quantified by the asymptotic attainability of Eq.~\ref{eq:AsymptoticAttainability}, 
\begin{align}
    R = \frac{1}{2\langle \hat{n}\rangle + 1},
\end{align}
which is reported in Eq.~\ref{eq:MainTextAsymptoticAttainability} of the main text. In the limit $\langle \hat{n} \rangle \rightarrow \infty$, $R \rightarrow 0$. The QCRB is therefore asymptotically attainable within a prefactor of at most two, which decreases to one in the limit of infinite spin-dependent squeezing. 

\subsubsection{Spin-dependent squeezed, real spin-dependent displaced states}
For purely real spin-dependent displaced and spin-dependent squeezed reference states, i.e. setting $\alpha,\zeta \in \mathbb{R}$ the QFIM is, 
\begin{align}
    \mathbf{Q}(\alpha \in \mathbb{R}) = \begin{pmatrix}
                4\sum_m |c_m|^2 e^{2 \zeta m} & 0 \\ 0 & 4\mathbf{K}_{\Re(\gamma_m),\Re(\gamma_m),\vec{w}} + 4\sum_m |c_m|^2 e^{-2 \zeta m} 
    \end{pmatrix}.
\end{align}
The surviving diagonal element of the covariance matrix, $4\mathbf{K}_{\Re(\gamma_m),\Re(\gamma_m),\vec{w}}$, is the variance and is therefore strictly positive. Increasing $\alpha$ therefore increases $\mathbf{Q}_{22}$, which results in improved metrological utility for sensing $\beta_{\rm im}$.
When $\alpha \in i \mathbb{R}$ the metrological utility is improved for sensing $\beta_{\rm re}$. Spin-dependent displacements can therefore be used to reduce the estimation uncertainty of either the real or imaginary component of a displacement signal. 

\subsubsection{Spin-dependent displaced states}
Finally, for reference states without spin-dependent squeezing, i.e. $\zeta = 0$, the QFI is $\bm{Q} = 4 \bm{P} \bm{K} \bm{P}$. For initial collective spin states with symmetric Dicke weights, $|c_m| = |c_{-m}|$, 
\begin{align}
    \bm{Q} = 4 \mathbb{1} + 16 \sum_m |c_m|^2 \begin{pmatrix}
     \Im(\alpha m)^2   &  \Im(\alpha m) \Re(\alpha m)  \\ \Im(\alpha m) \Re(\alpha m)  &  \Re(\alpha m)^2
    \end{pmatrix}. 
\end{align}
For single-parameter sensing of either $\beta_{\rm re}$ or $\beta_{\rm im}$, the QFI is maximized when the signal is orthogonal to the direction of the spin-dependent displacement that generated the reference state. The estimation of $\beta_{\rm re}$ therefore requires $\alpha \in i \mathbb{R}$, in which case, 
\begin{align}
    Q_{\beta_{\rm re}} = 4 + 16 \sum_m |c_m|^2 \Im(\alpha m)^2 = 16 \langle \hat{n} \rangle + 4, 
\end{align}
where we used the fact that the average mode occupation of a spin-dependent displaced state is $\langle \hat{n} \rangle = \sum_m |c_m|^2 |\alpha m|^2$. A similar result is obtained for single-parameter sensing of $\beta_{\rm im}$. Spin-dependent displaced states that are appropriately phase-aligned therefore saturate the Heisenberg limit for single-parameter estimation of either $\beta_{\rm re}$ or $\beta_{\rm im}$.

For multi-parameter sensing, in the limit $\beta \rightarrow 0$ and for initial symmetric Dicke weight collective spin states, the inverse of $\bm{Q}$ is,
\begin{align}
    \Tr(\bm{Q}^{-1}) = \frac{Q_{11} + Q_{22}}{-Q_{12}^2 + Q_{11} Q_{22}} = \frac{1}{16 \langle \hat{n} \rangle + 4} + \frac{1}{4}.
\end{align}
In the limit $\langle \hat{n} \rangle \rightarrow \infty$, we obtain $\lim_{\langle \hat{n} \rangle \rightarrow \infty} (\Tr(\bm{Q}^{-1})) = 1/4$. Therefore, the multi-parameter QCRB for spin-dependent displaced states is never small than a factor of two below the multi-parameter SQL of $1/2$ that is obtained by using coherent states, cf. Sec.~\ref{sec:SQL_HS}. 

\subsection{Phase-insensitive displacement magnitude sensing}
Next, we consider phase-insensitive sensing of a displacement's magnitude by changing the basis of the QFIM from Cartesian coordinates to polar coordinates~\cite{liuQuantumFisherInformation2020}, 
\begin{align}
    \bm{Q}[{\rho}(\bm{\theta})] = \bm{J}^T \bm{Q}[{\rho}(\bm{\theta'})] \bm{J}, 
\end{align}
where $\bm{J}$ is the Jacobian matrix whose elements are defined as $J_{ij} = \partial \theta'_i/\partial \theta_j$. In polar coordinates, the parameters to be estimated are $\bm{\theta'} = \{|\beta|,\arg(\beta)\}$. 
The Jacobian for Cartesian to polar coordinates is then, 
\begin{align}
    \bm{J} = \frac{1}{|\beta|^2 } \begin{pmatrix}
        |\beta| \beta_{\rm re}  & |\beta| \beta_{\rm im} \\ - \beta_{\rm im } & \beta_{\rm re}
    \end{pmatrix}. 
\end{align}
For reference states obtained by applying spin-dependent squeezing to symmetric-weight collective spin states, changing the basis of the QFIM of Eq.~\ref{eq:QFIM_SDS}, the QFIM in polar coordinates $\bm{Q'}$ is, 
\begin{align}
    \bm{Q'} = (8 \langle \hat{n} \rangle + 4) \begin{pmatrix}
    1 & 0 \\ 0 & |\beta|^2
    \end{pmatrix}. \label{eq:QFI_AbsBeta}
\end{align}
Note that $Q_{22}$, which corresponds to the estimation of the phase $\arg(\beta)$, depends on $|\beta|^2$. It vanishes in the limit of small displacements, $|\beta|\rightarrow 0$, because the phase in this regime is not well-defined. 
Focusing on displacement magnitude estimation, the QCRB is, 
\begin{align}
    V(|\beta|) \geq \frac{1}{Q_{|\beta|}} = \frac{1}{8 \langle \hat{n} \rangle + 4}, 
\end{align}
which is entirely independent of $\beta$, and saturates the Heisenberg limit, cf. Eq~\ref{eq:HS_Amplitude}. Reference states obtained by applying spin-dependent squeezing to symmetric Dicke weight spin states are therefore optimal for the phase-insensitive estimation of the displacement's amplitude $|\beta|$. 

\subsection{Connection to many-mode bosonic squeezing}
The usefulness of spin-dependent squeezed states for the joint estimation of $\beta_{\rm re}$ and $\beta_{\rm im}$ suggests a possible connection to other reference states. Here, we relate spin-dependent squeezing of $N$ spins to bosonic squeezing of $N+1$ modes by introducing hybrid spin-boson operators,
\begin{align}
    \hat{a}_m &= \hat{a} \ketbra{m}{m}_s, 
\end{align}
which are distinguished from purely bosonic operator by the $m$ subscript. The hybrid operators obey the commutation relations $[\hat{a}_m,\hat{a}_{m'}] = [\hat{a}_m^\dagger,\hat{a}_{m'}^\dagger] = 0,\; \forall \; m,m'$. However, the canonical commutation relations are modified due to a surviving spin projector, $[\hat{a}_m,\hat{a}_{m'}^\dagger] = \delta_{m,m'}\ketbra{m}_s$, which arises because spin-dependent squeezing can be viewed as a direct sum of $N+1$ bosonic modes, rather than their tensor product. 

Using the hybrid operators, spin-dependent squeezing can be written as an $(N+1)$-times product of single-mode squeezing operators, 
\begin{align}
     \hat{S}(\zeta \hat{J}_z) = \prod_{m=-N/2}^{N/2} \exp{\frac{m}{2}(\zeta^* \hat{a}_m^2 + \zeta \hat{a}_m^{\dagger2})}.
\end{align}
For $N = 1$, spin-dependent squeezed states correspond to two-mode squeezed states. Our metrological results can therefore be understood as a mapping from two-mode squeeze states~\cite{genoniOptimalEstimationJoint2013,bressaniniMultiparameterQuantumEstimation2024}. Indeed, the Uhlmann curvature of Eq.~\ref{eq:Uhlmann} and the QFIM of Eq.~\ref{eq:QFIM_SDS} for spin-dependent squeezed states with symmetric Dicke weights are equivalent to the expressions obtained for two-mode squeezed vacuum states~\cite{bressaniniMultiparameterQuantumEstimation2024}. For $N > 1$, our results can be used to determine the metrological performance of bosonic squeezed states of $N+1$ modes. 

Finally, we note that we can introduce hybrid quadrature operators $\hat{x}_m = \hat{a}_m^\dagger + \hat{a}_m$ and $\hat{p}_m = i(\hat{a}_m^\dagger - \hat{a}_m)$, which obey $[\hat{x}_m,\hat{p}_{m'}] = 2i \delta_{m,m'} \ketbra{m}{m}_s$. Because $[\hat{x}_m,\hat{p}_{m'}]=0, \; \forall \; m\neq m'$, the hybrid quadratures $\hat{x}_m$ and $\hat{p}_{m'}$ can be simultaneously measured, which is a useful property for multi-parameter displacement sensing.

\subsection{Analog bosonic-only states}

To further understand the metrological utility of spin-dependent squeezed states, it is instructive to plot the Wigner functions of purely bosonic states that are analogous to the spin-boson states that are obtained by applying spin-dependent squeezing to the GHZ spin state of Eq.~\ref{eq:psiGHZ} and the spin coherent state of Eq.~\ref{eq:psicoh}. Up to normalization, the analogue states are, 
\begin{subequations}
\begin{align}
    |\phi_{\rm GHZ}\rangle_b &\simeq \ket{-N\zeta/2}_b + \ket{N\zeta/2}_b, \\ 
    |\phi_{\rm SC}\rangle_b &\simeq \sum_m \sqrt{\binom{N}{m+N/2}} \ket{\zeta m}_b
\end{align}
\end{subequations}
We note that these bosonic analogue states are not obtained by tracing out the spin degree of freedom, which would produce a mixed state. 

\begin{figure}
    \centering
    \includegraphics{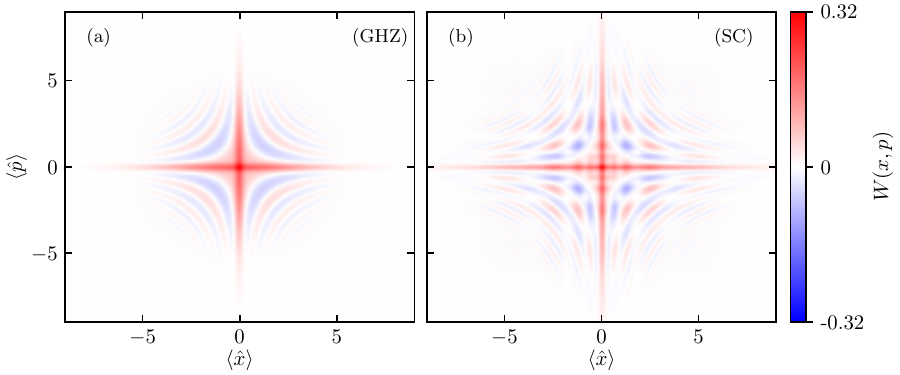}
    \caption{Wigner function $W(x,p)$ of the bosonic states (a) $|\phi_{\rm GHZ}\rangle_b \simeq \ket{-N\zeta/2}_b + \ket{N\zeta/2}_b$ and (b) $|\phi_{\rm SC}\rangle_b \simeq \sum_{m=-N/2}^{N/2} \sqrt{\binom{N}{m+N/2}} \ket{\zeta m}_b$, with $\zeta = 0.3$ and $N = 10$. Both Wigner functions feature simultaneous squeezing along orthogonal quadratures, sub-Planck structure, and $Z_4$ symmetry. There is more phase space interference in panel (b) due to the presence of additional states in the superposition.}
    \label{fig:sm_bosonic_analogues}
\end{figure}

In Fig.~\ref{fig:sm_bosonic_analogues} we plot the Wigner functions of each of these two states in panels (a) and panels (b), respectively. We observe simultaneous squeezing along orthogonal quadratures, and $Z_4$ symmetric sub-Planck structure. As discussed in Refs.~\cite{sandersSuperpositionTwoSqueezed1989,drechslerStateDependentMotional2020,armanGeneratingOverlapCompass2024}, these features indicate utility for the joint estimation of the real and imaginary components of a displacement signal, $\beta = \beta_{\rm re} + i \beta_{\rm im}$~\cite{toscanoSubPlanckPhasespaceStructures2006a}.

\subsection{Performance in terms of the squeezing parameter}

In the main text, we analyzed the QCRB for spin-dependent squeezed reference states as a function of the mode occupation, $\langle \hat{n} \rangle$. Here, we analyze the QCRB as a function of the squeezing parameter, $\zeta$. We focus on joint-parameter estimation of $\beta_{\rm re}$ and $\beta_{\rm im}$; the performance is similar for the estimation of $|\beta|$. 

\begin{figure}
    \centering
    \includegraphics[scale=1]{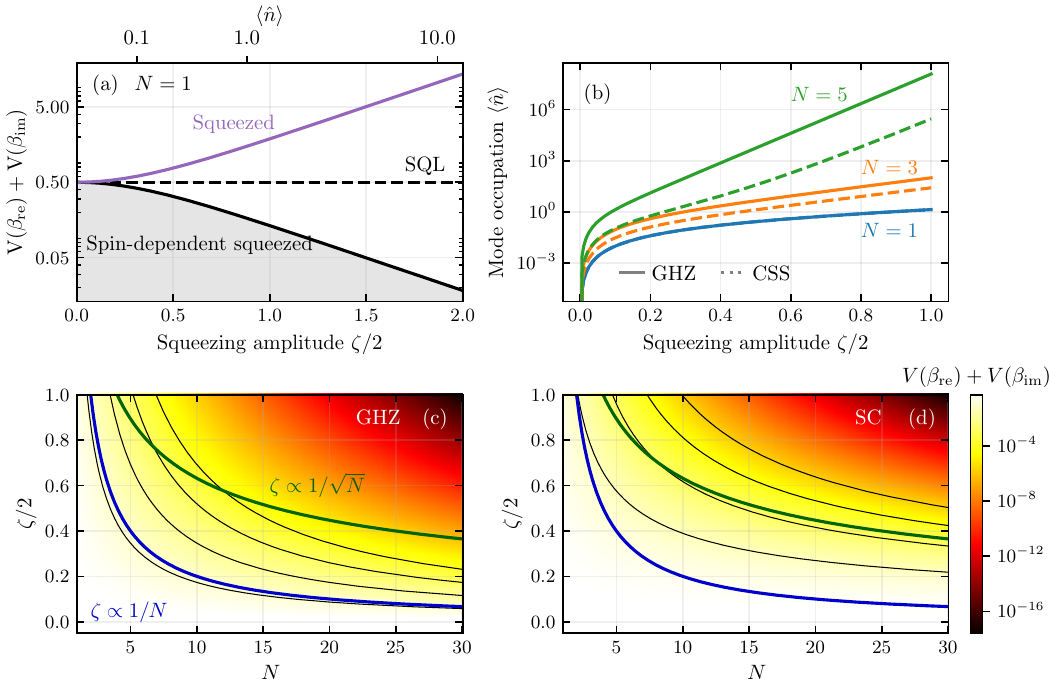}
    \caption{
    Metrological utility of spin-dependent squeezed states for the joint estimation of $\beta_{\rm re}$ and $\beta_{\rm im}$ versus the squeezing parameter, $\zeta/2$. \textbf{(a)} Case of $N = 1$ spin. The quantum Cram\'{e}r-Rao bound (QCRB) for a spin-dependent squeezed state (black line) always falls below the SQL (dashed line). The SQL is defined by the QCRB for bosonic coherent states, $\rm{SQL} = 1/2$. The QCRB for purely bosonic squeezed state (purple line) never falls below the SQL. 
    \textbf{(b)} Bosonic mode occupation of states obtained by applying spin-dependent squeezing to either GHZ spin states (solid lines) or spin coherent states (SC, dashed lines) for $N = 1$ (blue), $N = 3$ (orange) and $N = 5$ (green) spins. The occupation number grows with $\zeta$ and $N$. For fixed $\zeta$ and $N$, applying spin-dependent squeezing to GHZ states results in larger mode occupation than when applied to spin coherent states. 
    \textbf{(c)} QCRB for reference states obtained by applying spin-dependent squeezing to GHZ spin states. Black lines are contour lines. The QCRB decreases with both $N$ and $\zeta$. The QCRB is constant when $\zeta \propto 1/N$ (blue line)
    \textbf{(d)} 
    QCRB for reference states obtained by applying spin-dependent squeezing to spin coherent states. The QCRB is constant when $\zeta \propto 1/\sqrt{N}$ (green line). 
    }
    \label{fig:sm_phase_insensitive_metrology}
\end{figure}

In Fig.~\ref{fig:sm_phase_insensitive_metrology}(a) we set $N = 1$ and plot the QCRB (black line) of spin-dependent squeezed states versus $\zeta/2$ (bottom axis). For comparison, the top axis shows $\langle \hat{n} \rangle$. The QCRB is always below the SQL (dashed line). We also show the QCRB of purely bosonic squeezed states (purple line), which scales as $\propto \langle \hat{n} \rangle$ and therefore never falls below the SQL. 

In Fig.~\ref{fig:sm_phase_insensitive_metrology}(b), we plot the change in mode occupation, $\langle \hat{n} \rangle$, as a function of the squeezing amplitude, $\zeta/2$, for states obtained by applying spin-dependent squeezing to two different collective spin states: GHZ (solid) and coherent (dashed). The average mode occupation increases faster for the GHZ state because the amount of squeezing is proportional to the magnetization $m$, and GHZ states maximally populate the largest $m = \pm N/2$ spin-polarized Dicke states. The same $m$ dependence is responsible for the increasing mode occupations with higher $N$. 

In Fig.~\ref{fig:sm_phase_insensitive_metrology}(c) and (d), we plot the QCRB as a function of the squeezing parameter, $\zeta/2$, and the number of spins, $N$. We consider reference states obtained by applying spin-dependent squeezing to a GHZ spin state (panel (c)), and to a spin coherent state (panel (d)). 
At fixed $N$, the QCRB decreases when increasing $\zeta$ for both states. Similarly, at fixed $\zeta$, the QCRB decreases when increasing $N$. 
We note that in trapped ion systems, spin-dependent squeezing generated via second-order sidebands leads to a $\zeta \propto 1/N$ scaling (blue line)~\cite{katzBodyInteractionsTrapped2022b}. This results in a constant QCRB for GHZ states, but not for spin coherent states. In the latter case, $\zeta$ must scale approximately as $\zeta \propto 1/\sqrt{N}$ (green line) to obtain a constant QCRB. Note that, despite this requirement on $\zeta$ when using spin-dependent squeezed spin coherent states, scaling to $N$ ions still brings other advantages, namely the $\sqrt{N}$ amplification of the displacement signal~\cite{gilmoreQuantumenhancedSensingDisplacements2021} and the $\sqrt{N}$ reduction in protocol time achieved when using our protocol for preparing spin-dependent squeezed states.

\section{Measurement protocols and classical Fisher information}
Here, we analyze the time-reversal measurement schemes for estimating the displacement's magnitude, $|\beta|$, that were introduced in Sec.~\ref{sec:phase_insensitive_displacement_sensing} of the main text. The time-reversal sequence is composed of three steps,
\begin{align}
    \hat{S}(\zeta \hat{J}_z) \hat{D}(\beta) \hat{S}(- \zeta \hat{J}_z) = \hat{D}(\hat{\beta}),
\end{align}
where the equality is obtained by braiding the displacement and squeezing operators, and $\hat{\beta} = \beta \cosh\mathopen{(}\zeta \hat{J}_z\mathclose{)} - \beta^* \sinh\mathopen{(}\zeta \hat{J}_z\mathclose{)}$ is the spin-dependent displacement parameter. 

\subsection{Phase-insensitive amplitude estimation with a single spin}
For $N = 1$ spin initialized in $\ket{0}_b\ket{\psi_0}_s= \ket{0}_b(\ket{1/2}_s + \ket{-1/2}_s)/\sqrt{2}$, the final state after the time-reversal sequence is,
\begin{align}
    \ket{\psi_f} = \frac{1}{\sqrt{2}}[\hat{D}(\beta_-)\ket{0}_b \ket{-1/2}_s + \hat{D}(\beta_+) \ket{0}_b \ket{1/2}_s], 
\end{align}
where the spin-dependent displacement amplitudes are $\beta_- = \beta_{\rm re} e^{\zeta/2} + i \beta_{\rm im} e^{-\zeta/2}$ and $\beta_+ = \beta_{\rm re} e^{-\zeta/2} + i \beta_{\rm im} e^{\zeta/2}$. The state after a $\pi/2$ rotation about $\hat{J}_y$ becomes, 
\begin{align}
    e^{i \hat{J}_y \pi/2} \ket{\psi_f} = \frac{1}{2} \left[ \hat{D}(\beta_-) \ket{0}_b (\ket{-1/2}_s + \ket{1/2}_s) + \hat{D}(\beta_+) \ket{0}_b (\ket{1/2}_s - \ket{-1/2}_s \right].
\end{align}
The probability of measuring $\ket{1/2}_s$ in the Dicke basis is, 
\begin{align}\begin{split}
    P_{1/2} = \frac{1}{2} \left[ 1 + e^{-|\beta|^2 (\cosh(\zeta)-1)} \cos(2 \beta_{\rm re} \beta_{\rm im} \sinh(\zeta))\right] = 1 - \frac{1}{2} |\beta|^2 (\cosh(\zeta) -1) + \mathcal{O}(\beta^4),
\end{split}
\end{align}
where the second equality holds when $\beta \ll 1$, and is phase-insensitive to third order in $\beta$. 
Using the fact that $P_{-1/2} = 1- P_{1/2}$, the corresponding CFI is, 
\begin{align}
    F_{|\beta|} = \left(\frac{1}{P_{1/2}} + \frac{1}{1-P_{1/2}} \right)\left(\frac{dP_{1/2}}{d|\beta|}\right)^2 &= \frac{4 - 4\cosh(\zeta)}{|\beta|^2(\cosh(\zeta)-1)-2} \underset{\beta \rightarrow 0}{=} 4\langle \hat{n} \rangle, \label{eq:FBeta_N1}
\end{align}
where the last equality holds in the limit $\beta \rightarrow 0$. We also used the relation $\bra{\zeta/2}\hat{n} \ket{\zeta/2} = {\sinh^2(\zeta/2)} = {(\cosh(\zeta)-1)}/2$, which correspond to the mode occupation of the (spin-dependent squeezed) reference state. The CFI diverges in the limit $\langle \hat{n} \rangle \rightarrow 0$, because the time-reversal protocol reduces to a bosonic displacement only, so no information is mapped to the spin. For $\langle \hat{n} \rangle \gtrsim 1$ the CFI is a factor of two larger than the QFI, cf. Eq~\ref{eq:QFI_AbsBeta}, and therefore follows Heisenberg scaling. 

\subsection{Phase-insensitive amplitude estimation with GHZ spin states}
To perform phase-insensitive displacement amplitude estimation with GHZ spin states, we employ a one-axis twisting (OAT) unitary evolved for time $\pi/2$, 
\begin{align}
    \hat{U}_{\rm OAT} = \exp(-i \frac{\pi}{2} \hat{J}_x^2 ). 
\end{align}
When OAT acts on $\ket{-N/2}_s$, if the number of spins-$1/2$ is even it produces the GHZ state, 
\begin{align}
    \hat{U}_{\rm OAT} \ket{-N/2}_s = \frac{1}{\sqrt{2}} ( e^{-i\pi/4} \ket{-N/2}_s + e^{i \pi/4 + i \pi N/2} \ket{N/2}_s).
\end{align}
The final state after the time-reversal sequence, cf. Eq.~\ref{eq:braiding} of the main text, is,
\begin{align}
    \ket{\psi_f} = \frac{1}{\sqrt{2}}[e^{-i\pi/4} \hat{D}(\beta_-)\ket{0}_b \ket{-N/2}_s + e^{i \pi/4 + i \pi N/2} \hat{D}(\beta_+) \ket{0}_b \ket{N/2}_s], 
\end{align}
where $\beta_- = \beta_{\rm re} e^{N\zeta/2} + i \beta_{\rm im} e^{-N\zeta/2}$ and $\beta_+ = \beta_{\rm re} e^{-N\zeta/2} + i \beta_{\rm im} e^{N\zeta/2}$. Undoing the one-axis twisting that we used to prepare the GHZ state results in, 
\begin{align}
    \hat{U}_{\rm OAT}^\dagger \ket{\psi_f} = \frac{1}{\sqrt{2}}\left( \hat{D}(\beta_-) \ket{0}_b [ \ket{-N/2}_s + e^{-i \pi(N+1)/2} \ket{N/2}_s] + \hat{D}(\beta_+) \ket{0}_b [ \ket{-N/2}_s + e^{i \pi (N+1)/2 } \ket{N/2}_s ] \right), 
\end{align}
Using the fact that $e^{\pm i \pi (N+1)} = -1$ for even $N$, the probability of measuring the $\ket{N/2}_s$ state in the Dicke basis is then,
\begin{align}\begin{split}
    P_{N/2} &= \frac{1}{2} \left[ 1 - \Re \bra{0}\hat{D}^\dagger(\beta_-)\hat{D}(\beta_+)\ket{0} \right] = |\beta|^2 \sinh(N\zeta/2)^2 + \mathcal{O}(\beta^4),
\end{split}
\end{align}
where the second equality is obtained when $\beta \ll 1$, and is phase-insensitive to third order in $\beta$. In the limit $\beta \rightarrow 0$, the CFI for the estimation of $|\beta|$ is, 
\begin{align}\begin{split}
    F_{|\beta|} = \left(\frac{1}{P_{N/2}} + \frac{1}{1-P_{N/2}} \right)\left(\frac{dP_{N/2}}{d|\beta|}\right)^2 = 4\sinh(N\zeta/2)^2 = 4\langle \hat{n} \rangle,
\end{split}
\end{align}
which as a function of $\langle \hat{n} \rangle$ is the same sensitivity as for a single spin, cf. Eq.~\ref{eq:FBeta_N1}. For trapped ions, the advantage of using $N$ spins comes from the $\sqrt{N}$ reduction in preparation time of the reference state, as shown in Sec.~\ref{sec:SDSPreparation} of the main text, and the $\sqrt{N}$ amplification of the displacement signal~\cite{gilmoreQuantumenhancedSensingDisplacements2021}. 

\subsection{Phase-insensitive amplitude estimation with spin coherent states}
For $N>1$ spins initialized in a spin coherent state pointing along the $x$-axis of the collective Bloch sphere, $\ket{0}_b \ket{\psi_0}_s = \ket{0}_b \sum_m c_m \ket{m}_s$ where $c_m = 2^{-N/2} \sqrt{\binom{N}{m+N/2}}$, the final state after the time-reversal sequence is, 
\begin{align}
    \ket{\psi_f} = \sum_{m=-N/2}^{N/2} c_m \hat{D}(\beta_m) \ket{0}_b \ket{m}_s, 
\end{align}
where $\beta_m = \beta_{\rm re} e^{\zeta m} + i \beta_{\rm im} e^{- \zeta m}$ are the spin-dependent displacement amplitudes. 
We apply a $\pi/2$ rotation about $\hat{J}_y$, which, in the Dicke basis, is written as, 
\begin{align}
    e^{i \frac{\pi}{2} \hat{J}_y } = \sum_{m,m'=-N/2}^{N/2} d_{m',m}^{N/2}\left(\frac{\pi}{2}\right) \ketbra{m'}{m}_s.
\end{align}
The Wigner d-function, $d_{m',m}^{N/2}(\theta)$ for $\theta = \pi/2$ is, 
\begin{align}
    d_{m',m}^{N/2}\left(\frac{\pi}{2}\right) = 2^{-J} \sqrt{(J+m)! (J-m)! (J+m')! (J-m')!} \sum_{k=k_{\rm min}}^{k_{\rm max}} \frac{(-1)^{k-m+m'}}{(J+m'-k)!k!(J-m-k)!(k-m'+m)!}, 
\end{align}
where $k_{\rm min} = \rm{max}(0,m'-m)$, $k_{\rm max} = \rm{min}(J+m',J-m)$ and $J = N/2$. The rotated state is therefore, 
\begin{align}
    e^{i \hat{J}_y \pi/2} \ket{\psi_f} = \sum_{m,m'} c_m d_{m',m} \hat{D}(\beta_m) \ket{0}_b \ket{m'}_s,
\end{align}
where we dropped the Wigner d-function's $N/2$ superscript and $\pi/2$ dependence for notational simplicity. The probability of measuring $\ket{m'}_s$ in the Dicke basis is, 
\begin{align}
    P_{m'} = \left[\sum_{m=-N/2}^{N/2}  d_{m',m} c_m^* \bra{0,\beta_m} \right] \left[ \sum_{m=-N/2}^{N/2} d_{m',m} c_m \ket{\beta_m,0} \right]. \label{eq:Pm'}
\end{align}
We observe that for a spin coherent state pointing in the $x-y$ plane, the Dicke state coefficients obey the symmetry $c_m = c_{-m}$. Moreover, using the identity $d_{m',m}^{N/2}\left( \pi/2 \right) = (-1)^{J+m'}d_{m',-m}^{N/2}\left( \pi/2 \right)$, we can rewrite the sums in Eq.~\ref{eq:Pm'} as over positive $m$ values only to obtain, for odd $N$ (a similar expressions follows for $N$ even), 
\begin{align}\begin{split}
    P_{m'} = \sum_{m,n=1/2}^{N/2} d_{m',m} d_{m',n} c_m^* c_n [ \langle0|\hat{D}^\dagger(\beta_m)\hat{D}(\beta_n)|0\rangle + \langle0|\hat{D}^\dagger(\beta_{-m})\hat{D}(\beta_{-n})|0\rangle \\
    + (-1)^{N/2+m'}(\langle0|\hat{D}^\dagger(\beta_m)\hat{D}(\beta_{-n})|0\rangle + \langle0|\hat{D}^\dagger(\beta_{-m})\hat{D}(\beta_n)|0\rangle) ].
\end{split}
\end{align}
When $\beta \ll 1$, the expectation values involving displacement operators appearing in the first and second line can be written as, 
\begin{subequations}    
\begin{align}
    \langle0|\hat{D}^\dagger(\beta_m)\hat{D}(\beta_n)|0\rangle + \langle0|\hat{D}^\dagger(\beta_{-m})\hat{D}(\beta_{-n})|0\rangle &= 2 - |\beta|^2 \cosh((m+n)\zeta) \sinh^2((m-n)\zeta/2) + \mathcal{O}(\beta^4), \\ 
    \langle0|\hat{D}^\dagger(\beta_m)\hat{D}(\beta_{-n})|0\rangle + \langle0|\hat{D}^\dagger(\beta_{-m})\hat{D}(\beta_n)|0\rangle &= 2 - |\beta|^2 \left[ \cosh(2m\zeta) - 2 \cosh((m-n)\zeta) + \cosh(2n\zeta) \right] + \mathcal{O}(\beta^4).
\end{align}
\end{subequations}
Because these expressions are phase insensitive to third order in $\beta$, the probability $P_{m'}$ of Eq.~\ref{eq:Pm'} is also phase-insensitive to third order in $\beta$ for all values of $m'$. Therefore, the measurement outcomes encode phase-insensitive information about the displacement's magnitude. 

Computing the CFI for the estimation of $|\beta|$ requires summing over all $m'$ and calculating derivatives, which is challenging to perform exactly for general $N$ and $\zeta$. Instead, we calculate exact expressions for the CFI for odd numbers of spins up to $N = 7$, 
\begin{subequations}
\begin{align}
    F_{|\beta|}^{(N=1)} &= 2(\cosh(\zeta) - 1), \\ 
    F_{|\beta|}^{(N=3)} &= \frac{1}{2} [9 + 8 \cosh(\zeta) + 7\cosh(2\zeta)] \sinh(\zeta/2)^2, \\ 
    F_{|\beta|}^{(N=5)} &= \frac{1}{32}[165 + 204\cosh(\zeta) + 188 \cosh(2\zeta) + 52 \cosh(3\zeta) + 31 \cosh(4\zeta) ]\sinh(\zeta/2)^2, \\
    \nonumber F_{|\beta|}^{(N=7)} &= \frac{1}{512}[2954 + 4192\cosh(\zeta) + \\ & \quad\quad\quad\;\; 3953\cosh(2\zeta) + 1712\cosh(3\zeta) + 1158\cosh(4\zeta) + 240\cosh(5\zeta) + 127\cosh(6\zeta)] \sinh(\zeta/2)^2.
\end{align}
\end{subequations}
From the above results, we conjecture that the CFI for odd $N$ can be written in the general form,
\begin{align}
    F_{|\beta|} = \sinh(\zeta/2)^2 \sum_{i=0}^{N-1} a_i \cosh(i \zeta), 
\end{align}
with coefficients $a_i \in \mathbb{R}$. We report this expression in Eq.~\ref{eq:general_cfi_coherent_spin} of the main text. To compare the CCRB to the QCRB, we compute the ratio between the QFI and the CFI for odd $N$ up to $N=9$. From these exact expressions, we conjecture that, in the limit $\zeta \rightarrow \infty$, the ratio follows the function, 
\begin{align}
    Q_{|\beta|}/F_{|\beta|} = \frac{2^N}{2^N-1}.
\end{align}
This expression is reported in Eq.~\ref{eq:Q_F_Ratio}. We observe that in the limit $N \rightarrow \infty$, the CFI converges to the QFI. In this extreme regime, the CCRB of the time-reversal sequence using reference states obtained by applying spin-dependent squeezing to spin coherent states therefore saturates the Heisenberg limit. 

\subsection{Multi-parameter estimation with a single sensing spin}
Next, we propose a measurement sequence for jointly estimating $\beta_{\rm re}$ and $\beta_{\rm im}$. We consider $N = 1$ spin, and two ancilla qubits for readout. The measurement sequence, shown in Fig.~\ref{fig:multiparam_metrology} of the main text, requires wrapping the previous time-reversal sequence in bosonic displacements conditioned on the state of the ancillas, each with strength $g$. Note that for convenience, here we work with spin-dependent displacements that are $\hat{\sigma}_x$-conditioned, which are related by an ancilla basis rotation to the $\hat{\sigma}_z$-condition convention used in the main text, including in Fig.~\ref{fig:multiparam_metrology}. The measurement sequence unitary is, 
\begin{align}
    \hat{U}_p = \hat{D}(g \hat{\sigma}_x^{(2)}) \hat{D}(i g \hat{\sigma}_x^{(1)}) \hat{D}(\hat{\beta}) \hat{D}(-i g \hat{\sigma}_x^{(1)}) \hat{D}(-g \hat{\sigma}_x^{(2)}) &= \exp{2i \Im(i g \hat{\beta}^*\hat{\sigma}_x^{(1)} )} \exp{2i \Im(g \hat{\beta}^*\hat{\sigma}_x^{(2)}  )}  \hat{D}(\hat{\beta}), \label{eq:MPEstimation_Sequence}
\end{align}
where $\hat{\sigma}_x^{(i)}$ acts on the $i$th ancilla. The spin-dependent displacement parameter is $\hat{\beta} = \beta \cosh\mathopen{(}\zeta \hat{J}_z\mathclose{)} - \beta^* \sinh\mathclose{(}\zeta\hat{J}_z\mathclose{)}$. 
Braiding is achieved using $\hat{D}(a) \hat{D}(b) \hat{D}(-a) = e^{2i \Im(a b^*)} \hat{D}(a)$.
For a displacement signal $\beta = \beta_{\rm re} + i \beta_{\rm im}$, the ancilla rotations appearing in Eq.~\ref{eq:MPEstimation_Sequence} are, 
\begin{subequations}\begin{align}
    \Im (i g \hat{\sigma}_x^{(1)} \hat{\beta}^*) &= g \hat{\sigma}_x^{(1)} \begin{pmatrix}
        e^{- \zeta} \beta_{\rm re} & 0 \\ 0 & e^{\zeta} \beta_{\rm re}
    \end{pmatrix}, \\ 
     \Im (g \hat{\sigma}_x^{(2)} \hat{\beta}^*) &= - g \hat{\sigma}_x^{(2)}\begin{pmatrix}
        e^{\zeta} \beta_{\rm im} & 0 \\ 0 & e^{-\zeta} \beta_{\rm im}
    \end{pmatrix}.      
\end{align}\end{subequations}
In the regime where the squeezing is much larger than the conditional displacements, $\zeta \gg g$, and the displacement signal, $\zeta \gg \beta$, terms proportional to $e^{- \zeta}$ are negligible, and we can rewrite $\hat{U}_p$ as, 
\begin{align}
    \hat{U}_p \approx \exp{2i g e^\zeta \beta_{\rm re}  \ketbra{\downarrow}{\downarrow}_s\sigma_x^{(1)}} \exp{-2 ig e^\zeta \beta_{\rm im}   \ketbra{\uparrow}{\uparrow}_s\sigma_x^{(2)}}\hat{D}(\hat{\beta}). \label{eq:MPEstimation_SequenceLimit}
\end{align}

The system is initialized in $\ket{\psi_0} = \ket{0}_b\ket{+}_s\ket{\uparrow}_1 \ket{\uparrow}_2$, where the ancillas are denoted by numerical indices. The final state is, up to a normalization factor of $1/\sqrt{2}$,
\begin{align}
    \hat{U}_p \ket{\psi_0} \simeq  \hat{D}(\beta_\downarrow) \ket{0}_b \ket{\downarrow}_s [\cos(\phi_1) \ket{\uparrow}_1 + i \sin(\phi_1) \ket{\downarrow}_1] \ket{\uparrow}_2 + \hat{D}(\beta_\uparrow) \ket{0}_b \ket{\uparrow}_s \ket{\uparrow}_1 [\cos(\phi_2)\ket{\uparrow}_2 + i \sin(\phi_2) \ket{\downarrow}_2] , 
\end{align}
where we defined the rotation angles $\phi_1 = 2 g e^\zeta \beta_{\rm re}$ and $\phi_2 = -2 g e^{\zeta} \beta_{\rm im}$. The probability of measuring the $i$th ancilla in the state $\ket{\uparrow}$ is, 
\begin{align}
    P_{\uparrow}^{(i)} = \frac{1}{2}[1 + \cos^2(\phi_i)]. 
\end{align}
The measurement of the first ancilla reveals information about $\beta_{\rm re}$ only, while measurement of the second ancilla reveals information about $\beta_{\rm im}$ only. For the two independent measurement distributions $P_\uparrow^{(1)}$ and $P_\uparrow^{(2)}$, the metrological utility is determined by the classical Fisher information matrix,
\begin{align}
    F_{i,j} = \left(\frac{1} {P_\uparrow^{(1)}} + \frac{1}{1-P_\uparrow^{(1)}}\right) \frac{\partial P_{\uparrow}^{(1)}}{\partial \theta_i} \frac{\partial P_{\uparrow}^{(1)}}{\partial \theta_j} + \left(\frac{1} {P_\uparrow^{(2)}} + \frac{1}{1-P_\uparrow^{(2)}}\right) \frac{\partial P_{\uparrow}^{(2)}}{\partial \theta_i} \frac{\partial P_{\uparrow}^{(2)}}{\partial \theta_j}.
\end{align}
In the limit $\beta \rightarrow 0$ the CFIM is, 
\begin{align}
    \mathbf{F} =  8 g^2 e^{2 \zeta} \mathbb{1}.
\end{align}
For this measurement protocol, the multi-parameter variance is therefore lower bounded bounded according to the CCRB,  
\begin{align}
    V(\beta_{\rm re}) + V(\beta_{\rm im}) \geq \Tr(\mathbf{F}^{-1}) = \frac{1}{4 g^2 e^{2 \zeta}},
\end{align}
which we report in Eq.~\ref{eq:CCRB_MP} of the main text. The CCRB decreases as a function of the squeezing, $\zeta$, and also as a function of the spin-dependent displacement strength, $g$. Although increasing $g$ is a viable strategy to reduce the CCRB, maintaining the regime $\zeta \gg g$ which was used to derive Eq.~\ref{eq:MPEstimation_SequenceLimit} requires simultaneously increasing $\zeta$. The metrological performance of the protocol is studied in Fig.~\ref{fig:multiparam_metrology} of the main text.

\section{Further details on the spin-dependent squeezed state preparation protocol}
In this section, we provide additional derivations, error analysis and numerical simulations for our spin-dependent squeezed state preparation protocol of Sec.~\ref{sec:SDSPreparation} of the main text. 

\subsection{Derivation of Hamiltonian}
The Hamiltonian in Eq.~\ref{eq:Hamiltonian} of the main text is obtained by driving the first-order sidebands at four different frequencies: one pair of red- and blue- sideband interactions with detuning $\Delta(t)$, and another pair of red- and blue- sideband interactions with detuning $-\Delta(t)$. In the rotating frame with respect to the spin and the motion, ignoring all far-off-resonant motional modes, the four interaction Hamiltonians are, 
\begin{subequations}\begin{align}
    \hat{H}_{\rm r1}(t) &= \frac{g(t)}{2} ( \hat{a} \hat{J}_+ e^{-i \phi_{\rm r1}(t) - i t \Delta(t)} + \hat{a}^\dagger \hat{J}_- e^{i \phi_{\rm r1}(t) + i t \Delta(t)}), \\ 
    \hat{H}_{\rm r2}(t) &= \frac{g(t)}{2} ( \hat{a} \hat{J}_+ e^{-i \phi_{\rm r2}(t) + i t \Delta(t)} + \hat{a}^\dagger \hat{J}_- e^{i \phi_{\rm r2}(t) - i t \Delta(t)}), \\ 
    \hat{H}_{\rm b1}(t) &= \frac{g(t)}{2} ( \hat{a}^\dagger \hat{J}_+ e^{-i \phi_{\rm b1}(t) + i t \Delta(t)} + \hat{a} \hat{J}_- e^{i \phi_{\rm b1}(t) - i t \Delta(t)}), \\ 
    \hat{H}_{\rm b2}(t) &= \frac{g(t)}{2} ( \hat{a}^\dagger \hat{J}_+ e^{-i \phi_{\rm b2}(t) - i t \Delta(t)} + \hat{a} \hat{J}_- e^{i \phi_{\rm b2}(t) + i t \Delta(t)}), 
\end{align}\end{subequations}
where $\phi_{\rm r1}(t)$, $\phi_{\rm r2}(t)$, $\phi_{\rm b1}(t)$ and $\phi_{\rm b2}(t)$ are time-dependent phases. After setting $\phi_{\rm b1}(t) = -\phi_{\rm r1}(t)$, the sum of the Hamiltonians $\hat{H}_{\mathrm{r}_1}(t)$ and $\hat{H}_{\mathrm{b}_1}(t)$ gives, 
\begin{align}
    \hat{H}_{\rm r1}(t) + \hat{H}_{\rm b1}(t) = \frac{g(t)}{2} \left[ \hat{a} (\hat{J}_+ + \hat{J}_-) e^{-i \phi_{\rm r1}(t) - i t \Delta(t)} + \text{h.c.} \right].
\end{align}
Setting $\phi_{\rm r2}(t) \rightarrow \phi_{\rm r2}(t) + \pi/2$ and $\phi_{\rm b2}(t) \rightarrow -\phi_{\rm r2}(t) + \pi/2$, the sum of the Hamiltonians $\hat{H}_{\mathrm{r}_2}(t)$ and $\hat{H}_{\mathrm{b}_2}(t)$ is, 
\begin{align}
    \hat{H}_{r2}(t) + \hat{H}_{b2}(t) = \frac{g(t)}{2} \left[ \hat{a}(-i \hat{J}_+ + i \hat{J}_-) e^{-i \phi_{r2}(t) + i t \Delta(t)} + \text{h.c.} \right].
\end{align}
Replacing $(\hat{J}_+ + \hat{J}_-)/2 = \hat{J}_x$ and $(-i\hat{J}_+ + i\hat{J}_-)/2 = \hat{J}_y$, the sum of all four Hamiltonians is, 
\begin{align}
    \hat{H}(t) = g(t) \hat{a}\left[ \hat{J}_x e^{-i \phi_{r1}(t) - i t \Delta(t)} + \hat{J}_y e^{-i \phi_{r2}(t) + i t \Delta(t) }\right] + \rm{h.c.}.
\end{align}
Setting $\phi_{\rm r1}(t) = 0$ and relabeling $\phi_{\rm r2}(t) = \phi(t)$ yields the Hamiltonian reported in Eq.~\ref{eq:Hamiltonian} of the main text. 

\subsection{Errors due to higher-order Magnus expansion terms}

In Sec.~\ref{sec:SDSPreparation} of the main text, we neglected higher-order Magnus terms. Here, we provide additional justification for this step. 

\begin{figure}
    \centering
    \includegraphics[scale=1]{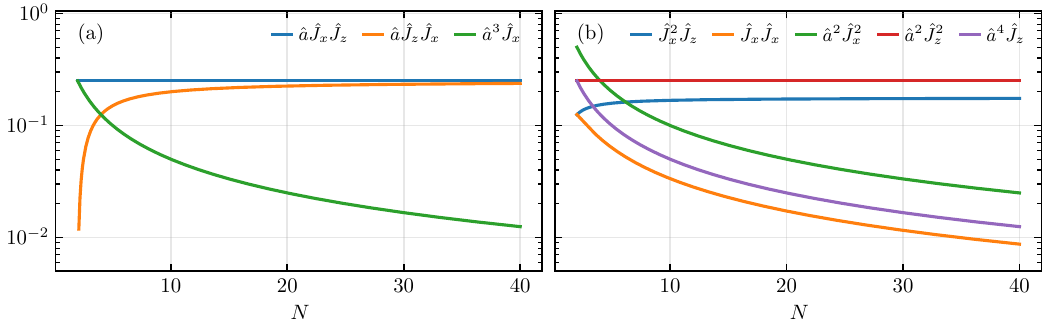}
    \caption{
    Magnitude of the spin operator matrix element that drives transitions out of the $\ket{N/2}_s$ Dicke state, for the terms that appear at third-order (panel (a)) and fourth-order (panel (b)) of the Magnus expansion. Although the higher-order Magnus terms include bosonic operators, which are included in the legend for clarity, the lines plotted are determined only by the collective spin operator's matrix elements. 
    All terms are either bounded, constant, or decreasing functions of $N$. The most deleterious terms for the bosonic mode, i.e. those that are cubic and quartic in $\hat{a}$,$\hat{a}^\dagger$, decrease in strength as $1/N$. 
    }
    \label{fig:sm_higher_order_Magnus_terms}
\end{figure}

We first note that third-order terms are of the form of $\hat{a}^3\hat{J}_x$ and $\hat{a}\hat{J}_x \hat{J}_z$, and fourth order terms are of the form $\hat{a}^4 \hat{J}_z$, $\hat{a}^2 \hat{J}_x^2$, $\hat{a}^2 \hat{J}_z$, $\hat{a}^2 \hat{J}_z \hat{J}_x$, $\hat{J}_x^2$ and $\hat{J}_z \hat{J}_x \hat{J}_y$. To investigate their effects, we plot, for each higher-order Magnus term as a function of $N$, the magnitude of the spin operator's matrix element that drives transitions out of the fully-polarized spin state (see Fig.~\ref{fig:sm_higher_order_Magnus_terms}). For example, for $\hat{J}_x$, the relevant matrix element is $ \bra{N/2-1}\hat{J}_-\ket{N/2}_s$, while higher-order spin operators will include powers of $\hat{J}_+$, $\hat{J}_-$ and the relevant element of $\hat{J}_z$. As another example, for $\hat{J}_x^2$, a relevant matrix element is $\bra{N/2-2}\hat{J}_-^2\ket{N/2}_s$. 
In Fig.~\ref{fig:sm_higher_order_Magnus_terms}, we observe that all terms are either bounded above, constant or decreasing functions of $N$. In particular, the terms with cubic or quartic powers of creation and annihilation operators (green line in panel (a), purple line in panel (b)) are the most deleterious terms for the bosonic mode. These are also the terms whose spin operator's matrix element decreases as $1/N$. As such, we expect that our protocol can be scaled to many ions. Our numerical simulations presented in Fig.~\ref{fig:sds_preparation_numerics} supports this analysis, where we observe a $\sqrt{N}$ reduction in protocol time when scaling to $N$ ions. 

\subsection{Numerical results: spin-polarized states and increased phase loops}

\begin{figure}
    \centering
    \includegraphics[scale=1]{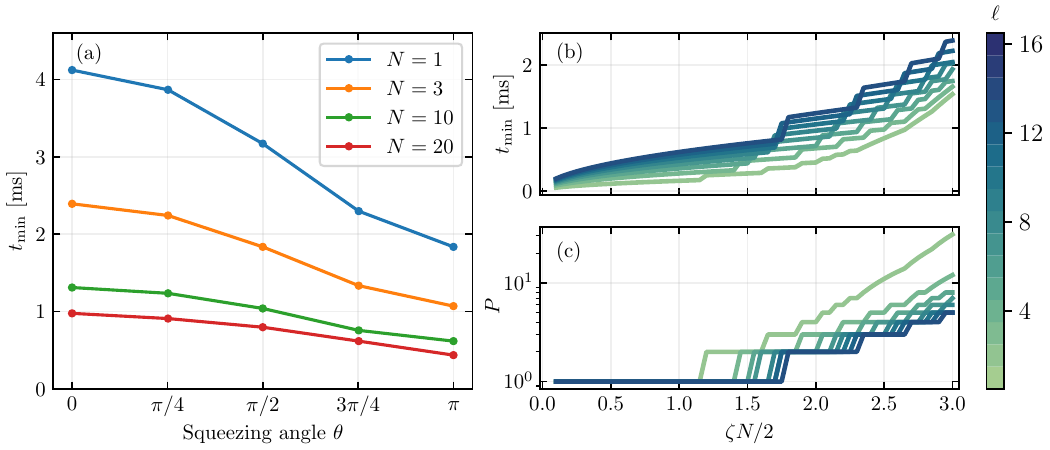}
    \caption{
    \textbf{(a)}
    Minimum protocol time, $t_{\rm min}$, to prepare the target spin-dependent squeezed state, $\ket{\psi_t} = \hat{S}(\zeta \hat{J}_z) \ket{0}_b \ket{N/2}_s$, with fidelity $\mathcal{F} > 0.99$ for various number of spins, $N$. The Hamiltonian parameters are the same as the main text, see Fig.~\ref{fig:sds_preparation_numerics}. Here, we fix the target squeezing parameter's magnitude, $|\zeta| = 2.5$, and vary the squeezing angle, $\theta = \arg(\zeta)$. The protocol time decreases with $N$, and is minimized when $\theta = \pi$. 
    \textbf{(b)} Minimum protocol time, $t_{\rm min}$, versus the amount of spin-dependent squeezing, $\zeta N/2$, for $\ell$ repetitions of a stroboscopic segment. Larger $\ell$ results in longer protocol times. $\textbf{(c)}$ Number of times, $P$, that the four-step stroboscopic sequence is repeated. Increasing the number of times $\ell$ that each stroboscopic segment is repeated allows for decreased $P$. 
    }
\label{fig:sm_sds_preparation_numerics}
\end{figure}

Applying spin-dependent squeezing to a spin-polarized state such as $\ket{N/2}_s$ produces a simple product state, $\ket{\zeta m/2}_b \ket{N/2}_s$. Studying the performance of our protocol starting from this state is still instructive, because it reveals a strong dependence on the target squeezing angle, as seen in Fig.~\ref{fig:sm_sds_preparation_numerics}. In panel (a), we plot the minimum protocol time, $t_{\rm min}$, to prepare a target spin-dependent squeezed state, $\ket{\psi_t} = \hat{S}(\zeta \hat{J}_z) \ket{0}_b \ket{N/2}_s$, with fidelity $\mathcal{F} > 0.99$ for varying target angles $\theta = \arg(\zeta)$. We observe a strong dependence on $\theta$, which can be understood from the first stroboscopic segment's third-order Magnus term, 
\begin{align}
    \hat{\Theta}_3[0,\tau] = \frac{3 g^3 \tau^3}{4\pi^2}\bigg[ -\hat{a}^3 (e^{-2i\phi} \hat{J}_x + e^{-i \phi} \hat{J}_y) + \hat{a}^\dagger \hat{a}^2 (\hat{J}_x + e^{-i \phi} \hat{J}_y ) + \hat{a} (\hat{J}_x + e^{-i \phi} \hat{J}_y + 2i \hat{J}_z \hat{J}_y - 2i e^{-i \phi} \hat{J}_z \hat{J}_x) \bigg] + \text{h.c.},
\end{align}
where $\phi$ is the phase in the Hamiltonian during the segment. Several terms, including those that are cubic in bosonic creation and annihilation operators, are independent of $\phi$. As such, the target squeezing angle should be chosen so that terms independent of $\phi$ act in the anti-squeezed quadrature, and hence their effect is masked by the increased variance. This occurs when $\theta = \pi$, as seen in Fig.~\ref{fig:sm_sds_preparation_numerics}, where the protocol's duration is significantly reduced compared to when $\theta = 0$. 

Next, we consider varying the number of phase-space loops, $\ell$, of each stroboscopic segment. In Fig.~\ref{fig:sm_sds_preparation_numerics}(b), we set $N = 20$ and vary $\ell$. The step-like behavior arises because a full phase-space loop must be completed, i.e. $t_f$ must be an integer multiple of $2\pi/|\Delta|$ where $\Delta$ is the detuning. Increasing $\ell$ results in longer $t_{\rm min}$ because the higher-order Trotter terms are larger with $\ell$. The benefit of increased $\ell$ is shown in panel (c), which plots the number of repetitions, $P$, of the four-step stroboscopic sequence. We observe that increasing $\ell$ results in smaller $P$, which means that fewer stroboscopic changes of the phase and detuning are required. This may be of benefit experimentally. 

\section{Metrological utility in the presence of spin dephasing}
In this section, we derive analytical expressions for the quantum and classical Fisher information for spin-dependent squeezed states in the presence of single-spin dephasing, i.e. dephasing $\hat{\sigma}_z$ that occurs identically but independently across all $N$ spins-$1/2$.  

\subsection{Quantum Fisher information matrix for displacement sensing}
Recall that for a mixed state, $\rho(\bm{\theta})$, the quantum Fisher information matrix (QFIM) is~\cite{liuQuantumFisherInformation2020},
\begin{align}
    Q_{i,j} = \sum_{k,l,\lambda_k + \lambda_l \neq 0} \frac{2 \Re(\bra{\lambda_k}  \partial_i \rho(\bm{\theta}) \ket{\lambda_l} \bra{\lambda_l} \partial_j \rho(\bm{\theta}) \ket{\lambda_k} ) }{\lambda_k + \lambda_l}, \label{eq:QFIM_Rho}
\end{align}
where $\ket{\lambda_k}$ and $\lambda_k$ are the eigenvectors and eigenvalues of $\rho(\bm{\theta})$, respectively. The derivative is denoted $\partial_j = \partial/\partial \theta_j$, and $\bm{\theta}$ is the vector of parameters that are to be estimated. For any density matrix $\rho(\bm{\theta})$ whose dependence on $\bm{\theta}$ enters unitarily, i.e. $\rho(\bm{\theta}) = \hat{U}(\bm{\theta}) \rho_{\rm ref} \hat{U}^\dagger(\bm{\theta})$, the derivative with respect to the $i$th parameter is, 
\begin{align}
    \partialD{\rho(\beta)}{\theta_i} = \partialD{\hat{U}(\bm{\theta})}{\theta_i} \rho_{\rm ref} \hat{U}^\dagger(\bm{\theta}) + \hat{U}(\bm{\theta}) \rho_{\rm ref}  \partialD{\hat{U}^\dagger(\bm{\theta})}{\theta_i}. 
\end{align}
Specifically for displacement sensing, $\hat{U} = \hat{D}(\beta) = {\exp(\beta \hat{a}^\dagger - \beta^* \hat{a})}$ is the displacement operator. For sensing in cartesian coordinates, the estimation parameters are $\bm{\theta} = \{\beta_{\rm re},\beta_{\rm im}\}$. In terms of these parameters the displacement operator can be written as, 
\begin{align}
    \hat{D}(\beta) = \exp(-i \beta_{\rm re} \beta_{\rm im}) \exp(i \beta_{\rm im} \hat{x}) \exp(-i \beta_{\rm re} \hat{p}). 
\end{align}
After computing the derivatives of $\hat{D}(\beta)$ with respect to $\beta_{\rm re}$ and $\beta_{\rm im}$, we arrive at the derivatives of the density matrix $\rho(\beta)$, 
\begin{align}
    \partialD{\rho(\beta)}{\beta_{\rm re}} &= -i[\hat{p},\rho(\beta)], \qquad \partialD{\rho(\beta)}{\beta_{\rm im}} = i[\hat{x},\rho(\beta)]. \label{eq:rhoDerivative}
\end{align}
We note that $\partial_i \rho(\beta)$ is Hermitian since $\hat{x}$, $\hat{p}$ and $\rho(\beta)$ are Hermitian, and the commutator of two Hermitian operators is anti-Hermitian. 

Without noise, the reference state obtained from spin-dependent squeezing is $\ket{\psi_{\rm ref}} = \hat{S}(\zeta \hat{J}_z) \ket{0}_b \ket{\psi_0}_s$. For the choice of initial spin state, we focus on GHZ spin states since we expect their metrological performance to be the most sensitive to spin dephasing. Accounting for single-spin dephasing during the spin-dependent squeezing of duration $t$ at a rate $\gamma_s$, the resulting reference state is, 
\begin{align}
    \rho_{\rm ref} = \frac{1}{2}(\ket{z}_b \ketbra{N/2}_s \bra{z}_b +  \ket{-z}_b \ketbra{-N/2}_s \bra{-z}_b + e^{- \gamma t N } \ket{z}_b \ketbra{N/2}{-N/2}_s \bra{-z}_b + \text{h.c.} ). 
\end{align}
For this reference state, the eigenvalues of $\rho(\beta)$ are, 
\begin{align}
    \{\lambda_1 = \frac{1+e^{-\gamma t N}}{2}, \lambda_2 = \frac{1-e^{-\gamma t N}}{2}\} \cup \{ \Lambda_c \}_{c=1}^\infty \cup \{ \Lambda_d\}_{c=1}^\infty, 
\end{align}
with two sets of infinitely-many degenerate zero eigenvalues, $\Lambda_c = \Lambda_d = 0 \; \forall \; c,d \geq 1$. We write a corresponding set of orthonormal eigenvectors of $\rho(\beta)$ using $2 \times 2$ matrix notation in the basis of spin states $\{\ket{N/2}_s, \ket{-N/2}_s \}$ as, 
\begin{subequations}
\begin{align}
    \ket{\lambda_1} &= \frac{1}{\sqrt{2}} \begin{pmatrix}
        \ket{\tilde{z}}_b \\ \ket{-\tilde{z}}
    \end{pmatrix}, \\ 
    \ket{\lambda_2} &= \frac{1}{\sqrt{2}} \begin{pmatrix}
        \ket{\tilde{z}}_b \\ -\ket{-\tilde{z}}
    \end{pmatrix}, \\ 
    \ket{\Lambda_c} &= \begin{pmatrix}
        \ket{\tilde{z},c} \\ 0
    \end{pmatrix}, \\ 
    \ket{\Lambda_d}& = \begin{pmatrix}
        0 \\ \ket{-\tilde{z},d}
    \end{pmatrix},
\end{align}
\end{subequations}
where the indices $c$ and $d$ are for all integers $c,d \geq 1$, and where we dropped the $b$ subscript on bosonic states to simplify the notation. We also introduced notation for displaced squeezed Fock states as $\ket{\tilde{z},n} = \hat{D}(\beta) \hat{S}(z) \ket{n}$ and $\ket{-\tilde{z},n} = \hat{D}(\beta) \hat{S}(-z) \ket{n}$. Similarly, displaced squeezed vacuum states are denoted $\ket{\tilde{z}} = \hat{D}(\beta) \hat{S}(z) \ket{0}$ and $\ket{-\tilde{z}} = \hat{D}(\beta) \hat{S}(-z) \ket{0}$. We constructed the set of orthogonal eigenvectors for the degenerate zero eigenvalues using the identity $\bra{n'} \hat{S}^\dagger(z) \hat{S}(z) \ket{n} = \delta_{n,n'}$ for any Fock states $\ket{n'},\ket{n}$. 

After writing the displaced reference state, $\rho(\beta) = \hat{D}(\beta) \rho_{\rm ref} \hat{D}(\beta)$, in the same spin matrix notation as, 
\begin{align}
    \rho(\beta) = \frac{1}{2} \begin{pmatrix} 
    \ketbra{\tilde{z}} & e^{- \gamma t N} \ketbra{\tilde{z}}{-\tilde{z}} \\ 
    e^{- \gamma t N} \ketbra{-\tilde{z}}{\tilde{z}} & \ketbra{-\tilde{z}}
    \end{pmatrix}, 
\end{align}
we can compute the overlaps that appear in the QFIM of Eq.~\ref{eq:QFIM_Rho}. For overlaps between $\ket{\lambda_1}$ and $\ket{\lambda_2}$ and their cross terms, we have, 
\begin{align}
    \bra{\lambda_1} \partial_i \rho \ket{\lambda_1} = \bra{\lambda_2} \partial_i \rho \ket{\lambda_2} = \bra{\lambda_1} \partial_i \rho \ket{\lambda_2} = 0,
\end{align}
which follows because, 
\begin{align}
    \bra{\lambda_1} \partial_1 \rho \ket{\lambda_1} &= \frac{-i}{2} \bra{\lambda_1}
    \left\{ \begin{pmatrix}
        \hat{p} \ketbra{\tilde{z}} & e^{- \gamma t N} \hat{p} \ketbra{\tilde{z}}{-\tilde{z}} \\ 
        e^{- \gamma t N} \hat{p} \ketbra{-\tilde{z}}{\tilde{z}} & \hat{p} \ketbra{\tilde{z}}
    \end{pmatrix} 
    -
    \begin{pmatrix}
        \ketbra{\tilde{z}}\hat{p}  & e^{- \gamma t N} \ketbra{\tilde{z}}{-\tilde{z}}\hat{p}  \\ 
        e^{- \gamma t N} \ketbra{-\tilde{z}}{\tilde{z}} \hat{p}  & \ketbra{\tilde{z}} \hat{p}
    \end{pmatrix} \right\} 
    \ket{\lambda_1} \\ 
    &= (1+e^{- \gamma t N})\bra{\tilde{z}} \hat{p} \ket{\tilde{z}} + (1+e^{- \gamma t N})\bra{-\tilde{z}} \hat{p} \ket{-\tilde{z}} - (1+e^{- \gamma t N})\bra{\tilde{z}} \hat{p} \ket{\tilde{z}} - (1+e^{- \gamma t N})\bra{-\tilde{z}} \hat{p} \ket{-\tilde{z}}  \\ 
    &= 0, 
\end{align}
and similarly for the other cases. Moreover, any terms involving overlaps between any combination of $\ket{\Lambda_c}$ and $\ket{\Lambda_d}$ must be zero because $\Lambda_c = \Lambda_d = 0 \; \forall \; c,d \geq 1$. The only surviving terms in the QFIM are therefore, 
\begin{align}\begin{split}
    Q_{i,j} = 2\sum_{c,d} &\frac{\Re(\bra{\lambda_1}  \partial_i \rho(\bm{\theta}) \ket{\Lambda_c} \bra{\Lambda_c} \partial_j \rho(\bm{\theta}) \ket{\lambda_1} ) }{\lambda_1 + \Lambda_c} 
    + \frac{\Re(\bra{\lambda_1}  \partial_i \rho(\bm{\theta}) \ket{\Lambda_d} \bra{\Lambda_d} \partial_j \rho(\bm{\theta}) \ket{\lambda_1} ) }{\lambda_1 + \Lambda_d} \\ 
    &+ \frac{\Re(\bra{\lambda_2}  \partial_i \rho(\bm{\theta}) \ket{\Lambda_c} \bra{\Lambda_c} \partial_j \rho(\bm{\theta}) \ket{\lambda_2} ) }{\lambda_2 + \Lambda_c} 
    + \frac{\Re(\bra{\lambda_2}  \partial_i \rho(\bm{\theta}) \ket{\Lambda_d} \bra{\Lambda_d} \partial_j \rho(\bm{\theta}) \ket{\lambda_2} ) }{\lambda_2 + \Lambda_d} \\ 
    &+  \frac{\Re(\bra{\Lambda_c}  \partial_i \rho(\bm{\theta}) \ket{\lambda_1} \bra{\lambda_1} \partial_j \rho(\bm{\theta}) \ket{\Lambda_c} ) }{\Lambda_c + \lambda_1} 
    + \frac{\Re(\bra{\Lambda_c}  \partial_i \rho(\bm{\theta}) \ket{\lambda_2} \bra{\lambda_2} \partial_j \rho(\bm{\theta}) \ket{\Lambda_c} ) }{\Lambda_c + \lambda_2} \\ 
    &+ \frac{\Re(\bra{\Lambda_d}  \partial_i \rho(\bm{\theta}) \ket{\lambda_1} \bra{\lambda_1} \partial_j \rho(\bm{\theta}) \ket{\Lambda_d} ) }{\Lambda_d + \lambda_1} +
    \frac{\Re(\bra{\Lambda_d}  \partial_i \rho(\bm{\theta}) \ket{\lambda_2} \bra{\lambda_2} \partial_j \rho(\bm{\theta}) \ket{\Lambda_d} ) }{\Lambda_d + \lambda_2},
\end{split}\end{align}
The surviving terms evaluate to,
\begin{subequations}\begin{align}
    \bra{\lambda_1} \partial_i \rho \ket{\Lambda_c} &= \frac{i \lambda_1}{\sqrt{2}} \bra{\tilde{z}} \hat{A}_i \ket{\tilde{z},c} \\ 
    \bra{\lambda_2} \partial_i \rho \ket{\Lambda_c} &= \frac{i\lambda_2}{\sqrt{2}} \bra{\tilde{z}} \hat{A}_i \ket{\tilde{z},c}, \\ 
    \bra{\lambda_1} \partial_i \rho \ket{\Lambda_d} &= \frac{i\lambda_1}{\sqrt{2}} \bra{-\tilde{z}} \hat{A}_i \ket{-\tilde{z},d}, \\ 
    \bra{\lambda_2} \partial_i \rho\ket{\Lambda_d} &= \frac{i\lambda_2}{\sqrt{2}} \bra{-\tilde{z}} \hat{A}_i \ket{-\tilde{z},d}, 
\end{align}\end{subequations}
where we introduced $\hat{A} = \{\hat{p},-\hat{x}\}$. These results follow from calculations such as, 
\begin{align}
    \bra{\Lambda_1} \partial_i \rho \ket{\lambda_c} &= \frac{-i}{2} \bra{\lambda_1}
    \left\{ \begin{pmatrix}
        \hat{A}_i \ketbra{\tilde{z}} & e^{- \gamma t N} \hat{A}_i \ketbra{\tilde{z}}{-\tilde{z}} \\ 
        e^{- \gamma t N} \hat{A}_i \ketbra{-\tilde{z}}{\tilde{z}} & \hat{A}_i \ketbra{\tilde{z}}
    \end{pmatrix} 
    -
    \begin{pmatrix}
        \ketbra{\tilde{z}} \hat{A}_i  & e^{- \gamma t N} \ketbra{\tilde{z}}{-\tilde{z}} \hat{A}_i  \\ 
        e^{- \gamma t N} \ketbra{-\tilde{z}}{\tilde{z}} \hat{A}_i  & \ketbra{\tilde{z}} \hat{p}
    \end{pmatrix} \right\} 
    \ket{\Lambda_c} \\ 
    &= 
    \frac{i}{2\sqrt{2}} \begin{pmatrix}
        \bra{\tilde{z}} & \bra{-\tilde{z}}
    \end{pmatrix} 
    \begin{pmatrix}
        \ket{\tilde{z}} \bra{\tilde{z}} \hat{A}_i \ket{\tilde{z},c} \\ 
        e^{- \gamma t N} \ket{-\tilde{z}} \bra{\tilde{z}} \hat{A}_i \ket{\tilde{z},c} \\ 
    \end{pmatrix} \\ 
    &= \frac{i}{2\sqrt{2}} (1+e^{- \gamma t N}) \bra{\tilde{z}} \hat{A}_i \ket{\tilde{z},c}, 
\end{align}
with the other cases following similarly. The QFIM is therefore,
\begin{align}\begin{split}
    Q_{i,j} &= \sum_{c=1}^{\infty} (\lambda_1 + \lambda_2) (\bra{\tilde{z}} \hat{A}_i \ket{\tilde{z},c}\bra{\tilde{z},c} \hat{A}_j \ket{\tilde{z}} + \bra{\tilde{z}} \hat{A}_j \ket{\tilde{z},c}\bra{\tilde{z},c} \hat{A}_i \ket{\tilde{z}}) \\ 
    &+ \sum_{d=1}^{\infty } (\lambda_1 + \lambda_2) (\bra{-\tilde{z}} \hat{A}_i \ket{-\tilde{z},d}\bra{-\tilde{z},d} \hat{A}_j \ket{-\tilde{z}} + \bra{-\tilde{z}} \hat{A}_j \ket{-\tilde{z},d}\bra{-\tilde{z},d} \hat{A}_i \ket{-\tilde{z}}), 
\end{split}\end{align}
which, after unifying the summation variables and using the fact that $\lambda_1 + \lambda_2 = 1$ becomes, 
\begin{align}
    Q_{i,j} = \sum_{n\geq1}^{\infty} \left[ \bra{\tilde{z}} \hat{A}_i \ket{\tilde{z},n}\bra{\tilde{z},n} \hat{A}_j \ket{\tilde{z}} + \bra{-\tilde{z}} \hat{A}_i \ket{-\tilde{z},n}\bra{-\tilde{z},n} \hat{A}_j \ket{-\tilde{z}} + {\rm h.c.} \right], 
\end{align}
Finally, using the resolution of identity of the Fock states, $\sum_{n \geq 1} \ketbra{n} = 1 - \ketbra{0}$, we rewrite the QFIM as, 
\begin{align}
    Q_{i,j} &= 2 \Re \left[ \bra{\tilde{z}} \hat{A}_i \hat{A}_j \ket{\tilde{z}} - \bra{\tilde{z}} \hat{A}_i \ket{\tilde{z}} \bra{\tilde{z}} \hat{A}_j \ket{\tilde{z}} + \bra{-\tilde{z}} \hat{A}_i \hat{A}_j \ket{-\tilde{z}} - \bra{-\tilde{z}} \hat{A}_i \ket{-\tilde{z}} \bra{-\tilde{z}} \hat{A}_j \ket{-\tilde{z}} \right]
\end{align}
The diagonal elements reduce to the variance, i.e. $Q_{ii} = 2(\Delta \hat{A}_i)_{\ket{\tilde{z}}}^2 + 2(\Delta \hat{A}_i)_{\ket{\tilde{-z}}}^2$. The quadrature variances are unchanged by a displacement, and therefore we obtain, 
\begin{align}
    Q_{1,1} = Q_{2,2} = 2e^{-2z} + 2e^{2z} = 4 \cosh(2z). 
\end{align}
Meanwhile, the off-diagonal elements vanish, $Q_{1,2} = Q_{2,1} = 0$, because the expectation value of $\langle \hat{x} \hat{p} \rangle$ is shifted by the displacement, with the real part canceled exactly by the $\langle \hat{x} \rangle \langle \hat{p} \rangle$ term. The QFIM is therefore, 
\begin{align}
    \bm{Q} = 4 \cosh(2z) \mathbb{1},
\end{align}
which is independent of the single-spin dephasing rate $\gamma_s$ and time, $t$. We therefore conclude that the metrological performance of spin-dependent squeezed states for displacement sensing is insensitive to spin dephasing. 

\subsection{Classical Fisher information for estimating a displacement amplitude}
Next, we calculate the classical Fisher information (CFI) of our time-reversal protocol for estimating a displacement amplitude, cf. Fig.~\ref{fig:phase_insensitive_schematic} of the main text. We consider an initial spin state that is prepared using one-axis twisting (OAT), cf. Eq.~\ref{eq:U_OAT}. We assume that spin dephasing acts during both the forward and reverse spin-dependent squeezing operations. After the spin-dependent squeezing, unknown displacement, and reverse spin-dependent squeezing, the final state is, 
\begin{align}
    \rho_{\rm f} = \frac{1}{2} (  \ket{-N/2} \ketbra{\beta_-} \bra{-N/2}  + \ket{N/2} \ket{\beta_+}\bra{\beta_+} \bra{N/2} + e^{- \gamma t N - i \pi (N+1)/2} \ket{-N/2} \ketbra{\beta_-}{\beta_+} \bra{N/2}  + \text{h.c.}  ),
\end{align}
Note that in the limit $\gamma t N \rightarrow \infty$ the off-diagonal coherences vanish, and therefore the resulting state is not spin-boson entangled. 

Before measurement in the computational basis, the one-axis twisting is reversed, i.e. $\hat{U}_{\rm OAT}^\dagger \rho_f \hat{U}r_{\rm OAT}$. To write the state after this step, we use the following identities, 
\begin{subequations}
    \begin{align}
        \hat{U}_{\rm OAT}^\dagger \ket{-N/2} &= \frac{1}{\sqrt{2}} \left( e^{i \pi/4} \ket{-N/2} + e^{-i \pi/4 - i \pi N/2} \ket{N/2} \right) \\ 
        U_{\rm OAT}^\dagger \ket{N/2} &= \frac{1}{\sqrt{2}} \left( -e^{i \pi/4 - i \pi N/2} \ket{-N/2} + e^{i \pi/4} \ket{N/2} \right). 
    \end{align}
\end{subequations}
Immediately after the application of $\hat{U}^\dagger_{\rm OAT}$ we apply the projector $\ketbra{N/2}$. Only keeping terms that survive this projection, the probability of measuring $\ket{N/2}$ is therefore, 
\begin{align}
    \tr{\ketbra{N/2} \hat{U}_{\rm OAT}^\dagger \rho_{\rm f} \hat{U}_{\rm OAT} } &= \frac{1}{4} \tr{ \ket{\beta_-}\bra{\beta_-} + \ket{\beta_+} \bra{\beta_+} + e^{- \gamma t N} e^{-2i \pi (N+1)/2} \ket{\beta_-} \bra{\beta_+} + \text{h.c.}} \\ 
    &= \frac{1}{2} (1 - e^{- \gamma t N} \Re( \bra{0} \hat{D}^\dagger(\beta_+) \hat{D}(\beta_-)\ket{0} ), 
\end{align}
where for even $N$, we used the identity ${\exp(-2i \pi(N+1)/2)} = -1$. The CFI for the estimation of $|\beta|$ is therefore, 
\begin{align}
    F_{|\beta|}&[\rho(\beta)] = \frac{4 |\beta|^2 e^{2 |\beta|^2}}{e^{2 \left(|\beta|^2 \cosh (2 z)+\gamma_s t N \right)}-e^{2 |\beta|^2} \cos^2(b)} \left[\sinh(2z) \sin(2\phi) \sin(b) + 2 \sinh ^2(z) \cos(b)\right]^2,
\end{align}
where we defined $b = |\beta|^2 {\sinh(2z)} \sin(2\phi)$. This expression is reported in Eq.~\ref{eq:Noisy_CFI} of the main text. We note that, unlike the QFIM, the CFI is sensitive to single-spin dephasing. In Sec.~\ref{sec:NoisySensing} of the main text we discuss the behavior of $F_{|\beta|}[\rho(\beta)]$ in the limit of vanishing displacement amplitudes and vanishing dephasing rates. Moreover, in Fig.~\ref{fig:noisyspin_phase_insensitive_metrology} of the main text, we use the $\beta \propto \sqrt{N}$ displacement amplitude amplification and $t \propto 1/\sqrt{N}$ scaling of the spin-dependent squeezed state preparation time that are available in trapped ions, and identify parameter regimes where, even in the presence of spin dephasing, the classical Fisher information increases with the number of ions, $N$. Therefore, there are parameter regimes where the metrological utility of our protocol increases with $N$, even with spin dephasing occurring during the spin-dependent squeezing operation. 

\end{document}